\documentclass[normalheadings,abstracton,11pt]{scrartcl}
 
\usepackage{amsmath}
\usepackage{amsfonts}
\usepackage{amssymb}

\usepackage[thmmarks,amsmath,amsthm]{ntheorem}

\usepackage[mathscr]{eucal}
\usepackage{graphicx}

\theoremstyle{plain}
\newtheorem{thm}{Theorem}[section]

\newtheorem{lemma}[thm]{Lemma}
\newtheorem{cor}[thm]{Corollary}

\newtheorem{fact}[thm]{Fact}

\theoremstyle{remark}
\newtheorem{claim}{Claim}

\theoremsymbol{\ensuremath{_\square}}
\renewtheorem*{proof}{Proof\setcounter{claim}{0}}

\theoremsymbol{\ensuremath{_\dashv}}
\newtheorem*{clproof}{Proof}

\let\epsilon=\varepsilon

\newcommand{\m}[1]{\mathcal{#1}}
\newcommand{\mfam}[1]{\mathbf{#1}} %
\newcommand{\mset}[1]{\left\lbrace#1\right\rbrace} %

\newlength{\ppshort}
\newcommand{\hash}{\#}%

\newcommand{\Nat}{ \mathbb{N}}

\newcommand{\Real}{ \mathbb{R}}

\newcommand{\scalp}[1]{\ensuremath{\langle #1\rangle}}

\newcommand{\Tle}{ \ensuremath{\le} }  %
\newcommand{\Tequiv}{ \ensuremath{\equiv} }  %
\newcommand{\thick}[2]{\ensuremath{T_{#1}#2}} %
\newcommand{\str}[2]{\ensuremath{S_{#1}#2}} %
\newcommand{\rank}[1]{\ensuremath{\mathrm{rank}\, #1}} %
\newcommand{\abs}[1]{\ensuremath{\mathrm{abs}(#1)}} %

\newcommand{\tr}[1]{\ensuremath{\mathrm{tr}(#1)}} %

\newcommand{\PP}{\textup{P}}

\newcommand{\eval}[1]{\textup{EVAL}(#1)}
\newcommand{\evaleven}[1]{\textup{EVALeven}(#1)}

\newcommand{\gquad}{\ensuremath{g_{\alpha,\beta,\gamma}}}
\newcommand{\gphi}{\ensuremath{g^R\phi^R}}
\newcommand{\gphiquad}{\ensuremath{g^{R}\phi^R_{a,b,\gamma}}}

\newcommand{\gft}{\ensuremath{\mathbb{F}_2}} %
\renewcommand{\vec}[1]{\mathbf{#1}}

\newcommand{\nodew}{\dot{\omega}}  %

\newenvironment{condition}[1]
{\begin{list}{} {\setlength{\leftmargin}{40pt} 
                \setlength{\itemindent}{-6pt}}
\item[\textbf{#1}]} 
{\end{list}}

\newcommand{\cond}[1]{\textup{\textbf{(#1)}}}

\newcommand{\absent}{*}%
\newcommand{\row}[1]{_{#1,\absent}} %
\newcommand{\col}[1]{_{\absent,#1}} %

\newcommand{\margin}[1]{}
\newcommand{\leslie}[1]{{\tiny \margin{\textbf{L: }#1}}}
\newcommand{\martin}[1]{{\tiny \margin{\textbf{MG: }#1}}}

\numberwithin{equation}{section}

\newcommand{\var}{\textup{var}}

\newcommand{\twinred}[1]{\mathcal{T}(#1)}

\newcommand{\algR}{\mathbb{R}_{\mathbb{A}}}
\begin{document}

\title{A complexity dichotomy for partition functions with mixed signs\thanks{Partly funded by the EPSRC
grant ``The complexity of counting in constraint satisfaction problems''.
and by the Deutsche Forschungsgemeinschaft within the research training group 'Methods for Discrete Structures' (GRK 1408)}}
\author{Leslie~Ann~Goldberg\thanks{Department of Computer Science, University of Liverpool, Liverpool L69 3BX, UK} 
\and Martin~Grohe\thanks{Institut f\"{u}r Informatik, Humboldt-Universit\"at zu Berlin, 10099 Berlin, Germany}
\and Mark~Jerrum\thanks{School of Mathematical Sciences, Queen Mary, University of London, Mile End Road, London E1 4NS, UK}
\and Marc~Thurley\thanks{Institut f\"{u}r Informatik, Humboldt-Universit\"at zu Berlin, 10099 Berlin, Germany}
}
\maketitle

\begin{abstract}
   \emph{Partition functions}, also known as \emph{homomorphism
    functions}, form a rich family of graph invariants that contain
  combinatorial invariants such as the number of $k$-colourings or the
  number of independent sets of a graph and also the partition
  functions of certain ``spin glass'' models of statistical physics
  such as the Ising model.

  Building on earlier work by Dyer and Greenhill~\cite{dyegre00} and
  Bulatov and Grohe~\cite{bulgro05}, we completely classify the
  computational complexity of partition functions. Our main result is
  a dichotomy theorem stating that every partition function is either
  computable in polynomial time or \#P-complete. Partition functions
  are described by symmetric matrices with real entries, and we prove
  that it is decidable in polynomial time in terms of the matrix
  whether a given partition function is in polynomial time or
  \#P-complete. 

  While in general it is very complicated to give an explicit
  algebraic or combinatorial description of the tractable cases, for
  partition functions described by a Hadamard matrices --- these turn
  out to be central in our proofs --- we obtain a simple algebraic
  tractability criterion, which says that the tractable cases are
  those ``representable'' by a quadratic polynomial over the field $\gft$. 
\end{abstract}
\thispagestyle{empty}

\newpage

\pagenumbering{arabic}

\section{Introduction}
 
We study the complexity of a family of graph invariants known as
\emph{partition functions} or \emph{homomorphism functions} (see, for example,
\cite{frelovschri07,lov06,lovschri08+}). Many natural graph
invariants can be expressed as homomorphism functions, among them the number
of $k$-colourings, the number of independent sets, and the number of
nowhere-zero $k$-flows of a graph. The functions also appear as the partition
functions of certain ``spin-glass'' models of statistical physics such as the
Ising model or the $q$-state Potts model.

Let $A\in\mathbb R^{m\times m}$ be a symmetric real matrix with entries
$A_{i,j}$. 
The \emph{partition function $Z_A$} associates with every
graph $G=(V,E)$ the real number
\[
Z_A(G)=\sum_{\xi:V\to[m]}\prod_{\{u,v\}\in E}A_{\xi(u),\xi(v)}.
\]
We refer to the row  and column indices of the
matrix, which are elements of $[m]:=\{1,\ldots,m\}$, as \emph{spins}.
We 
use the term \emph{configuration} to refer to a
mapping $\xi:V\to[m]$ assigning a spin to each vertex of the graph. To avoid
difficulties with models of real number computation, throughtout this paper
we restrict our attention to algebraic numbers. Let $\algR$ denote the set of
algebraic real numbers.\footnote{There is a
  problem with the treatment of real numbers in \cite{bulgro05}, but all
  results stated in \cite{bulgro05}
  are valid for algebraic real numbers. We use a standard representation of
  algebraic numbers by polynomials and standard Turing machines as
  our underlying model of computation.}

Our main result is a dichotomy theorem stating that for every symmetric %
matrix $A\in\algR^{m\times m}$ the partition function $Z_A$ is either
computable in polynomial time or \#P-hard.  This extends earlier results by
Dyer and Greenhill~\cite{dyegre00}, who proved the dichotomy for 0-1-matrices, and
Bulatov and Grohe~\cite{bulgro05}, who proved it for nonnegative
matrices. Therefore, in this paper we are mainly interested in matrices with
negative entries.

\subsection*{Examples}
In the following, let $G=(V,E)$ be a graph with $N$ vertices.
Consider the matrices
\[
S=
\begin{pmatrix}
  0&1\\
  1&1
\end{pmatrix}
\quad
\text{and}
\quad
C_3=
\begin{pmatrix}
  0&1&1\\
  1&0&1\\
  1&1&0
\end{pmatrix}.
\]
It is not hard to see that $Z_S(G)$ is the number of independent sets of a
graph $G$ and $Z_{C_3}(G)$ is the number of 3-colourings of $G$. More
generally, if $A$ is the adjacency matrix of a graph $H$ then $Z_A(G)$ is the
number of homomorphisms from $G$ to $H$. Here we allow $H$ to have loops
and parallel edges;
the entry $A_{i,j}$ in the adjacency matrix is the number of
edges from vertex $i$ to vertex $j$.

Let us turn to matrices with negative entries. Consider
\begin{equation}
  \label{eq:h2}
H_2=\left(\begin{array}{r r}
  1&1\\
  1&-1
\end{array}\right).
\end{equation}
Then $\frac{1}{2}Z_{H_2}(G)+2^{N-1}$ is the number of induced subgraphs of $G$
with an even number of edges. Hence up to a simple transformation, $Z_{H_2}$
counts induced subgraphs with an even number of edges. To see this,
observe
that for every configuration $\xi:V\to[2]$ the 
term $\prod_{\{u,v\}\in E}A_{\xi(u),\xi(v)}$ is
$1$ if the subgraph of $G$ induced by $\xi^{-1}(2)$ has an even number of
edges and $-1$ otherwise. Note that $H_2$ is the simplest nontrivial Hadamard
matrix. Hadamard matrices will play a central role in this paper.  Another
simple example is the matrix
\[
U=\left(\begin{array}{r r}
  1&-1\\
  -1&1
\end{array}\right).
\]
It is a nice exercise to verify that for connected $G$ the number $Z_U(G)$
is $2^N$ if $G$ is Eulerian and $0$ otherwise.

A less obvious example of a counting function that can be expressed in terms
of a partition function is the number of nowhere-zero $k$-flows of a graph. It
can be shown that the number of nowhere-zero $k$-flows of a graph $G$ with $N$
vertices is 
$k^{-N} \cdot Z_{F_k}(G)$, 
where $F_k$ is the $k\times k$~matrix with $(k-1)$s on the diagonal
and $-1$s everywhere else. This is a special case of a more general connection
between partition functions for matrices $A$ with diagonal entries $d$ and off
diagonal entries $c$ and certain values of the Tutte polynomial.
This well-known connection can be derived by establishing certain contraction-deletion
identities for the partition functions.
For example, it follows 
from~\cite[Equations~(3.5.4)]{wel93} and
\cite[Equation~(2.26) and (2.9)]{Sokal05}

\subsection*{Complexity}
Like the complexity of graph polynomials
\cite{bladel07,goljer07,jaeverwel90,lotmak04} and constraint satisfaction
problems
\cite{barkozniv08,bul06,bul08,buldal03,dyegolpat07,golkelpat02,helnes90},
which are both closely related to our partition functions, the complexity of
partition functions has already received quite a bit of a attention. Dyer and
Greenhill \cite{dyegre00} studied the complexity of counting homomorphisms
from a given graph $G$ to a fixed graph $H$ without parallel edges.
(Homomorphisms from $G$ to $H$ are also known as \emph{$H$-colourings} of $G$.)
They proved that the problem is in polynomial time if every connected
component of $H$ is either a complete graph with a loop at every vertex or a
complete bipartite graph, and the problem is \#P-hard otherwise. Note that, in
particular, this gives a complete classification of the complexity of
computing $Z_A$ for symmetric 0-1-matrices $A$.  Bulatov and
Grohe~\cite{bulgro05} extended this to symmetric nonnegative matrices. To
state the result, it is convenient to introduce the notion of a \emph{block}
of a matrix $A$. To define the blocks of $A$, it is best to view $A$ as the
adjacency matrix of a graph with weighted edges; then each non-bipartite
connected component of this graph corresponds to one block and each bipartite
connected component corresponds to two blocks. A formal definition will be
given below. Bulatov and Grohe \cite{bulgro05} proved that computing the
function $Z_A$ is in polynomial time if the row rank of every block of $A$ is
$1$ and $\#P$-hard otherwise. The problem for matrices with negative entries
was left open. In particular, Bulatov and Grohe asked for the complexity of
the partition function $Z_{H_2}$ for the matrix $H_2$ introduced in
\eqref{eq:h2}.  Note that $H_2$ is a matrix with one block of row rank $2$. As
we shall see, $Z_{H_2}$ is computable in polynomial time. Hence the complexity
classification of Bulatov and Grohe does not extend to matrices with negative
entries. Nevertheless, we obtain a dichotomy, and this is our main result.

\subsection*{Results and outline of the proofs}

Our main theorem is the following.

\begin{thm}[Dichotomy Theorem]
  \label{thm:dichotomy}
  Let $A\in\algR^{m\times m}$ be a symmetric matrix. Then the function
  $Z_{A}$ either can be computed in polynomial time or is \#P-hard.

  Furthermore, there is a polynomial time algorithm that, given the matrix
  $A$, decides whether $Z_{A}$ is in polynomial time or \#P-hard.
\end{thm}

Let us call a matrix $A$ \emph{tractable} if $Z_A$ can be computed in
polynomial time and \emph{hard} if computing $Z_A$ is \#P-hard. Then the
Dichotomy Theorem states that every symmetric matrix with entries in $\algR$ is either tractable
or hard.  The classification of matrices into tractable and hard ones can be
made explicit, but is very complicated and does not give any real insights.
Very roughly, a matrix $A$ is tractable if each of its blocks can be written
as a tensor product of a positive matrix of row rank 1 and a tractable
Hadamard matrix. Unfortunately, the real classification is not that simple, but for now
let us focus on tractable Hadamard matrices. Recall that a Hadamard matrix is
a square matrix $H$ with entries from $\{-1,1\}$ such that $H\cdot H^T$ is a
diagonal matrix. Let $H\in\{-1,1\}^{n\times n}$
be a symmetric $n\times n$ Hadamard matrix with
$n=2^k$. Let
$\rho:\gft^k\to[n]$ be a bijective mapping, which we call an \emph{index mapping}. We say that a multivariate
polynomial $h(X_1,\ldots,X_k,Y_1,\ldots,Y_k)$ over $\gft$ 
\emph{symmetrically represents $H$
  with respect to $\rho$} if, for all $\vec x=(x_1,\ldots,x_k),\vec y=(y_1,\ldots,y_k)\in\gft^k$,
it holds that
\[
h(x_1,\ldots,x_k,y_1,\ldots,y_k)=1\iff H_{\rho(\vec x),\rho(\vec y)}=-1.
\]

For example, the \gft-polynomial $h_2(X_1,Y_1)=X_1\cdot Y_1$ 
symmetrically represents the
matrix $H_2$  with respect to the index mapping $\rho(x_1)=x_1+1$. The
\gft-polynomial $h_4(X_1,X_2,Y_1,Y_2)=X_1\cdot Y_2\oplus X_2\cdot Y_1$
symmetrically represents the matrix
\[
H_4=
\left(\begin{array}{r r r r}
  1&1&1&1\\
  1&1&-1&-1\\
  1&-1&1&-1\\
  1&-1&-1&1
\end{array}\right)
\]
with respect to the index mapping $\rho(x_1,x_2)=2\cdot x_1+x_2+1$.
The qualifier ``symmetrically'' 
in ``symmetrically represents'' indicates that the same index mapping
is applied to both $\vec x$ and $\vec y$. We will need to
consider asymmetric representations later.
Note that we can only represent a matrix $H\in\{-1,1\}^{n\times n}$ by an
\gft-polynomial in this way if $n$ is a power of $2$. In this case, for every
index mapping $\rho$ there is a unique $\gft$-polynomial symmetrically representing $h$ with
respect to $\rho$.
We say that $H$ has a \emph{quadratic representation} if there is an index
mapping $\rho$ and an \gft-polynomial $h$ of degree at most 2 that symmetrically represents $H$ with
respect to $\rho$.
Our dichotomy theorem for Hadamard matrices is as follows.

\begin{thm}[Complexity Classification for Hadamard Matrices]
  \label{thm:hadamard}\sloppy
  A symmetric Hada\-mard matrix $H$ is tractable if it has a quadratic
  representation and hard otherwise.
\end{thm}

Hence, in particular, the matrices $H_2$ and $H_4$ are tractable.
The tractability part of Theorem~\ref{thm:hadamard} is an easy consequence of
the fact that counting the number of solutions of a quadratic equation over
$\gft$ (or any other finite field) is in polynomial time (see
\cite{EK90,lidnie97}).
The
difficulty in proving the hardness part is that the degree of a polynomial
representing a Hadamard matrix is not invariant under the choice of the index
mapping $\rho$. However, for \emph{normalised} Hadamard matrices, that is,
Hadamard matrices whose first row and column consists entirely of $+1$s, we
can show that either they are hard or they can be written as an iterated
tensor product of the two simple Hadamard matrices $H_2$ and
$H_4$. This gives us a
canonical index mapping and hence a canonical representation by a quadratic
\gft-polynomial. Unfortunately, we could not find a direct reduction from
arbitrary to normalised Hadamard matrices. 
To get a reduction, we first need
to work with a generalisation of partition functions. If we view the matrix
$A$ defining a partition function as an edge-weighted graph, then this is the
natural generalisation to graphs with edge and vertex weights. Let
$A\in\algR^{m\times m}$ be a symmetric matrix and $D\in\algR^{m\times
  m}$ a diagonal matrix, which may be viewed as assigning the weight $D_{i,i}$
to each vertex $i$. We define the \emph{partition function} $Z_{A,D}$ by
\[
Z_{A,D}(G)=\sum_{\xi:V\to[m]}\prod_{\{u,v\}\in
  E}A_{\xi(u),\xi(v)}\cdot\prod_{v\in V}D_{\xi(v),\xi(v)},
\]
for every graph $G=(V,E)$. As a matter of fact, we need a further
generalisation that takes into account that vertices of even and odd degree
behave differently when it comes to negative edge weights. For a symmetric matrix
$A\in\algR^{m\times m}$ and two diagonal matrices $D,O\in\algR^{m\times
  m}$ we let
\[
Z_{A,D,O}(G)=\sum_{\xi:V\to[m]}\prod_{\{u,v\}\in
  E}A_{\xi(u),\xi(v)}\cdot\prod_{\substack{v\in V\\\deg(v)\text{ is even}}}D_{\xi(v),\xi(v)}\cdot\prod_{\substack{v\in V\\\deg(v)\text{ is odd}}}O_{\xi(v),\xi(v)},
\]
for every graph $G=(V,E)$. We call $Z_{A,D,O}$ the \emph{parity-distinguishing partition
  function} (pdpf) 
defined by $A,D,O$.
We show that the problem of computing $Z_{A,D,O}(G)$ is always either polynomial-time solvable
or \#P-hard, and we call a
triple $(A,D,O)$
\emph{tractable} or \emph{hard} accordingly.
Obviously, if $D=O=I_m$ are identity matrices,
then we have $Z_A=Z_{A,D}=Z_{A,D,O}$.

Returning to the outline of the
proof of Theorem~\ref{thm:hadamard}, we can show that, for
every Hadamard matrix $H$, either $H$ is hard or there is a normalised
Hadamard matrix $H'$ and diagonal matrices $D',O'$ such that computing $Z_H$
is polynomial time equivalent to computing $Z_{H',D',O'}$. Actually, it turns out that we may
assume $D'$ to be an identity matrix and $O'$ to be a diagonal matrix with
entries $0,1$ only. For the normalised matrix $H'$ we have a canonical index
mapping, and we can use this to represent the matrices $D'$ and $O'$ over
$\gft$. Then we obtain a tractability criterion that essentially says that
$(H',D',O')$ is tractable if the representation of $H'$ is quadratic and that
of $O'$ is linear (remember that $D'$ is an identity matrix, which we do not
have to worry about).

For the proof of the Dichotomy Theorem~\ref{thm:dichotomy}, we actually need
an extension of Theorem~\ref{thm:hadamard} that states a dichotomy for parity-distinguishing
partition functions 
$Z_{A,D,O}$, where $A$ is a ``bipartisation'' of a Hadamard matrix
(this notion will be defined later).
The proof
sketched above can be generalised to give this extension. Then to prove the Dichotomy
Theorem, we first reduce the problem of computing $Z_A$ to the problem of
computing $Z_{C}$ for the connected components $C$ of $A$.
The next step is to
eliminate duplicate rows and columns in the matrix, which can be done at the
price of introducing vertex weights. Using the classification theorem for
nonnegative matrices and some gadgetry, from there we get the desired reduction
to parity-distinguishing partition functions for bipartisations of Hadamard matrices.

Let us finally mention that our proof shows that the Dichotomy Theorem
not only holds
for simple partition functions $Z_A$, but also for
vertex-weighted and parity-distinguishing partition functions.

\subsection*{Preliminaries}
Let $A\in\algR^{m\times n}$ be an $(m\times n)$-matrix.
The entries of $A$ are denoted by $A_{i,j}$. The $i$th row of $A$ is denoted by
$A\row i$, and the $j$th column by $A\col j$.  By $\abs{A}$ we denote the
matrix obtained from $A$ by taking the absolute value of each entry in $A$.

Let $I_m$ be the $m\times m$ identity matrix and let
$I_{m;\Lambda}$ be the $m\times m$ matrix that is all zero except
that $I_{j,j}=1$ for $j\in \Lambda$.

The \emph{Hadamard} product~$C$ of two $m\times n$ matrices~$A$ and~$B$,
written $C = A \circ B$, is the $m\times n$ component-wise product 
in which $C_{i,j} = A_{i,j} B_{i,j}$.
$-A$ denotes the Hadamard product of~$A$ and the matrix in which every
entry is~$-1$.

We write $\langle u,v \rangle$ to denote the inner product
(or dot product) of two vectors in $\algR^n$.

Recall  that the \emph{tensor product}
(or \emph{Kronecker product}) of an $r\times s$ matrix~$B$
and an $t\times u$ matrix $C$ is an $r t \times s u$ matrix
$B \otimes C$. For $k\in[r]$, $i\in[t]$, $\ell\in[s]$ and $j\in[u]$,
we have $(B\otimes C)_{(k-1)t+i,(\ell-1)u+j}=B_{k,\ell}C_{i,j}$.
It is sometimes useful to think of the product in terms
of 
$r s$ ``blocks'' or ``tiles'' of size $t\times u$.
$$
B \otimes C = \left(\begin{array}{c c c} 
                       B_{11}C & \ldots & B_{1s}C \\
                        \vdots & \ddots & \vdots \\
                       B_{r1}C & \ldots & B_{rs}C
                   \end{array}\right)
$$

For index sets $I\subseteq [m], J\subseteq[n]$, we let $A_{I,J}$ be the
$(|I|\times|J|)$-\emph{submatrix} with entries $A_{i,j}$ for $i\in I$, $j\in
J$.  The matrix $A$ is \emph{indecomposable} if there are no index sets
$I\subseteq [m], J\subseteq[n]$ such that
$(I,J)\not=(\emptyset,\emptyset)$,
$(I,J)\neq([m],[n])$ and $A_{i,j}=0$
for all $(i,j)\in \big(([m]\setminus I)\times
J\big)\cup\big(I\times([n]\setminus J)\big)$. Note that, in particular, an
indecomposable matrix has at least one nonzero entry. The \emph{blocks} of a
matrix are the maximal indecomposable submatrices.
For
every symmetric matrix
$A\in\mathbb R^{n\times n}$ we can define a graph $G$ with vertex set
$[n]$ and edge set $\big\{\{i,j\}\;\big|\;A_{i,j}\neq 0\big\}$. We call the
matrix $A$
\emph{bipartite} if the graph $G$ is bipartite. We call $A$
\emph{connected} if the graph $G$ is connected. 
The \emph{connected
  components} of $A$ are the maximal submatrices $A_{C,C}$ 
such that $G[C]$, the subgraph of~$G$ induced by $C\subseteq[n]$, is
a connected component. If the connected
component $G[C]$ is not bipartite then $A_{C,C}$ is a block of~$A$. If the connected
component $G[C]$ is bipartite and
contains an edge then $A_{C,C}$ has the form
$
\begin{pmatrix}
  0&B\\
  B^T&0
\end{pmatrix}$, where $B$ is a block of $A$. Furthermore, all blocks of $A$
arise from connected components in this way.

For two Counting Problems $f$ and $g$, we write $f \Tle g$ if there is a
polynomial time Turing reduction from $f$ to $g$. If $f \Tle g$ and $g \Tle f$
holds, we write $f \Tequiv g$. For a symmetric matrix $A$ and diagonal
matrices $D,O$ of the same size, $\eval{A,D,O}$ ($\eval{A,D}$,
$\eval{A}$) denotes  the problem of computing $Z_{A,D,O}(G)$
($Z_{A,D}(G)$, $Z_{A}(G)$, respectively) for an input
graph~$G$ (which need not be a simple graph - it may have loops and/or multi-edges).

\section{Hadamard matrices}
\label{sec:hadamard}

The main focus of this section is to prove Theorem
\ref{thm:bipolar_dichotomy} below which is a strengthened version of
Theorem \ref{thm:hadamard}.  Suppose that $H$ is an $n\times n$
Hadamard matrix and that $\Lambda^R$ and $\Lambda^C$ are subsets
of~$[n]$.  It will be useful to work with the \emph{bipartisation}
$M,\Lambda$ of $H$, $\Lambda^R$ and $\Lambda^C$ which we define as
follows.  Let $m=2n$ and let $M$ be the $m\times m$ matrix defined by
the following equations for $i,j\in[n]$:
$M_{i,j}=0$, $M_{i,n+j}=H_{i,j}$, $M_{n+i,j}=H_{j,i}$, and
$M_{n+i,n+j}=0$.  The matrix $M$ can be broken into four ``tiles'' as
follows.
\[
  M = \left( \begin{array}{c c} 0 & H \\ H^T & 0 \end{array}\right).
\]
Let $\Lambda = \Lambda^R \cup \{n+j\mid j\in \Lambda^C\}$.  Note that
the matrix $I_{m;\Lambda}$ can be decomposed naturally in terms of the
tiles $I_{n;\Lambda^R}$ and $I_{n;\Lambda^C}$.
\[
  I_{m;\Lambda} = \left( \begin{array}{c c} I_{n;\Lambda^R} & 0 \\ 0 & I_{n;\Lambda^C} \end{array}\right).
\]
We identify a set of conditions on $H$, $\Lambda^R$ and~$\Lambda^C$
that determine whether or not the problem $\eval{M,I_m,I_{m;\Lambda}}$
can be computed in polynomial time.  We will see how this implies
Theorem~\ref{thm:hadamard}. 

\paragraph*{The Group Condition.} 

For an $n\times n$ matrix $H$ and
a row index $l\in[n]$, let
\[
G(H,l) : = \mset{
H\row{i}  
\circ H\row{l} \mid  i \in[n]}
\cup
 \mset{
- H\row{i}
\circ H\row{l} \mid  i \in[n]}.
\] 
The \emph{group condition for $H$} is:

\begin{condition}{(GC)}
 For all $l \in[n]$, both
$G(H,l) = G(H,1)$ and $G(H^T,l) = G(H^T,1)$.
\end{condition}

The group condition gets its name from the fact that
the condition implies that $G(H,l)$ is an Abelian group (see
Lemma~\ref{lem:isgroup}).
As all elements of this group have order 2, the group condition gives us some information about
the order of such matrices, as the following lemma (which we prove later in Section~\ref{sec:appendix_hadamard}) shows:

\begin{lemma}\label{lem:bipolar_Hadamard_group}
Let $H$ be an $n \times n$ Hadamard matrix.
If $H$ satisfies \cond{GC} then
$n = 2^k$ for some integer $k$.
\end{lemma}

\paragraph*{The Representability Conditions.}
We describe Hadamard matrices $H$ satisfying \cond{GC} by
$\gft$-polynomials. By Lemma \ref{lem:bipolar_Hadamard_group} these
matrices have order $n=2^k$.  We extend our notion of
``symmetric representation'':  Let $\rho^R:\gft^k\to[n]$ and
$\rho^C:\gft^k\to[n]$ be index mappings (i.e. bijective mappings) and
$X = (X_1,\ldots,X_k)$ and $Y= (Y_1, \ldots,Y_k)$. A polynomial
$h(X,Y)$ over $\gft$ \emph{represents $H$ with respect to $\rho^R$ and
  $\rho^C$} if for all $\vec x,\vec y \in\gft^k$ it holds that
\[
h(\vec x, \vec y)=1\iff H_{\rho^R(\vec x),\rho^C(\vec y)}=-1.
\]
So a symmetric representation is just a representation with
$\rho^R=\rho^C$.  We say that the set $\Lambda^R$ is \emph{linear with
  respect to $\rho^R$} if there is a linear subvectorspace $L^R \subseteq
\gft^k$ a such that $\rho^R(L^R) = \Lambda^R$.  Note that, if
$\Lambda^R$ is linear, then $\vert \Lambda^R \vert = 2^l$ for some
$l\le k$.  We may therefore define a \emph{coordinatisation of $\Lambda^R$
  (with respect to $\rho^R$)} as a linear map $\phi^R: \gft^l
\rightarrow \gft^k$ such that $\phi^R(\gft^l) = L^R$, that is
$\Lambda^R$ is just the image of the concatenated mapping $\rho^R\circ
\phi^R$.  We define the notion of linearity of $\Lambda^C$ with
respect to $\rho^C$ and the coordinatisation of $\Lambda^C$ with respect to
$\rho^C$ similarly.  For a permutation $\pi \in S_k$ we use the
shorthand $X_{\pi}\cdot Y := \bigoplus_{i=1}^k X_{\pi(i)}\cdot Y_i$.

The following conditions stipulate the representability~\cond{R} of $H$ by $\gft$-polynomials, the linearity~\cond{L} of the sets $\Lambda^R$ and $\Lambda^C$, and the appropriate
degree restrictions on the associated polynomials~\cond{D}.

\begin{condition}{(R)} There are index mappings 
$\rho^R:\gft^k\to[n]$ and
$\rho^C:\gft^k\to[n]$ 
and a permutation $\pi \in S_k$ such that (w.r.t. $\rho^R$ and $\rho^C$) the matrix $H$ is represented by a polynomial of the form
\begin{equation}
\label{eq:conditionR}
h(X,Y) = X_{\pi}\cdot Y \oplus g^R(X) \oplus g^C(Y).
\end{equation}
Moreover, if $\Lambda^R$ is non-empty, then $\rho^R(0) \in \Lambda^R$.
Similarly, if $\Lambda^C$ is non-empty, then $\rho^C(0) \in \Lambda^C$.
Finally, if $H$ is symmetric and $\Lambda^R=\Lambda^C$, then $g^R=g^C$ and $\rho^R=\rho^C$.

\end{condition}
\begin{condition}{(L)}
$\Lambda^R$ and $\Lambda^C$ are linear with respect to $\rho^R$ and $\rho^C$ respectively. 
\end{condition}
\begin{condition}{(D)}
Either $\Lambda^R$ is empty or there is a  coordinatisation $\phi^R$ of $\Lambda^R$ w.r.t $\rho^R$ such that the
polynomial $g^R \circ \phi^R$ has degree at
most $2$. 
Similarly,
either $\Lambda^C$ is empty or there is a  coordinatisation $\phi^C$ of $\Lambda^C$ w.r.t $\rho^C$ such that the
polynomial $g^C \circ \phi^C$ has degree at
most $2$. 
Finally, if $H$ is symmetric and $\Lambda^R=\Lambda^C$ is nonempty
then $\phi^R=\phi^C$.
\end{condition}
Actually, it turns out that condition (D) is invariant under the
choice of the coordinatisations $\phi^R,\phi^C$. However, the
conditions are not invariant under the choice of the representation
$\rho^R,\rho^C$, and this is a major source of technical
problems.
 
Before we can apply the conditions \cond{R}, \cond{L} and \cond{D}
we deal with one technical issue.  
Let $H$ be an $n \times n$ Hadamard matrix 
and let
$\Lambda^R,\Lambda^C \subseteq  [n]$ be subsets of indices.
Let $M,\Lambda$ be the bipartisation of $H$, $\Lambda^R$ and $\Lambda^C$. We say that $H$ is \emph{positive} for $\Lambda^R$ and $\Lambda^C$
if there is an entry $H_{i,j}=+1$
such that (1) $i\in\Lambda^R$ or $\Lambda^R=\emptyset$,
(2) $j\in\Lambda^C$ or $\Lambda^C=\emptyset$, and
(3) If $H$ is symmetric and $\Lambda^R=\Lambda^C$ then $i=j$.
Otherwise, note that
$-H$ is positive for $\Lambda^R$ and $\Lambda^C$.
Since
$Z_{M,I_m,I_{m;\Lambda}}(G) = {(-1)}^{|E(G)|}Z_{-M,I_m,I_{m;\Lambda}}(G)$, the
problems
$\eval{M,I_m,I_{m;\Lambda}}$ and
$\eval{-M,I_m,I_{m;\Lambda}}$ 
have equivalent complexity,
so we lose no generality by
restricting attention to the
positive case, which is helpful for a technical reason.

We can now state the theorem which is proved in this section.

\begin{thm}\label{thm:bipolar_dichotomy}
Let $H$ be an $n \times n$ Hadamard matrix 
and let
$\Lambda^R,\Lambda^C \subseteq  [n]$ be subsets of indices.
Let $M,\Lambda$ be the bipartisation of $H$, $\Lambda^R$ and $\Lambda^C$ and let $m=2n$.
If $H$ is positive for $\Lambda^R$ and $\Lambda^C$
then $\eval{M,I_m,I_{m;\Lambda}}$ is polynomial-time computable if, and only if, $H$ $\Lambda^R$
and $\Lambda^C$ satisfy the group condition \cond{GC} and conditions \cond R, \cond L, and \cond D.
Otherwise $\eval{M,I_m,I_{m;\Lambda}}$ is $\#\PP$-hard.
If  $H$ is not positive for $\Lambda^R$ and $\Lambda^C$
then $\eval{M,I_m,I_{m;\Lambda}}$ is polynomial-time computable if, and only if, $-H$ $\Lambda^R$
and $\Lambda^C$ satisfy the group condition \cond{GC} and conditions \cond R, \cond L, and \cond D.
Otherwise $\eval{M,I_m,I_{m;\Lambda}}$ is $\#\PP$-hard.
There is a polynomial-time algorithm that takes input $H$, $\Lambda^R$
and $\Lambda^C$ and decides 
whether $\eval{M,I_m,I_{m;\Lambda}}$ is polynomial-time computable
or $\#\PP$-hard.
\end{thm}

The theorem is proved using a sequence of lemmas.
Proof sketches of these lemmas will be given in this section and full proofs will be given later
in Section~\ref{sec:appendix_hadamard}.

\begin{lemma}
[Group Condition Lemma]
\label{lem:bipolar_hard_nonGC}
Let $H$ be an $n \times n$ Hadamard matrix 
and let
$\Lambda^R,\Lambda^C \subseteq  [n]$ be subsets of indices.
Let $M,\Lambda$ be the bipartisation of $H$, $\Lambda^R$ and $\Lambda^C$ and let $m=2n$.
If $H$ does not satisfy \cond{GC} then $\eval{M, I_m, I_{m;\Lambda}}$ is $\#\PP$-hard.
There is a polynomial-time algorithm that takes  
determines whether $H$ satisfies
\cond{GC}.
\end{lemma}
 
\noindent\textit{Proof sketch.} 
For any integer $p$ and a symmetric non-negative
matrix $C^{[p]}$, which depends upon~$H$,
the proof uses gadgetry to 
transform an  
input to $\eval{C^{[p]}}$ into an input to $\eval{M, I_m, I_{m;\Lambda}}$.
The fact that $H$ does not satisfy \cond{GC} is used to 
show that, as long as~$p$
is sufficiently large with respect to~$M$,
then $C^{[p]}$ has a block of rank greater than one. By a result
of Bulatov and Grohe, $\eval{C^{[p]}}$ is \#P-hard, so
$\eval{M, I_m, I_{m;\Lambda}}$ is $\#\PP$-hard.

\martin{we have to fix the margins, here and elsewhere}
\begin{lemma}[Polynomial Representation Lemma]\label{lem:poly_rep}
Let $H$ be an $n \times n$ Hadamard matrix and $\Lambda^R,\Lambda^C \subseteq  [n]$ subsets of indices.
Suppose that $H$ satisfies \cond{GC} and that $H$ is positive for
$\Lambda^R$ and $\Lambda^C$.
Then the Representability Condition \cond{R} is satisfied.  
There is a polynomial-time algorithm that  
computes the representation. 
\end{lemma}

\noindent\textit{Proof sketch.} 
The representation is constructed inductively.
First, permutations are used to transform $H$
into a normalised matrix $\hat{H}$, that is, a
Hadamard matrix $\hat{H}$ whose first row and column consist 
entirely of $+1$s,
which still satisfies \cond{GC}. 
We then show that there is a permutation of $\hat{H}$ which can be
expressed as the tensor product of a simple Hadamard matrix (either $H_2$ or $H_4)$ and a smaller normalised symmetric Hadamard matrix~$H'$.
By induction, we construct a representation for $H'$  and use this to
construct a representation for the normalised matrix $\hat{H}$
of the form $X_\pi \cdot Y$ for a permutation $\pi\in S_k$.
We use this to construct a representation for $H$.

\begin{lemma}[Linearity Lemma]\label{lem:bipolar_lam_linear}
Let $H$ be an $n \times n$ Hadamard matrix and $\Lambda^R,\Lambda^C \subseteq  [n]$ subsets of indices.
Let $M,\Lambda$ be the bipartisation of $H$, $\Lambda^R$ and $\Lambda^C$ and let $m=2n$.
Suppose that 
\cond{GC} and  \cond{R} are satisfied.
Then the problem $\eval{M,I_m,I_{m;\Lambda}}$ is $\#\PP$-hard unless the Linearity condition \cond{L} holds.
There is a polynomial-time algorithm that 
determines whether \cond{L} holds.
\end{lemma}

\noindent\textit{Proof sketch.} 
For a symmetric non-negative matrix $C$, which depends upon~$H$,
the proof uses gadgetry to transform an input to $\eval{C,I_m,I_{m;\Lambda}}$ to
an input of $\eval{M,I_m,I_{m;\Lambda}}$.
By  \cond{R},
there are bijective index mappings 
$\rho^R:\gft^k\to[n]$ and
$\rho^C:\gft^k\to[n]$ 
and a permutation $\pi \in S_k$ such that (w.r.t. $\rho^R$ and $\rho^C$) the matrix $H$ is represented by a polynomial of the appropriate form.
Let 
$\tau^R$ be the inverse of $\rho^R$ and $\tau^C$ be the inverse of $\rho^C$. 
Let $L^C=\tau^C(\Lambda^C)$ and $L^R=\tau^R(\Lambda^R)$.
We show that
either
$\eval{C,I_m,I_{m;\Lambda}}$ is \#P-hard or \cond{L} is satisfied.
In particular, the assumption that $\eval{C,I_m,I_{m;\Lambda}}$.
is not \#P-hard means that its blocks all have rank~1 by the result of Bulatov and Grohe.
We use this fact to show that $L^R$ is a linear subspace of $\Lambda^R$ and
that $L^C$ is a linear subspace of $L^C$.
To show that $L^R$ is a linear space of $\Lambda^R$, we use $L^R$ to construct an appropriate linear subspace
and compare Fourier coefficients to see that it is in fact $L^R$ itself.

\begin{lemma}[Degree Lemma]\label{lem:degree_two}
Let $H$ be an $n \times n$ Hadamard matrix and $\Lambda^R,\Lambda^C \subseteq  [n]$ subsets of indices.
Let $M,\Lambda$ be the bipartisation of $H$, $\Lambda^R$ and $\Lambda^C$ and let $m=2n$.
Suppose that  
\cond{GC},\cond{R} and \cond{L} are satisfied. 
Then $\eval{M,I_m,I_{m;\Lambda}}$ is $\#\PP$-hard unless the Degree Condition \cond{D} holds.
There is a polynomial-time algorithm that determines whether 
\cond{D} holds.
\end{lemma}

\noindent\textit{Proof sketch.} 
For any (even) integer $p$  
and a symmetric non-negative matrix $C^{[p]}$, which depends upon~$H$,
the proof uses gadgetry to transform an input to
$\eval{C^{[p]}}$ into an input to $\eval{M, I_m, I_{m;\Lambda}}$.
Using the representation of $H$, a coordinatisation $\phi^R$ with respect to $\Lambda^R$,
and a coordinatisation $\phi^C$ with respect to $\Lambda^C$,
some of the entries $C^{[p]}_{a,b}$ of the matrix $C^{[p]}$ may be expressed
as sums, over elements in $\gft^\ell$, for some $\ell$, of appropriate powers of $-1$.
We study properties of polynomials $g(X_1,\ldots,X_k)\in \gft[X_1,\ldots,X_k]$, 
discovering that the number of roots of 
a certain polynomial $\gquad(X_1,\ldots,X_k)$, which is derived from $g(X_1,\ldots,X_k)$,
depends upon the degree of~$g$.
From this we can show that if \cond{D} does not hold then there is an even $p$ such
that $\eval{C^{[p]}}$
is \hash P-hard.

\leslie{COULD DO: We chose the names $\Lambda^R$
and $L^R$ before introducing the roman/greek distinction, so they should really be the other way around, but fixing this could introduce errors (so I don't think we should fix it)}

\begin{proof}[Proof of Theorem \ref{thm:bipolar_dichotomy}]

By the equivalence of the problems $\eval{M,I_m,I_{m;\Lambda}}$ and
$\eval{-M,I_m,I_{m;\Lambda}}$ we can assume that $H$ is
positive for $\Lambda^R$ and $\Lambda^C$.
The hardness part follows directly from the Lemmas above. We shall give the proof for the tractability part.
Given $H$, $\Lambda^R$ and $\Lambda^C$ satisfying \cond{GC}, \cond{R}, \cond{L} and \cond{D},
we shall 
show how to compute  $Z_{M,I_m,I_{m;\Lambda}}(G)$ for an input graph $G$ in polynomial time.

Note first that $Z_{M,I_m,I_{m;\Lambda}}(G) = 0$ unless $G$ is bipartite. If $G$ has connected components $G_1,\ldots G_c$, then $$Z_{M,I_m,I_{m;\Lambda}}(G) = \prod_{i=1}^c Z_{M,I_m,I_{m;\Lambda}}(G_i).$$
Therefore, it suffices to give the proof for connected bipartite graphs.
Let $G= (V,E)$ be such a graph with vertex bipartition $U \,\dot \cup\, W = V$. 
Let $V_o \subseteq V$ be the set of odd-degree vertices in $G$ and let $U_o = W \cap V_o$ and 
$W_o = W \cap V_o$ be the corresponding subsets of $U$ and $W$. 
Let $U_e=U\setminus U_o$ and $W_e=W\setminus W_o$.
We have
\begin{eqnarray*}
Z_{M,I_m,I_{m;\Lambda}}(G) &=& \sum_{\xi:V \rightarrow [m]} \prod_{\{u,w\} \in E} M_{\xi(u),\xi(w)} \prod_{v \in V_o } (I_{m;\Lambda})_{\xi(v),\xi(v)} 
= \sum_{\substack{\xi:V \rightarrow [m] \\ \xi(V_o) \subseteq \Lambda}} \prod_{\{u,w\} \in E} M_{\xi(u),\xi(w)}.
\end{eqnarray*}
\newcommand{\zback}{Z^{\leftarrow}}
\newcommand{\zforth}{Z^{\rightarrow}}
As $G$ is bipartite and connected this sum splits into $Z_{M,I_m,I_{m;\Lambda}}(G) = \zforth  +  \zback$
for values 
$$
\zforth = \sum_{\substack{\xi:U \rightarrow [n] \\ \xi(U_o) \subseteq \Lambda^R}}
    \sum_{\substack{\zeta:W \rightarrow [n] \\ \zeta(W_o) \subseteq \Lambda^C}}
    \prod_{\substack{\{u,w\} \in E\\ u \in U}} 
H_{\xi(u),\zeta(w)}
\quad \text{ and } \quad
\zback = \sum_{\substack{\xi:U \rightarrow [n] \\ \xi(U_o) \subseteq \Lambda^C}}
    \sum_{\substack{\zeta:W \rightarrow [n] \\ \zeta(W_o) \subseteq \Lambda^R}}
    \prod_{\substack{\{u,w\} \in E\\u \in U}} H_{\zeta(w),\xi(u)}
$$
We will show how to compute $\zforth$. The computation of the value $\zback$ is similar.

Fix configurations $\xi : U \rightarrow [n]$ and $\zeta : W \rightarrow [n]$ and let $\rho^R,\rho^C$ be the index mappings and $h$ the $\gft$-polynomial representing $H$ as given in condition \cond{R}. 
Let $\tau^R$ be the inverse of $\rho^R$ and let $\tau^C$ be the inverse of $\rho^C$.
Let $L^R=\tau^R(\Lambda^R)$ and $L^C=\tau^C(\Lambda^C)$.
Then $\xi$ and $\zeta$ induce a configuration  $\varsigma: V \rightarrow \gft^k$ defined by
$$\varsigma(v) := \left\lbrace \begin{array}{l l}
                                \tau^R ( \xi(v)) &, \text{ if } v \in U\\
                                \tau^C (\zeta(v)) &, \text{ if } v \in W                              \end{array}\right.
$$
which implies, for all $u \in U, w \in W$ that $h(\varsigma(u),\varsigma(w)) = 1$ iff $H_{\xi(u),\zeta(w)} = -1$.
Let $\phi^R$ and $\phi^C$ be  coordinatisations of $\Lambda^R$ and $\Lambda^C$ w.r.t. $\rho^R$ and $\rho^C$ satisfying \cond{L} and \cond{D}. 
We can simplify
\begin{eqnarray*}
\zforth &=& \sum_{\substack{\xi:U \rightarrow [n] \\ \xi(U_o) \subseteq \Lambda^R}}
    \sum_{\substack{\zeta:W \rightarrow [n] \\ \zeta(W_o) \subseteq \Lambda^C}}
    \prod_{\substack{\{u,w\} \in E\\u \in U}} (-1)^{h( \tau^R (\xi(u)), \tau^C ( \zeta(w)))} \\
&=& \sum_{\substack{\varsigma: V \rightarrow \gft^k\\ 
\varsigma(U_o) \subseteq L^R
            \\ 
\varsigma(W_o) \subseteq L^C
}}
    (-1)^{\bigoplus_{\{u,w\} \in E: u \in U} h(\varsigma(u),\varsigma(w))}
\end{eqnarray*}
Define, for $a \in \gft$, sets
\begin{equation}
   s_a := \left\vert \left\lbrace{\varsigma: V \rightarrow \gft^k \mid  \varsigma(U_o) \subseteq
L^R,\;
                                             \varsigma(W_o) \subseteq L^C,\;
         \bigoplus_{\substack{\{u,w\} \in E\\ u \in U}} h(\varsigma(u),\varsigma(w)) =  a}
         \right\rbrace \right\vert.
\end{equation}

Then   $\zforth = s_0 - s_1$. Therefore, it
remains to show how to compute the values $s_a$.
Define, for each $v \in V$, a tuple $X^v = (X^v_1,\ldots,X^v_k)$ and let $h_G$ be the $\gft$-polynomial
\begin{equation}\label{eq:def_h_G}
h_G := 
\bigoplus_{\substack{\{u,w\} \in E\\u \in U}} h(X^u,X^w) 
= \bigoplus_{\substack{\{u,w\} \in E\\u \in U}} (X^u)_\pi \cdot X^w \oplus \bigoplus_{u \in U_o} g^R(X^u) \oplus \bigoplus_{w \in W_o} g^C(X^w). \\
\end{equation}
Here the second equality follows from the definition of the polynomial $h$ given in condition \cond{R} and the fact that the terms $g^R(X^u)$ and $g^C(X^w)$ in the definition of $h$ appear exactly $\deg(u)$ and $\deg(w)$ many times in $h_G$. Therefore, these terms cancel for all even degree vertices.

\newcommand{\vtvar}[1]{\mathbf{X}(#1)}

Let $\var(h_G)$ denote the set of variables in $h_G$ and for mappings $\chi: \var(h_G) \rightarrow \gft$ we use the expression $\chi(X^v) := (\chi(X^v_1),\ldots,\chi(X^v_k))$ as a shorthand and define the $\gft$-sum $h_G(\chi) := \bigoplus_{\{u,w\} \in E: u \in U} h(\chi(X^u),\chi(X^w))$. We find that $s_a$ can be expressed by
\begin{equation}\label{eq:s_a_second}
   s_a = \left\vert \left\lbrace{ \chi: \var(h_G) \rightarrow \gft \mid  \begin{array}{l l}
                                   \chi(X^u) \in L^R & \text{ for all } u\, \in\,U_o,\\
                                   \chi(X^w) \in L^C & \text{ for all } w \in W_o, \end{array}
         \;h(\chi) =  a)}
         \right\rbrace \right\vert 
\end{equation}

By equation \eqref{eq:s_a_second} we are interested only in those assignments $\chi$ of the variables of $h_G$ which satisfy $\chi(X^u) \in  L^R$ and $\chi(X^w) \in  L^C$ for all $u \in U_o$ and $w\in W_o$. With $\vert  \Lambda^R \vert = 2^{\ell^R}$ and $\vert \Lambda^C \vert = 2^{\ell^C}$ for some appropriate $\ell^R, \ell^C$, we introduce variable vectors $Y^u = (Y^u_1,\ldots,Y^u_{\ell^R})$ and $Z^w = (Z^w_1,\ldots,Z^w_{\ell^C})$ for all $u \in U_o$ and $w \in W_o$.
If $u\in U_o$ or $w\in W_o$ then 
we can express the term $  (X^u)_\pi \cdot X^w$ in $h_G$ in 
terms of these new variables. In particular,
let
\begin{eqnarray*}
h''_G &=& 
\phantom{\oplus} \bigoplus_{\substack{\{u,w\} \in E\\u \in U_o,w\in W_o}} (\phi^R(Y^u))_\pi \cdot \phi^C(Z^w)
\oplus
\bigoplus_{\substack{\{u,w\} \in E\\u \in U_e,w\in W_e}}  (X^u)_\pi \cdot X^w  \\
& & \oplus
\bigoplus_{\substack{\{u,w\} \in E\\u \in U_e,w\in W_o}} (X^u)_\pi \cdot \phi^C(Z^w) 
\oplus
\bigoplus_{\substack{\{u,w\} \in E\\u \in U_o,w\in W_e}} (\phi^R(Y^u) )_\pi \cdot X^w.
\end{eqnarray*}

Let
\begin{equation}%
h'_G =  h''_G \oplus \bigoplus_{u \in U_o} g^R(
\phi^R(Y^u)) \oplus \bigoplus_{w \in W_o} g^C(\phi^C(Z^w)) \\
\end{equation}

We therefore have 
\begin{equation} s_a = \left\vert \left\lbrace{ \chi: \var(h'_G) \rightarrow  \gft \mid 
h'_G(\chi) = a)} 
\right\rbrace \right\vert 
.\end{equation}

By condition \cond{D}, the polynomials $g^R \circ \phi^R$ and $g^C \circ \phi^C$ are of degree at most $2$ and therefore $h'_G$ is a polynomial of degree at most $2$. Furthermore, we have expressed $s_a$ as the number of solutions to a polynomial equation over $\gft$. Therefore, the proof now follows by the following well-known fact.
\begin{fact}\label{fact:gft_count}
The number of solutions to polynomial equations of degree at most $2$ over $\gft$ can be computed in polynomial time.
\end{fact}
This is a direct consequence of Theorems 6.30 and 6.32 in \cite{lidnie97} (see also \cite{EK90}).
\end{proof}

\section{The General Case}\label{sec:general_case}

In this section we will prove Theorem \ref{thm:dichotomy}. Before we can give the proof some further results have to be derived, which then enable us to extend Theorems \ref{thm:hadamard} and \ref{thm:bipolar_dichotomy}.
It will be convenient to focus on connected components. This is expressed by the following lemma, which will
be proved later in Section~\ref{sec:appendix_general_case}.

\begin{lemma}\label{lem:component_pinning}
Let $A$ be a symmetric real-valued matrix with components $A_1, \ldots, A_c$. Then the following holds
\begin{itemize}
\item[(1)] If $\eval{A_i}$ is $\#\PP$-hard for some $i \in [c]$ then $\eval{A}$ is $\#\PP$-hard.
\item[(2)] If $\eval{A_i}$ is \textup{PTIME} computable for all $i \in [c]$ then $\eval{A}$ is \textup{PTIME} computable.  
\end{itemize}
\end{lemma}
 
Recall that for each connected symmetric matrix $A$ there is a block
$B$ such that either $A=B$ or, up to permutation of the rows and columns,
$
  A = \left(\begin{array}{c c} 0 & B \\ B^T & 0 \end{array}\right).
$
We call $B$ the block \emph{underlying} $A$.
For such connected $A$ we furthermore see that the evaluation problem is either $\#\PP$-hard or we can reduce it to
the evaluation problem on bipartisations of Hadamard matrices.
This is expressed in the following lemma, which will be proved later in Section~\ref{sec:appendix_general_case}.
 
\begin{lemma}\label{lem:decomp}
Suppose that $A$ is
a symmetric connected matrix.
Then either $\eval{A}$ is $\#\PP$-hard or the following holds.
\begin{itemize}
 \item[(1)] If $A$ is not bipartite there is a symmetric $r \times r$ Hadamard matrix $H$ and a set $\Lambda^R \subseteq [r]$ such that
\leslie{Used $\Lambda^R$ rather than $\Lambda$ here to avoid
clashes}
       $$
            \eval{A} \Tequiv \eval{H,I_r,I_{r;\Lambda^R}}.
        $$ 
 \item[(2)] If $A$ is bipartite then there is an $r \times r$ Hadamard matrix $H$, sets $\Lambda^R,\Lambda^C \subseteq [r]$  and a bipartisation $M,\Lambda$ of $H, \Lambda^R$ and $\Lambda^C$ such that
        $$
            \eval{A} \Tequiv \eval{M,I_{2r},I_{2r;\Lambda}}.
        $$
\end{itemize}

Furthermore it can be decided in time polynomial in the size of $A$ which of
the three alternatives ($\#\PP$-hardness, (1), or (2)) holds.\martin{rephrased this}
\end{lemma}

We are now able to prove the main Theorem.

\begin{proof}[Proof of Theorem \ref{thm:dichotomy}]
Given a symmetric matrix $A \in \algR^{m \times m}$. By Lemma \ref{lem:component_pinning} we may assume that the matrix $A$ is connected. By Lemma \ref{lem:decomp}, Theorem \ref{thm:bipolar_dichotomy} and Corollary \ref{cor:nonbip} the problem $\eval{A}$ is either polynomial time computable or $\#\PP$-hard. The existence of a polynomial time algorithm for deciding which of the two possibilities holds, given a matrix $A$, follows directly by these results.
\end{proof}

\section{Outline of the paper}

The rest of the paper is organised as follows.
Section~\ref{sec:related} describes some generalisations of the partition function
evaluation problem, including related work and open problems.
The remaining sections contain the proofs of the lemmas which have been stated
without proof. In particular, Section~\ref{sec:tech} develops some technical tools which we will
use. Section~\ref{sec:appendix_hadamard} proves the lemmas that are stated in Section~\ref{sec:hadamard}.
Finally, Section~\ref{sec:appendix_general_case} proves the lemmas that are
stated in Section~\ref{sec:general_case}.

\section{Related Work}
\label{sec:related}
There are several natural directions in which the work in this paper could be extended.
First,  the matrix $A$ could be extended to include algebraic \emph{complex} numbers and not merely
algebraic real numbers. This extension has been carried out, subsequent to 
this paper,
in an impressive 111-page paper by Cai, Chen and Lu~\cite{CCL}.
The work could also be extended by allowing the matrix $A$ to be \emph{asymmetric}.
A recent breakthrough by Bulatov~\cite{Bulatov} establishes the existence of a dichotomy
theorem for counting satisfying assignments in constraint satisfaction.
This implies that a dichotomy exists for the case in which $A$ is a 0-1 matrix
(which is not necessarily symmetric) --- in this case computing the partition function
corresponds to counting homomorphisms to a directed graph, in particular to the directed
graph with adjacency matrix $A$. Bulatov's dichotomy is not known to be \emph{effective} in the
sense that it is not known 
to be decidable given a matrix $A$ whether it is tractable or not.
An effective dichotomy was given by Dyer, Goldberg and Paterson~\cite{DGP} for the special
case in which the directed graph with adjacency matrix $A$ is \emph{acyclic} but no
effective dichotomy is currently known for the case of general 0-1 matrices $A$.
A generalisation of Bulatov and Grohe's dichotomy 
for symmetric non-negative matrices to
symmetric non-negative functions of \emph{arbitrary arity} was
given recently by Dyer, Goldberg, and Jerrum \cite{HyperHom}.
However, nothing is known about how how to handle functions of larger arity in the presence of mixed signs.

\section{Technical Tools}
\label{sec:tech}

\subsection{Stretchings and Thickenings}

We introduce some fundamental relations which will be used in most of our
reductions.  Let $G=(V,E)$ be a graph. The \emph{s-stretch} of $G$ is the
graph $\str{s}{G}$ obtained from $G$ by replacing each edge by a path on $s$
edges. The \emph{$t$-thickening} of $G$ is the graph $\thick{t}{G}$ obtained
from $G$ by replacing each edge by $t$ parallel edges. Let $A^{(t)}$ denote
the matrix obtained from $A$ by taking each of its entries to the power of
$t$.

\begin{lemma}[\cite{dyegre00}]
  For a symmetric matrix $A \in \algR^{m \times m}$ and a diagonal $m \times
  m$ matrix $D$ we have, for all $s,t \in \Nat$
  \[
  \eval{A(DA)^{s-1},D} \Tle  \eval{A,D} \quad \text{ and } \quad \eval{A^{(t)},D} \Tle  \eval{A,D} 
  \]
These reducibilities hold as
\[
  Z_{A(DA)^{s-1},D}(G) = Z_{A,D}(S_sG) \quad \text{ and } \quad Z_{A^{(t)},D}(G) = Z_{A,D}(T_tG).
  \]
\end{lemma}

\subsubsection{Twin Reduction} 
We need some extensions of Lemma 3.5 in \cite{dyegre00}.  For a symmetric $m
\times m$ matrix $A$ we say that two rows $A\row i$ and $A \row j$ are
\emph{twins} iff $A \row i = A \row j$. This induces an equivalence relation
on the rows (and by symmetry on the columns) of $A$. Let $I_1, \ldots I_n$ be
a partition of the row indices of $A$ according to this relation. The
\emph{twin-resolvent} of $A$ is the matrix defined, for all $i,j \in [n]$, by
\[
\twinred A_{i,j} := A_{\mu,\nu} \text{ for some } \mu \in I_i, \nu \in I_j.
\]
The definition of the classes $I_i$ implies that $A_{\mu,\nu} = A_{\mu',\nu'}$ for all $\mu,\mu' \in I_i$ and $\nu,\nu' \in I_j$ and therefore the matrix $\twinred A$ is well-defined.

The above definition furthermore give rise to a mapping $\tau: [m] \rightarrow
[n]$ defined by $\mu \in I_{\tau(\mu)}$ that is $\tau$ maps $\mu \in [m]$ to
the class $I_j$ it is contained in.  Therefore, we have $\twinred
A_{\tau(i),\tau(j)} = A_{i,j}$ for all $i,j \in [m]$. We call $\tau$ the
\emph{twin-resolution mapping} of $A$.

\begin{lemma}[Twin Reduction Lemma]\label{lem:twin_red}
  Let $A$ be a symmetric $m \times m$ matrix and $D$ a diagonal $m \times m$
  matrix of vertex weights. Let $I_1,\ldots,I_n$ be a partition of the row
  indices of $A$ according to the twin-relation. Then
  \[
  Z_{A,D}(G) = Z_{\twinred A,\Delta}(G) \text{ for all graphs } G
  \]
  where $\Delta$ is a diagonal $n \times n$ matrix defined by $\Delta_{i,i} =
  \sum_{\nu \in I_i} D_{\nu,\nu}$ for all $i \in [n]$.
\end{lemma}

\begin{proof}
Let $\tau$ be the twin-resolution mapping of $A$. Then
\begin{eqnarray*}
Z_{A,D}(G) &=& \sum_{\xi:V \rightarrow [m]} \prod_{\{u,v\} \in E} A_{\xi(u),\xi(v)} \prod_{v \in V} D_{\xi(v),\xi(v)} \\
&=& \sum_{\xi:V \rightarrow [m]} \prod_{\{u,v\} \in E} \twinred A_{\tau\circ\xi(u),\tau \circ \xi(v)} \prod_{v \in V} D_{\xi(v),\xi(v)}
\end{eqnarray*}
where the second equality follows from the definition of $\tau$. As for all $\xi: V \rightarrow [m]$ we have $\tau \circ \xi : V \rightarrow [n]$, we can partition the $\xi$ into classes according to their images under concatenation with $\tau$ and obtain:
\begin{eqnarray*}
Z_{A,D}(G) &=& \sum_{\psi:V \rightarrow [n]}\sum_{\substack{\xi:V \rightarrow [m] \\ \tau \circ \xi = \psi}} \prod_{\{u,v\} \in E} \twinred A_{\psi(u), \psi(v)} \prod_{v \in V} D_{\xi(v), \xi(v)} \\
           &=& \sum_{\psi:V \rightarrow [n]} \prod_{\{u,v\} \in E} \twinred A_{\psi(u), \psi(v)} \left(\sum_{\substack{\xi:V \rightarrow [m] \\ \tau \circ \xi = \psi}} \prod_{v \in V} D_{\xi(v), \xi(v)}\right)
\end{eqnarray*}
Fix some $\psi: V \rightarrow [n]$. For $\xi: V \rightarrow [m]$ we have $\tau
\circ \xi = \psi$ if and only if $\psi^{-1}(\{i\}) =
\xi^{-1}(I_i)$  for all $i \in
[n]$. Define $V_i := \psi^{-1}(\{i\})$ for all $i \in [n]$ which yields a
partition of $V$.  Thus
\begin{eqnarray*}
\sum_{\substack{\xi:V \rightarrow [m] \\ \tau \circ \xi = \psi}} \prod_{v \in V} D_{\xi(v),\xi(v)}
&=&\sum_{\substack{\xi: V \rightarrow [m] \\ \forall \; i \in [n]:\;\xi(V_i) \subseteq I_i }} \prod_{v \in V} D_{\xi(v),\xi(v)} \\
&=& \prod_{i=1}^n \sum_{\xi_i:V_i \rightarrow I_i} \prod_{v \in V_i}  D_{\xi(v), \xi(v)}\\
 &=&\prod_{i=1}^n \prod_{v \in V_i} \sum_{\nu \in I_i}   D_{\nu, \nu} \\
&=&\prod_{v \in V} \Delta_{\psi(v), \psi(v)}
\end{eqnarray*}
Hence 
\begin{eqnarray*}
Z_{A,D}(G) &=& \sum_{\psi:V \rightarrow [n]} \prod_{\{u,v\} \in E} \twinred A_{\psi(u), \psi(v)} 
                        \prod_{v \in V} \Delta_{\psi(v), \psi(v)}.
\end{eqnarray*}
\end{proof}

\subsection{Basic Tractability and $\#\PP$-hardness}

The following Lemma is a straightforward extension of Theorem 6 in \cite{bulgro05}.
\begin{lemma}\label{lem:rank1_FP}
Let $A \in \algR^{m\times m}$ be a symmetric matrix and $D$ a diagonal $m\times m$ matrix.
If each component of $A$ either has row rank $1$ or is bipartite and has rank $2$ then $\eval{A,D}$ is polynomial time computable.
\end{lemma}
\begin{proof}
Let $G=(V,E)$ be a given graph with components $G_1, \ldots, G_c$ and let $A_1, \ldots,A_l$ be the components of $A$ and $D_{1},\ldots,D_{l}$ the submatrices of $D$ corresponding to these components. 
Then
$$
Z_{A,D}(G) = \prod_{i=1}^c \sum_{j=1}^l Z_{A_j,D_j}(G_i).
$$
Therefore the proof follows straightforwardly from the special case of connected $G$ and $A$.
Assume therefore that both $G$ and $A$ are connected.

We will prove the following claim, which holds for directed graphs.

\begin{claim}\label{cl:polytime_rank1}
  Let $B^{m \times m}$ be a (not necessarily symmetric) matrix of row rank $1$
  and $D'$ a diagonal matrix. Then for every directed graph $G$ the value
  \[
  Z^*_{B,D'}(G) = \sum_{\xi : V \rightarrow [m]} \prod_{(u,v) \in E} B_{\xi(u),\xi(v)} \prod_{v \in V}D'_{\xi(v),\xi(v)}
\]
can be computed in polynomial time. 
\end{claim}

\proof
Let $G = (V,E)$ be a directed graph and for every vertex $v \in V$ denote by $\text{outdeg}(v)$ and $\text{indeg}(v)$ the number of outgoing and incoming edges incident with $v$. There are vectors $a, b \in \algR^m$ such that $B = a b^T$.
Then, for every configuration $\xi: V \rightarrow [m]$,
$$
\prod_{(u,v) \in E} B_{\xi(u),\xi(v)} = \prod_{(u,v) \in E} a_{\xi(u)} b_{\xi(v)} = \prod_{v \in V} a^{\textrm{outdeg}(v)}_{\xi(v)} b^{\textrm{indeg}(v)}_{\xi(v)}
$$
and therefore
\begin{eqnarray*}
Z^*_{B,D'}(G) &=& \sum_{\xi : V \rightarrow [m]} \prod_{(u,v) \in E} B_{\xi(u),\xi(v)} \prod_{v \in V}D'_{\xi(v),\xi(v)} \\
           &=& \sum_{\xi : V \rightarrow [m]} \prod_{v \in V} a^{\textrm{outdeg}(v)}_{\xi(v)} b^{\textrm{indeg}(v)}_{\xi(v)} D'_{\xi(v),\xi(v)} \\
           &=& \prod_{v \in V} \sum_{i = 1}^m  a^{\textrm{outdeg}(v)}_{i} b^{\textrm{indeg}(v)}_{i} D'_{i,i} 
\end{eqnarray*}
And the terms in the last line can be evaluated in polynomial time.
This completes the proof of the claim.

\medskip
With this claim we are now able to prove the Lemma. Recall that $A$ is connected and symmetric. If $A$ is non-bipartite then $A$ has rank $1$. For a given connected graph $G$ let $G'$ be a directed graph obtained from $G$ by orienting its edges arbitrarily. We have $Z_{A,D}(G) = Z^*_{A,D}(G')$ and the value $Z^*_{A,D}(G')$ can be computed by Claim~\ref{cl:polytime_rank1}.

Otherwise, if $A$ is bipartite then we have (up to permutation of the rows/columns of $A$)
$$
A = \left( \begin{array}{c c} 0 & B\\ B^T & 0 \end{array}\right)
$$
for a block $B$ of rank $1$. Let $A'$ be the matrix 
$$
A' = \left( \begin{array}{c c} 0 & B\\ 0 & 0 \end{array}\right)
$$
which has rank $1$ because $B$ has. Note furthermore, that $Z_{A,D}(G) = 0$ unless $G$ is bipartite. Assume therefore that $G= (U,W,E)$ is a bipartite graph and let the graphs $G_{UW}$, $G_{WU}$ be obtained from $G$ by directing all edges from $U$ to $W$ ($W$ to $U$, resp.).
Then
$$
A_{A,D}(G) = Z_{A',D}(G_{UW}) + Z_{A',D}(G_{WU})
$$
and the terms of the right hand side are polynomial time computable by Claim~\ref{cl:polytime_rank1}.
\end{proof}

The following $\#\PP$-hardness result will be the basis of all our proofs of intractability.

\begin{lemma}\label{lem:bipolar_block2_hard}
Given a symmetric matrix $A$ of order $n$ and diagonal $n \times n$ matrices $D,O$ such that $D$ is a non-singular matrix of non-negative integers.
If $\abs{A}$ contains a block of row rank at least $2$ then $\eval{A, D, O}$ is \#P-hard.
\end{lemma}

\begin{proof}
  Observe that by $2$-thickening we have $\eval{A^{(2)},D} \Tle \eval{A, D,
    O}$. We can form a matrix $A'$ from $A^{(2)}$ by \emph{introducing twins
    according to $D$} that is, doing the inverse operation of Lemma
  \ref{lem:twin_red}.  More precisely, let $n_i : = D_{i,i}$ for all $i \in
  [n]$ and define $m := n\cdot \left(\sum_{i=1}^n n_i\right)$.  To define the
  $m \times m$ matrix $A'$ we consider its row and column indices as pairs and
  define
\begin{equation}\label{eq:def_aprime}
A'_{(\kappa,i),(\lambda,j)} := A^{(2)}_{\kappa ,\lambda} \text{ for all } \kappa,\lambda \in [n],\, i \in n_\kappa,\, j \in n_\lambda.
\end{equation}
By the definition of $A'$ we see that Application of the Twin Reduction Lemma \ref{lem:twin_red} to $A'$ yields 
$$Z_{A'}(G) = Z_{A^{(2)},D}(G) \text{ for every graph } G.$$
and thus $\eval{A'} \Tequiv \eval{A^{(2)},D}$.
By equation \eqref{eq:def_aprime} the matrix $A'$ contains a block of row rank at least $2$ iff $A^{(2)}$ does which in turn is the case iff $\abs A$ contains such a block. The proof now follows from the result of Bulatov and Grohe \cite{bulgro05}.
\end{proof}

\subsection{Interpolation Lemma}
In the next chapters we will make extensive use of the following lemma which is an analogue of the interpolation technique as used for example in \cite{dyegre00}.

\begin{lemma}\label{lem:bij_coeff_power}
Let $x_1, \ldots, x_n \in \Real_{>0}$ be pairwise distinct and let $\mathcal{P}$ and $\mathcal{N}$ two finite multisets of real numbers with $\vert \mathcal{P} \vert = \vert \mathcal{N} \vert = n$. Then the following are equivalent
\begin{itemize}
\item[(1)] $ \mathcal{P} = \mathcal{N}$
\item[(2)] there is an ordering of the elements in $\mathcal{P}$ and $\mathcal{N}$ such that for arbitrarily large $p$, we have 
       $$ \sum_{a_i \in \mathcal{P}} x_i^pa_i = \sum_{b_i \in \mathcal{N}} x_i^pb_i.$$
\end{itemize}
\end{lemma}
\begin{proof}
The forward direction is trivial. Hence, assume that (2) holds but not (1). With the given ordering of $\mathcal{P}$ and $\mathcal{N}$ we have $\mathcal{P} = \mset{a_1, \ldots,a_n}$ and $\mathcal{N} = {b_1, \ldots,b_n}$. We may assume that there is no $i \in [n]$ such that $a_i = b_i$ because otherwise, we might delete this pair from $\mathcal{P}$ and $\mathcal{N}$.
Hence, let $k \in [n]$ be such that $x_k  = \max_{i\in[n]} x_i$. Assume w.l.o.g. that $a_k > b_k$ then, for a constant $c \neq 0$
\begin{eqnarray*}
 0 & = & \sum_{a_i \in \mathcal{P}} x_i^pa_i - \sum_{b_i \in \mathcal{N}} x_i^pb_i 
    =  x_k^p(a_k-b_k) + \sum_{i \in [n]\setminus\mset{k}} x_i^p(a_i - b_i) \\
\iff 0 &  = &  c + \sum_{i \in [n]\setminus\mset{k}} \left(\dfrac{x_i}{x_k}\right)^p(a_i - b_i).
\end{eqnarray*}
By
$\lim_{p \rightarrow \infty} \sum_{i \in [n]\setminus\mset{k}}
\left(\dfrac{x_i}{x_k}\right)^p(a_i - b_i) = 0$, this yields a contradiction.
\end{proof}

\begin{lemma}\label{lem:bij_coeff_new}
Let $x_1, \ldots, x_n \in \Real_{>0}$ be pairwise distinct and let $a_1,\ldots,a_n \in \Real$ and $b_1,\ldots,b_n \in \Real$. 
There is a $p_0 \in \Nat$ such that for all $p \ge p_0$, the equation
$$ \sum_{i = 1}^n x_i^pa_i = \sum_{i = 1}^n x_i^pb_i$$
holds if, and only if, $a_i = b_i$ for all $i \in [n]$.
\end{lemma}
\begin{proof}
Note first that backward direction is trivial. It remains therefore to prove the following. For each $I \subseteq [n]$ there is a $p_I \in \Nat$ such that for all $p \ge p_I$, if
\begin{equation}\label{eq:polysum_eq_zero}
\sum_{i \in I} x_i^p(a_i - b_i) = 0 
\end{equation}
then $a_i = b_i$ for all $i \in I$.
We will give the proof by induction on the cardinality of $I$. For empty $I$ there is nothing to be shown. Assume therefore that $I \neq \emptyset$ let $k \in I$ be such that $x_k  = \max_{i\in I} x_i$ and define $I' = I \setminus \{k\}$.

\begin{claim}
There is a $p_k \in \Nat$ such that for all $p \ge p_k$, if equation \eqref{eq:polysum_eq_zero} is satisfied then $a_k = b_k$.
\end{claim}
\begin{clproof}
Assume for contradiction that $a_k \neq b_k$ but equation \eqref{eq:polysum_eq_zero} holds for all $p \in \Nat$.
This implies
\begin{equation}\label{eq:zero_coeff_new_B}
0 =  (a_k - b_k) + \sum_{i \in I'} \left(\dfrac{x_{i}}{x_k}\right)^p(a_i - b_i).
\end{equation}
As $i \in I'$ with $a_i= b_i$ do not contribute to the above sum we may further assume that $a_i \neq b_i$ for all $i \in I'$. If $I' = \emptyset$ we already have a contradiction. If otherwise $I' \neq \emptyset$, let $k'$ be such that $x_{k'} = \max_{i\in I'} x_i$. We find that
$$
\left\vert \sum_{i \in I'} \left(\dfrac{x_i}{x_k}\right)^p(a_i - b_i) \right\vert \le \left(\dfrac{x_{k'}}{x_k}\right)^p \sum_{i \in I'} \vert a_i - b_i\vert. 
$$
In particular equation \eqref{eq:zero_coeff_new_B} does not hold if
$
\left(\dfrac{x_{k'}}{x_k}\right)^p \sum_{i \in I'} \vert a_i - b_i\vert < \vert a_k - b_k \vert
$
which, as $x_k > x_{k'}$ is the case for all
$$
p > \dfrac{\log \vert a_k - b_k \vert -  \log \sum_{i \in I'} \vert a_i - b_i\vert}{\left(\log x_{k'} - \log x_k\right)}.
$$
in contradiction to our assumption. 
\end{clproof}

By the induction hypothesis there is a $p_{I'}$ such that for all $p \ge p_{I'}$
$$
\sum_{i \in I'} x_i^p(a_i - b_i) = 0 
$$
implies $a_i = b_i$ for all $i \in I'$. Let $p_k$ be defined as in Claim~1 then the proof follows with $p_I = \max\{p_k,p_{I'}\}$.
\end{proof}

\section{The Proofs for Section~\ref{sec:hadamard}}
\label{sec:appendix_hadamard}
 
\subsection{Notation and Preliminaries}
For $x=(x_1,\ldots,x_n),y=(y_1,\ldots,y_n)\in\mathbb R^n$, by $\langle
x,y\rangle$ we denote the inner product $\sum_{i=1}^nx_iy_i$ of $x$ and
$y$. It may be a source of confusion that we work over two different fields,
$\mathbb R$ and $\gft$. Addition in $\gft$ is denoted by $\oplus$, and for $\alpha=(\alpha_1,\ldots,\alpha_n),\beta=(\beta_1,\ldots,\beta_n)\in\gft^k$,
$\alpha \cdot \beta$ is
the dot product 
$\bigoplus_{i=1}^k \alpha_{i} \beta_{i}$.
Similarly, for $\pi\in S_k$,
$\alpha_\pi \cdot \beta$ denotes
$\bigoplus_{i=1}^k \alpha_{\pi(i)} \beta_{i}$.
$\alpha \oplus \beta$ denotes the
element
$(\alpha_{1} \oplus \beta_{1}, \cdots,
  \alpha_{k} \oplus \beta_{k})$ in
$\gft^k$.
Similarly,
for $\pi\in S_k$,
$\alpha_\pi \oplus \beta$ denotes the
element
$(\alpha_{\pi(1)} \oplus \beta_{1}, \cdots,
  \alpha_{\pi(k)} \oplus \beta_{k})$.
Similar notation applies to variables, so 
if $X=(X_1,\ldots,X_k)$ and $Y=(Y_1,\ldots,Y_k)$
then $X_\pi \cdot Y$ 
denotes $\bigoplus_{i=1}^k X_{\pi(i)} Y_i$.
For $I\subseteq [k]$, let
Let $X\setminus I$ be the tuple  
containing, in order,  all variables in
$\{X_1,\ldots,X_k\}$ other than those with indices in~$I$.
For example, $X\setminus \{2,3\}$
denotes the tuple
$(X_1,X_4,\ldots,X_k)$.

\subsection{The Group Condition}

\begin{lemma}
\label{lem:isgroup}
Let $H$ be  an $n \times n$ Hadamard matrix.
If $H$ satisfies (GC) then $G(H,1)$ forms an Abelian group under
the Hadamard product.
\end{lemma}
\begin{proof}
Commutativity and associativity follow from the definition of
the Hadamard product.
To show closure, we consider two elements in $G(H,1)$ and
show that their Hadamard product is also in $G(H,1)$.
First, consider $H\row{i}\circ H\row{1}$
and $H\row{j}\circ H\row{1}$.
Their Hadamard product is
$H\row{i}\circ H\row{1} \circ H\row{j} \circ H\row{1}
=
H\row{i}\circ H\row{j}$ which is in $G(H,j)$ by the definition of
$G(H,j)$ and therefore in $H\row{1}$ by (GC).
Similarly, we fined that the product of $-H\row{i}\circ H\row{1}$
and $H\row{j}\circ H\row{1}$ is in $G(H,1)$ and also the
product of $-H\row{i}\circ H\row{1}$ and $-H\row{j}\circ H\row{1}$
is in $G(H,1)$.
From closure, it follows that the product of
$H\row{1} \circ H \row{1}$  and itself is in $G(H,1)$ and
this row (the all ones row) is the identity element in the group.
\end{proof}

\begin{proof}[Proof of Lemma \ref{lem:bipolar_Hadamard_group}]
By Lemma~\ref{lem:isgroup}, $G(H,1)$ forms an Abelian group under
the Hadamard product. All elements of this group have order $2$, and thus it
follows from elementary algebra that the order of the group is a power of $2$.
\end{proof}

\begin{proof}[Proof of Lemma \ref{lem:bipolar_hard_nonGC}, the Group Condition lemma]
It is clear from the definition of 
the Group Condition that there is a polynomial-time algorithm that  
determines whether $H$ satisfies
\cond{GC}.
We focus on the \#P-hardness result.
Let
$\evaleven{A}$ denote the problem of computing $Z_{A}(G)$ for an input
graph~$G$ in which every vertex of~$G$ has even degree.
 
Let $H$, $n$, $M$, 
$\Lambda$ and~$m$ be defined as in the statement of 
the lemma.
Let~$p$ be an even number.
We will show how to transform any graph~$G$ into
a graph~$G_p$ 
with all even-degree vertices
so
that 
$Z_{C^{[p]}}(G)=Z_M(G_p)$ for a matrix~$C^{[p]}$ which we will define below.
The definition of~$C^{[p]}$ depends upon~$M$ but not upon~$G$.
Thus, we will have
$\eval{C^{[p]}} \leq \evaleven{M} \leq \eval{M,I_m,I_{m;\Lambda}}$.

To finish the proof, we will show that, as long as~$p$
is   sufficiently large with respect to~$M$,
then $\eval{C^{[p]}}$ is \#P-hard.

We start by giving the transformation from $G=(V,E)$
into 
$G_p=(V_p,E_p)$:
\begin{eqnarray*}
 V_p & := & V \, \cup \, \mset{v_e, v_{e^{\alpha}}, v_{e,1} \ldots, v_{e,p} \, \vert \, e \in E} \\
 E_p & := & \;\;\; \mset{\,\mset{u,v_{e,1}}, \ldots,\mset{u,v_{e,p}} \, \vert \, e=\mset{u,v} \in E}\\
     &    & \cup \,\mset{\,\mset{v,v_{e,1}}, \ldots,\mset{v,v_{e,p}} \, \vert \, e=\mset{u,v} \in E}\\
     &    & \cup \,\mset{\,\mset{v_{e,1},v_e}, \ldots,\mset{v_{e,p},v_e} \, \vert \, e \in E} \\
     &    & \cup \,\mset{\,\mset{v_{e,1},v_{e^{\alpha}}}, \ldots,\mset{v_{e,p},v_{e^{\alpha}}} \, \vert \, e \in E}
\end{eqnarray*}
Essentially, every edge $e=\{u,v\}$ in~$G$ is replaced by a distinct gadget.
Figure \ref{fig:gc_gadget} illustrates this gadget for $p =4$.
Since $p$ is even, it is clear that all vertices of~$G_p$ have even degree.
\begin{figure}
\begin{center} 
\resizebox{5cm}{!}{
\includegraphics{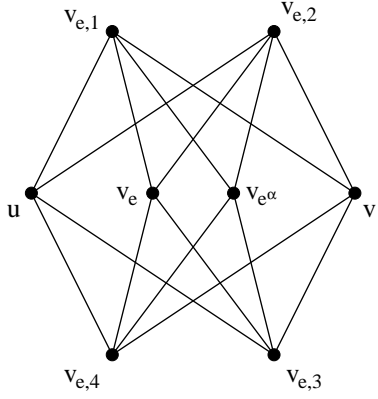}%
}
\end{center}
\caption{The gadget for $p=4$}
\label{fig:gc_gadget}
\end{figure}

Let us now construct the matrix~$C^{[p]}$. Let $\Gamma$ denote the graph with
vertices~$u$ and~$v$ and a single edge between them.
Clearly 
$C^{[p]}_{i,j}$ is equal to the contribution to $Z_M(\Gamma^p)$ corresponding
to those configurations~$\xi$ with $\xi(u)=i$ and $\xi(v)=j$. 
Thus, 
\begin{equation}
\label{eq:Cdef}
C^{[p]}_{i,j} =
\sum_{a=1}^m \sum_{b=1}^m
{
\left(
\sum_{c=1}^m M_{i,c} M_{j,c} M_{a,c} M_{b,c}
\right)
}^p,\end{equation}
where $a$ denotes the choice of spin for~$v_e$ and
$b$ denotes the choice of spin for~$v_{e^\alpha}$
and $c$ denotes the choice of spin for a vertex $v_{e,\ell}$.

To finish the proof we must show that, as long as~$p$
is  sufficiently large with respect to~$M$,
then $\eval{C^{[p]}}$ is \#P-hard.
From the definition of~$M$, we see that, for
$i\in [n]$, $j\in\{n+1,\ldots,2n\}$, we have
$C^{[p]}_{i,j}=C^{[p]}_{j,i}=0$.
Also, for all $i,j\in[n]$, we have the following.

\begin{eqnarray*}
C^{[p]}_{i,j} &=&  \sum_{a=1}^{n}\sum_{b=1}^{n}
{\langle
H\row{i} \circ H\row{j},
H\row{a} \circ H\row{b}
\rangle}^p, \mbox{ and}
\\
C^{[p]}_{n+i, n+j} &=&  
\sum_{a=1}^{n}\sum_{b=1}^{n}
{\langle
H\col{i} \circ H\col{j},
H\col{a} \circ H\col{b}
\rangle}^p.
\end{eqnarray*}

Now,
for all $i,j \in [n]$ and 
$x \in \mset{0, \ldots,n}$ 
let
$s_{i,j}^{[x]}$
be the number of pairs~$(a,b)$ such that
$|{\langle
H\row{i} \circ H\row{j},
H\row{a} \circ H\row{b}
\rangle}
| = x$
and similarly let
$s_{n+i,n+j}^{[x]}$ be the number
of pairs $(a,b)$ such that
$
|{\langle
H\col{i} \circ H\col{j},
H\col{a} \circ H\col{b}
\rangle}|=x$.
Then for all $i,j\in[n]$ we have
\begin{equation}
\label{seq}
C^{[p]}_{i,j}  = \sum_{x = 0}^{n} s_{i,j}^{[x]} x^p
\mbox{ and }
C^{[p]}_{n+i,n+j}  = \sum_{x = 0}^{n} s_{n+i,n+j}^{[x]} x^p.
\end{equation}

The pair $(a,b)=(i,j)$ contributes one towards 
$s_{i,j}^{[n]}$ and one towards $s_{n+i,n+j}^{[n]}$
so, for all
$i,j\in [n]$, we have $C^{[p]}_{i,j}>0$ and $C^{[p]}_{n+i,n+j}>0$ (remember
that $p$ is even).

Since $H$ is Hadamard, $s_{i,i}^{[n]}=n$ for every $i\in[n]$
and, for every $x\in\{1,\ldots,n-1\}$, $s_{i,i}^{[x]}=0$
so $C_{i,i}^{[p]} = n^{p+1}$.
Also, since $H$ is Hadamard, $H H^T = n I$, so
$H^T/n$ is the right inverse, hence also the left inverse, of $H$,
so $(1/n) H^T H = I$, so $H^T$ is also Hadamard.
It follows that $s_{n+i,n+i}^{[n]}=n$ and,
for every $x\in\{1,\ldots,n-1\}$, $s_{n+i,n+i}^{[x]}=0$
so $C_{n+i,n+i}^{[p]} = n^{p+1}$.

We will prove that $\eval{C^{[p]}}$ is \#P-hard for
some
sufficiently large even~$p$. We will assume for contradiction that,
for every even~$p$,
$\eval{C^{[p]}}$ is not \#P-hard.
Equation~(\ref{eq:Cdef}) indicates that $C^{[p]}$ is symmetric, so
by Lemma
\ref{lem:bipolar_block2_hard} (due to Bulatov and Grohe), for every even~$p$,  both blocks of $C^{[p]}$ have rank~$1$.
This means that every principal  
$2 \times 2$ submatrix 
in the blocks has
a zero determinant. So, for $i,j\in [n]$, we have
$(C^{[p]}_{i,i})^2 - (C^{[p]}_{i,j})^2 = 0$ and
$(C^{[p]}_{n+i,n+i})^2 - (C^{[p]}_{n+i,n+j})^2 = 0$, so
\begin{equation}
\label{seq2}
C^{[p]}_{i,j} = C^{[p]}_{i,i} 
\mbox{ and }
C^{[p]}_{n+i,n+j} = C^{[p]}_{n+i,n+i}.
\end{equation}
Since equations~(\ref{seq})
and~(\ref{seq2}) hold for all even~$p$ and
all $i,j\in[n]$, 
Lemma~\ref{lem:bij_coeff_power} allows us to deduce that,
for all $i,j\in[n]$ and~$x\in\{0,\ldots,n\}$,
$s_{i,j}^{[x]} = s_{i,i}^{[x]}$ and
$s_{n+i,n+j}^{[x]} = s_{n+i,n+i}^{[x]}$.
Thus, for all $i,j\in [n]$,

\begin{equation}
\label{eq:fail_cond}
s_{i,j}^{[1]} = \cdots = s_{i,j}^{[n-1]} = 
s_{n+i,n+j}^{[1]} = \cdots = s_{n+i,n+j}^{[n-1]} = 0
\mbox{ and } 
s_{i,j}^{[n]} = s_{n+i,n+j}^{[n]}=n.
\end{equation}
 
From the statement of the lemma, we assume that $H$
does not satisfy \cond{GC}. 
There are two similar cases.

{\bf Case~1: }Suppose there are $i,j \in[n]$ such that $G(H,i) \neq G(H,j)$. 
Fix such a pair $i,j$. 
Fix $a\in[n]$ such that
$H \row a\circ H\row i$ 
\martin{I removed ``and $- H\row a \circ H \row i$'' here. I don't think it
  should be there.}
is not in $G(H,j)$,
Now consider any $b\in [n]$. 
If it were
the case that
$\vert \scalp{H\row a\circ H\row i,H\row b\circ H\row j}\vert = n$
then we would know that either
$H_{a ,v}H_{i,v} = H_{b,v} H_{j,v}$ for all~$v$ or
$H_{a,v}H_{i,v}= -H_{b,v} H_{j,v}$ for all~$v$.
Either of these would imply $H\row a\circ H\row i \in G(H,j)$ 
which is not the case.
So we conclude that 
$\vert \scalp{H\row a\circ H\row i,H\row b\circ H\row j}\vert 
< n$.
\martin{removed $= \scalp{H\row i\circ H\row j,H\row a\circ H\row b}$, as it
  doesn't really matter}

Furthermore, there is some  $b \in [n]$ such that
$\vert \scalp{H\row a\circ H\row i,H\row b\circ H\row j}\vert \neq 0$.
Otherwise, 
\[
\mset{H\row 1\circ H\row j, \ldots, H\row n\circ H\row j, H\row a\circ H\row
  i}
\]
would be a set of $n+1$ linearly independent vectors, which is impossible.

But this implies that for some $x \in [n-1]$ we have $s_{i,j}^{[x]}
\neq 0$ contradicting equation \eqref{eq:fail_cond}.

{\bf Case~2:} Suppose there are $i,j\in [n]$ such that
$G(H^T,i) \neq G(H^T,j)$.
As in Case~1, we can deduce that
$\vert \scalp{H^T\row a\circ H^T\row i,H^T\row b\circ H^T\row j}\vert < n$.
Furthermore, there is some  $b \in [n]$ such that
$\vert \scalp{H^T\row a\circ H^T\row i,H^T\row b\circ H^T\row j}\vert \neq 0$.
But this implies that for some $x \in [n-1]$ we have $s_{n+i,n+j}^{[x]}
\neq 0$ contradicting equation \eqref{eq:fail_cond}.
\end{proof}

\subsection{Polynomial Representation}

For an $n\times n$ matrix $H$ and
a row index $l\in[n]$, let
$R(H) : = \mset{
H\row{i}  \mid  i \in[n]}
$. 
The \emph{Extended Group Condition for $H$} is:

\begin{condition}{(EGC)}
$R(H)$ is an Abelian group under the Hadamard product.
\end{condition}

The following lemmas are useful preparation for the proof of
Lemma~\ref{lem:poly_rep}, the Polynomial Representation Lemma.
We say that a   Hadamard matrix is \emph{normalised} if
its first row and column consists entirely of +1s.

\begin{lemma}
\label{lem:ext_GC}
Let $H$ be a normalised $n \times n$ Hadamard matrix. 
If $G(H,1)$ is closed under the Hadamard product then
$R(H)$ is closed under the Hadamard product.
\end{lemma}
\begin{proof}

Fix $i,j\in[n]$. Since $G(H,1)$ is closed under the Hadamard product, 
and
$H\row i \circ H\row 1 \in H(G,1)$
and $H\row j \circ H \row 1 \in H(G,1)$,
we have $H\row i \circ H \row j \in H(G,1)$.
Thus, there is a $\ell\in [n]$
such that either 
$H\row i \circ H \row j = H\row \ell \circ H \row 1 = H \row \ell$
(using the fact that the first row of $H$ is all ones)
or $H\row i \circ H \row j = - H \row \ell \circ H \row 1 = - H \row \ell$.
The latter is equivalent to $H\row i \circ H \row \ell = - H \row j$.
And since $H_{j,1}=1$ (since the first column of $H$ is positive)
this implies that one of $H_{i,1}$ and $H_{\ell,1}$ is negative, a contradiction.
We conclude that 
$H\row i \circ H \row j   = H \row \ell$.
\end{proof}

\begin{cor}
Let $H$ be a normalised $n \times n$ Hadamard matrix. 
If $H$ satisfies the group condition then $H$ satisfies the extended group
condition.
\label{cor:ext_GC}
\end{cor}
\begin{proof}
Suppose that $H$ satisfies the group condition. By Lemma~\ref{lem:isgroup},
$G(H,1)$ is an Abelian group under the Hadamard product. The identity
is the all ones row, which is in $R(H)$,and every element is its own
inverse.
Closure of $R(H)$ follows from Lemma~\ref{lem:ext_GC}.
\end{proof}

\begin{lemma}\label{lem:tensor_EGC}
Suppose that $B$ is an $r\times r$ matrix with entries in $\{-1,+1\}$
and that $C$ is an $t\times t$ matrix with entries in $\{-1,+1\}$. 
Suppose that the tensor product $H=B \otimes C$ is a Hadamard
matrix.
Then $B$ and $C$ are Hadamard.
If $H$ is symmetric then so are $B$ and $C$.
If $H$ and $B$ are normalised and $H$ satisfies (EGC), 
then $B$ and $C$ satisfy (EGC) and $C$ is normalised.
\end{lemma}
\begin{proof}
 
Since $H$ is Hadamard, we know that for any such $k\in[r]$
and distinct $i$ and $i'$ in  $[t]$,
the inner product 
$\langle 
H\row{(k-1)t + i}, H\row{(k-1)t + i'}
\rangle$ is zero.
But this inner product is
\begin{align*}
\sum_{\ell\in [r]}\sum_{j\in [t]} 
H_{(k-1)t+i,(\ell-1)t+j} H_{(k-1)t+i',(\ell-1)t+j} &
=\sum_{\ell\in [r]}\sum_{j\in [t]} B_{k,\ell}C_{i,j} B_{k,\ell} C_{i',j}\\
& =\sum_{\ell\in[r]} B_{k,\ell}^2 \langle C\row{i}, C\row{i'}\rangle\\
& = r \langle C\row{i}, C\row{i'}\rangle,
\end{align*}
so $C$ is Hadamard. 
Similarly,
for any distinct $k,k'\in[r]$ and any $i\in[t]$,
\begin{align*}
0 = \langle 
H\row{(k-1)t + i}, H\row{(k'-1)t + i}
\rangle
&= \sum_{\ell\in [r]}\sum_{j\in [t]} 
H_{(k-1)t+i,(\ell-1)t+j} H_{(k'-1)t+i,(\ell-1)t+j} \\
& =\sum_{\ell\in [r]}\sum_{j\in [t]} B_{k,\ell}C_{i,j} B_{k',\ell} C_{i,j}\\
& =\sum_{j\in[t]} C_{i,j}^2 \langle B\row{k}, B\row{k'}\rangle\\
& = t \langle B\row{k}, B\row{k'}\rangle,
\end{align*}
so $B$ is Hadamard. 
If $H$ is symmetric then it is easy to see that $B$ and $C$
are symmetric as well.
Also, if $H$ and $B$ are normalised, then it is easy to see
that $C$ is normalised as well.

Suppose now that $H$ and $B$ are normalised and $H$ satisfies (EGC).
We first show that C satisfies (EGC).
Then we will finish by showing that B satisfies (EGC).
 
To show that $R(C)$ is an Abelian group under the Hadamard
product we just need to show closure. (Commutativity and
Associativity come from the definition of the Hadamard product,
the identity element is the row of all ones, and every element
is its own inverse.)
Since $R(H)$ is closed under the Hadamard product,
we know that, for any 
distinct $i,i'\in[t]$, 
$ 
H\row{i} \circ  H\row{i'}\in R(H)$.
But the first $t$ elements of this row are
$H_{i,1}H_{i',1},\ldots,H_{i,t}H_{i',t} =
B_{1,1}C_{i,1}B_{1,1}C_{i',1},\ldots,B_{1,1}C_{i,t}B_{1,1}C_{i',1}$ which is  
equal to 
$C\row i \circ C\row{i'}$. 
This shows that $C\row i \circ C\row{i'} \in G(C,1)$.
Now use lemma
\ref{lem:ext_GC} to show that $R(C)$ is closed under the Hadamard
product.

Similarly, to show that $R(B)$ is closed under the Hadamard product,
note that for any distinct $k,k'\in[r]$,
$H\row{(k-1)t+1} \circ H\row{(k'-1)t+1}\in R(H)$.
But the  elements of this row are
$$H_{(k-1)t+1,(\ell-1)t+j} H_{(k'-1)t+1,(\ell-1)t+j},$$ for $\ell\in [r]$, $j\in[t]$,
and taking those with $\ell=1$ (which occur every $r$ elements along the
row) we get
$B_{k,1}C_{1,j}B_{k',1}C_{1,j}$. 
Thus, the sub-row of these elements is the Hadamard product of
$B\row k$ and $B\row{k'}$. This shows that
$B\row k \circ B \row {k'}\in G(B,1)$. Now
use lemma \ref{lem:ext_GC} to show that $R(B)$ is closed under the
Hadamard product.
\end{proof}

Given an $n\times n$ matrix $H$ and permutations~$\Sigma$ and~$\Pi$
in $S_n$, let $H_{\Sigma,\Pi}$ denote the matrix with
$(H_{\Sigma,\Pi})_{i,j} = H_{\Sigma(i),\Pi(j)}$.

\begin{lemma}
\label{lem:assymmetric}
Let $H$ be a normalised $n\times n$ Hadamard matrix
with $n\geq 2$
that satisfies \cond{GC}. 
Then there are permutations $\Sigma,\Pi$ in
$S_n$ with $\Sigma(1)=1$ and $\Pi(1)=1$ 
and a normalised Hadamard matrix $H'$ satisfying \cond{GC}
such that
$H_{\Sigma,\Pi} = H_2 \oplus H'$.
$\Sigma$, $\Pi$, and $H'$ can be constructed in polynomial time.
\end{lemma}

\begin{proof}

By Lemma~\ref{lem:bipolar_Hadamard_group} we know $n$ is a power of~$2$,
say $n=2^{k+1}$. 
The lemma is trivial for $k=0$ since $H=H_2$ and $\Sigma$ and $\Pi$
can be taken to be the identity. So suppose $k\geq 1$.
Let~$\nu=2^k$.
 
{\bf Part 1:} Choose $\Sigma'$ and $\Pi'$ in $S_n$
with $\Sigma'(1)=1$ and $\Pi'(1)=1$ 
so that $(H_{\Sigma',\Pi'})_{\nu+1,\nu+1}=-1$. 

How to choose $\Sigma'$ and $\Pi'$:
$H$ is Hadamard, so some entry
$H_{i,j}=-1$. The indices $i$ and $j$ are not $1$ because $H$ is normalised.
Let $\Sigma'$ be the transposition $(i,\nu+1)$
and let $\Pi'$ be the transposition $(j,\nu+1)$.

{\bf Part 2C:} Choose $\pi$ in $S_n$
with $\pi(1)=1$  and $\pi(\nu+1)=\nu+1$
so that, for $\ell\in[\nu]$,
\begin{equation}
\label{eq:sigmaprime}
{(H_{\Sigma',\Pi''})}_{\nu+1,\ell}=
+1
\mbox{ and }
{(H_{\Sigma',\Pi''})}_{\nu+1,\nu+\ell}=
-1,
\end{equation}
where $\Pi''$ denotes the composition of first $\Pi'$ then $\pi$.

How to choose $\pi$:
We construct a sequence of permutations 
$\pi_1,\ldots,\pi_\nu$ where $\pi_1$ is the identity and we  
let $\pi=\pi_\nu$.
Let $H^j$ denote $H_{\Sigma',\pi_j \Pi'}$.
For $j\in\{2,\ldots \nu\}$, we define $\pi_j$ as follows.
If $H^{j-1}_{\nu+1,\nu+j}=-1$ then 
$\pi_j=\pi_{j-1}$.
Otherwise,
there is an $1<\ell<\nu+1$ with 
$H^{j-1}_{\nu+1,\ell}=-1$. 
So $\pi'_j$ is the composition of first applying $\pi'_{j-1}$ 
and then transposing $\nu+j$ and $\ell$.
To see that such an $\ell$ exists, note that
$H$ is Hadamard, so $\langle H\row 1,H\row{\nu+1}\rangle=0$.
But $H\row 1$ is positive, so $H\row{\nu+1}$ has exactly $\nu$ ones.
$\ell>1$ because $\pi_{j-1}\Pi'(1)=1$.

{\bf Part 2R:} Choose $\sigma$ in $S_n$ 
with $\sigma(1)=1$  and $\sigma(\nu+1)=\nu+1$
so that, for $\ell\in[\nu]$,
\begin{equation}
\label{eq:piprime}
{(H_{\Sigma'',\Pi''})}_{\ell,\nu+1}=
+1
\mbox{ and }
{(H_{\Sigma'',\Pi''})}_{\nu+\ell,\nu+1}=
-1,
\end{equation}
where $\Sigma''$ denotes the composition of first $\Sigma'$ then $\sigma$.

How to choose $\sigma$:
This is symmetric to how we chose $\pi$.

Since $\sigma(\nu+1)=\nu+1$, we have
${(H_{\Sigma',\Pi''})}_{\nu+1,\ell}
={(H_{\Sigma',\Pi''})}_{\sigma(\nu+1),\ell}
={(H_{\Sigma'',\Pi''})}_{\nu+1,\ell}
$
 for every $\ell\in[n]$,
so Equations~(\ref{eq:sigmaprime}) and~(\ref{eq:piprime})
give
\begin{equation}
\label{eq:bothprime}
{(H_{\Sigma'',\Pi''})}_{\nu+1,\ell}={(H_{\Sigma'',\Pi''})}_{\ell,\nu+1}=
+1
\mbox{ and }
{(H_{\Sigma'',\Pi''})}_{\nu+1,\nu+\ell}=
{(H_{\Sigma'',\Pi''})}_{\nu+\ell,\nu+1}=
-1,
\end{equation}

{\bf Part 3C:} Choose $\pi'$ in $S_n$ with
$\pi'(1)=1$ 
and $\pi'([\nu])=[\nu]$
so that, for $j,\ell\in[\nu]$
\begin{equation}
\label{eq:sigmadouble} 
{(H_{\Sigma'',\Pi})}_{\ell,j}
=
{(H_{\Sigma'',\Pi})}_{\ell,\nu+j} \mbox{ and }
{(H_{\Sigma'',\Pi})}_{\nu+\ell,j}
=-
{(H_{\Sigma'',\Pi})}_{\nu+\ell,\nu+j},
\end{equation}
where $\Pi$ denotes the composition of first $\Pi''$ then $\pi'$.

How to choose $\pi'$:
Note that $H$ satisfies \cond{EGC}
by Corollary~\ref{cor:ext_GC} hence
$H_{\Sigma'',\Pi''}$ satisfies \cond{EGC} (permuting does not
change \cond{ECG}).
Start with $\pi'(1)=1$ and $\pi'(\nu+1)=\nu+1$.
Note that for $j=1$, we have, by normalisation
and Equation (\ref{eq:piprime}),
\begin{equation}
\label{eq:working}
\forall \ell\in[\nu],
{(H_{\Sigma'',\Pi''})}_{\ell,\pi'(j)}
=
{(H_{\Sigma'',\Pi''})}_{\ell,\pi'(\nu+j)} \mbox{ and }
{(H_{\Sigma'',\Pi''})}_{\nu+\ell,\pi'(j)}
=-
{(H_{\Sigma'',\Pi''})}_{\nu+\ell,\pi'(\nu+j)},
\end{equation}
Now for $j\in\{2,\ldots,\nu\}$ we define $\pi'(j)$ and
$\pi'(\nu+j)$ to satisfy (\ref{eq:working}) as follows.
Choose any $i\in[\nu]$ such that
${\pi'}^{-1}(i)$ is undefined and set $\pi'(j)=i$.
By \cond{EGC} there is a unique $i'$ with
\begin{equation}
\label{eq:doublestar}
(H_{\Sigma'',\Pi''})\row i \circ (H_{\Sigma'',\Pi''}) \row {\nu+1} = 
(H_{\Sigma'',\Pi''}) \row {i'}.\end{equation}
Also, $i'$ is not in $[\nu]$ since by (\ref{eq:doublestar})
$(H_{\Sigma'',\Pi''})_{i,\nu+1}  (H_{\Sigma'',\Pi''})_{\nu+1,\nu+1} = 
(H_{\Sigma'',\Pi''}){i',\nu+1}$,
and the left-hand-side is~$-1$ by Equation~(\ref{eq:piprime}).
Finally, 
${\pi'}^{-1}(i')$ is undefined 
since no other $i$ satisfies (\ref{eq:doublestar}).
So set $\pi'(\nu+j)=i'$.

{\bf Part 3R:} Choose $\sigma'$ in $S_n$ with
$\sigma'(1)=1$ 
and $\sigma'([\nu])=[\nu]$
so that, for $j,\ell\in[\nu]$
\begin{equation}
\label{pi}
{(H_{\Sigma,\Pi})}_{\ell,j}
=
{(H_{\Sigma,\Pi})}_{\nu+\ell,j} \mbox{ and }
{(H_{\Sigma,\Pi})}_{\ell,\nu+j}
=-
{(H_{\Sigma,\Pi})}_{\nu+\ell,\nu+j},
\end{equation}
where $\Sigma$ denotes the composition of first $\Sigma''$ then $\sigma'$.

How to choose $\sigma'$: This is symmetric to how we chose $\pi'$.

Now, since $\sigma'([\nu])=[\nu]$, Equation~\ref{eq:sigmadouble} implies
\begin{equation*}
{(H_{\Sigma'',\Pi})}_{\sigma(\ell),j}
=
{(H_{\Sigma'',\Pi})}_{\sigma(\ell),\nu+j} \mbox{ and }
{(H_{\Sigma'',\Pi})}_{\sigma(\nu+\ell),j}
=-
{(H_{\Sigma'',\Pi})}_{\sigma(\nu+\ell),\nu+j},
\end{equation*}
or equivalently
\begin{equation}
\label{eq:sigma} 
{(H_{\Sigma,\Pi})}_{\ell,j}
=
{(H_{\Sigma,\Pi})}_{\ell,\nu+j} \mbox{ and }
{(H_{\Sigma,\Pi})}_{\nu+\ell,j}
=-
{(H_{\Sigma,\Pi})}_{\nu+\ell,\nu+j}.
\end{equation}
By Equations~(\ref{pi}) and~(\ref{eq:sigma})
we can take $H'$ to be the first $\nu$ rows and columns of $H_{\Sigma,\Pi}$.
\end{proof}

\begin{lemma}
\label{lem:symmetric}
Let $H$ be a normalised symmetric $n\times n$ Hadamard matrix
with $n\geq 2$ that has an entry $-1$ on the diagonal
and satisfies \cond{GC}. 
Then there is a permutation $\Sigma$ in
$S_n$ with $\Sigma(1)=1$  
and a normalised symmetric Hadamard matrix $H'$ satisfying \cond{GC}
such that
$H_{\Sigma,\Sigma} = H_2 \oplus H'$.
$\Sigma$ and $H'$ can be constructed in polynomial time.
\end{lemma}

\begin{proof}
In the proof of Lemma~\ref{lem:assymmetric} note that we can
ensure $\Pi=\Sigma$.
If $H_{a,a}=-1$ then $i=j=1$ in Part~1. 
\end{proof}

Define $H_4$ as follows.
$$H_4 =  \left( \begin{array}{r r r r}
            + & + & + & + \\
            + & + & - & - \\
            + & - & + & - \\
            + & - & - & + 
           \end{array}\right) 
$$

\begin{lemma}\label{lem:worsepermuting}
Let $H$ be a normalised symmetric $n\times n$ Hadamard matrix
with $n > 2$. Suppose that~$H$ has a positive diagonal
and satisfies \cond{GC}.
Then there is a permutation $\Sigma\in S_n$
with $\Sigma(1)=1$ 
and a normalised symmetric Hadamard matrix $H'$ satisfying \cond{GC}
such that
$H_{\Sigma,\Sigma} = H_4 \oplus H'$.\martin{$H_{\Sigma,\Pi}\to H_{\Sigma,\Sigma}$}
$\Sigma$ and $H'$ can be constructed in polynomial time.
\end{lemma}

\begin{proof}

By Lemma~\ref{lem:bipolar_Hadamard_group} we know $n$ is a power of~$2$,
say $n=2^{k+2}$. 
The lemma is trivial for $k=0$ since $H=H_4$ and $\Sigma$ 
can be taken to be the identity. So suppose $k\geq 1$.
Let~$\nu=2^k$.

{\bf Part 1:} Choose $\Sigma'$ in $S_n$ with
$\Sigma'(1)=1$ 
and $\Sigma'(\nu+1)=\nu+1$
so that, for $j\in[2\nu]$.
\begin{equation}
\label{eq:partone}
{(H_{\Sigma',\Sigma'})}_{\nu+1,j}=+1
\mbox{ and }
{(H_{\Sigma',\Sigma'})}_{\nu+1,2\nu+j}=-1.\end{equation}
 
How to choose $\Sigma'$: 
We construct a sequence of permutations 
$\sigma_0,\ldots,\sigma_{2\nu}$ where $\sigma_0$ is the identity and we  
let $\Sigma'=\sigma_{2\nu}$.
Let $H^j$ denote $H_{\sigma_j ,\sigma_j}$.
For $j\in\{1,\ldots 2\nu\}$, we define $\sigma_j$ as follows.
If $H^{j-1}_{\nu+1,2\nu+j}=-1$ then 
$\sigma_j=\sigma_{j-1}$.
Otherwise,
there is an $1<\ell<2\nu+1$ with $\ell\neq \nu+1$  with 
$H^{j-1}_{\nu+1,\ell}=-1$. 
So $\sigma_j$ is the composition of first applying $\sigma_{j-1}$ 
and then transposing $2\nu+j$ and $\ell$.
To see that such an $\ell$ exists, note that
$H\row{\nu+1}$ has exactly $2\nu$ ones. However, since $H$ is normalised and has a positive diagonal,
$H_{\nu+1,1}=H_{\nu+1,\nu+1}=+1$ 
so $H^{j-1}_{\nu+1,1}=H^{j-1}_{\nu+1,\nu+1}=+1$.

Observation: Since $H_{\Sigma',\Sigma'}$ is Hadamard, 
${(H_{\Sigma',\Sigma'})}\row
{2\nu+1}$ has $2\nu$ positive
entries (since its dot product with row~1 is~0).
Also, half of these are in the first $2\nu$ columns (since its
dot product with row~$\nu+1$ is~0).

{\bf Part 2:} Choose $\sigma'$ in $S_n$ with
$\sigma'(1)=1$,
$\sigma'(\nu+1)=\nu+1$,
$\sigma'(2\nu+1)=2\nu+1$ and
$\sigma'([2\nu])=[2\nu]$
so that, for $j\in[\nu]$,
\begin{equation}
\label{eq:parttwoA}
{(H_{\Sigma'',\Sigma''})}_{2\nu+1,j}=
{(H_{\Sigma'',\Sigma''})}_{2\nu+1,2\nu+j}=
+1
\mbox{ and }
{(H_{\Sigma'',\Sigma''})}_{2\nu+1,\nu+j}=
{(H_{\Sigma'',\Sigma''})}_{2\nu+1,3\nu+j}=
-1,\end{equation}
where $\Sigma''$ is the composition of $\Sigma'$ then $\sigma'$.

How to choose $\sigma'$: 
We construct a sequence of permutations 
$\sigma'_1,\ldots,\sigma'_{2\nu}$ where $\sigma'_1$ is the identity and we  
let $\sigma'=\sigma'_{2\nu}$.
Let $H^j$ denote $H_{\sigma'_j \Sigma' ,\sigma'_j \Sigma'}$.
Note that $H^1_{2\nu+1,\nu+1}=-1$ by (\ref{eq:partone}) and symmetry 
of~$H^1$.
For $j\in\{2,\ldots \nu\}$, we define $\sigma'_j$ as follows.
If $H^{j-1}_{2\nu+1,\nu+j}=-1$ then 
$\sigma'_j=\sigma'_{j-1}$.
Otherwise, by the observation at the end of Part~1,
there is an $1 < \ell < \nu + 1$ with 
$H^{j-1}_{2\nu+1,\ell}=-1$. 
So $\sigma'_j$ is the composition of first applying $\sigma'_{j-1}$ 
and then transposing $\nu+j$ and $\ell$.
For $j\in\{\nu+1,\ldots 2\nu\}$, we define $\sigma_j$ as follows.
If $H^{j-1}_{2\nu+1,2\nu+j}=-1$ then 
$\sigma'_j=\sigma'_{j-1}$.
Otherwise, by the observation at the end of Part~1,
there is an $2\nu+1<\ell<3\nu+1$ with 
$H^{j-1}_{2\nu+1,\ell}=-1$. 
So $\sigma'_j$ is the composition of first applying $\sigma'_{j-1}$ 
and then transposing $2\nu+j$ and $\ell$.
(The reason that $\ell>2\nu+1$ is that the diagonal is positive.)

Note that $\Sigma''(1)=1$. Since $\sigma'(\nu+1)=\nu+1$ and $\sigma'([2\nu])=[2\nu]$,
$${(H_{\Sigma'',\Sigma''})}_{\nu+1,j}=
{(H_{\Sigma',\Sigma'})}_{\sigma'(\nu+1),\sigma'(j)}=
{(H_{\Sigma',\Sigma'})}_{\nu+1,\sigma'(j)},$$ so
Equation~(\ref{eq:partone}) gives us
\begin{equation}
\label{eq:parttwoB}\forall j\in[2\nu],
{(H_{\Sigma'',\Sigma''})}_{\nu+1,j}=+1
\mbox{ and }
{(H_{\Sigma'',\Sigma''})}_{\nu+1,2\nu+j}=-1.\end{equation}
Equations~(\ref{eq:parttwoA}) and (\ref{eq:parttwoB}) are 
summarised by the following picture, which takes into account
the symmetry of~$H_{\Sigma'',\Sigma''}$.

\begin{small}$$
H_{\Sigma'',\Sigma''} = \left( \begin{array}{r c r|r c r|r c r|r c r}
            +      &\ldots & + & + & \ldots & + & + & \ldots & + & + & \ldots & + \\
            \vdots &       &   & \vdots  &      &   &\vdots   &    &   &   &        &   \\
            +      &       &   & +  &        &   &  + &        &   &   &        &   \\
            \hline
            +      & \ldots & + & + & \ldots & + & - & \ldots & - & - & \ldots & - \\
            \vdots &        &   & \vdots  &      &   & \vdots  &        &   &   &        &   \\
            +      &        &   & + &        &   &  - &        &   &   &        &   \\
            \hline
            +      & \ldots & + & -      & \ldots & - & +      & \ldots & + & - & \ldots & - \\
            \vdots &        &   & \vdots  &        &    & \vdots &        &   &   &        &   \\
            +      &        &   & -      &        &    & +      &        &   &   &        &   \\
           \hline
            +      &        &  & -     &         &    & -     &        &   &   &        &  \\
           \vdots &        &  & \vdots &         &    & \vdots &        &   &   &        &   \\
           +      &        &  & -     &         &    & -     &        &   &   &        &   \\
           \end{array}\right)
$$  \end{small}

{\bf Part 3:} Choose $\sigma''$ in $S_n$ with
$\sigma''(1)=1$,
$\sigma''(\nu+1)=\nu+1$,
$\sigma''(2\nu+1)=2\nu+1$,
$\sigma''([\nu])=[\nu]$,
$\sigma''(\{\nu+1,\ldots,2\nu\}) = \{\nu+1,\ldots,2\nu\}$
and
$\sigma''(\{2\nu+1,\ldots,3\nu\}) = \{2\nu+1,\ldots,3\nu\}$
so that, for $j\in[\nu]$, we have the following, where
$\Sigma$ denotes the composition of  $\Sigma''$ then $\sigma''$.

\begin{equation}
\label{eq:afterEGCone} 
{(H_{\Sigma,\Sigma})}
\row j 
\circ
{(H_{\Sigma,\Sigma})}
\row {2\nu+j}
=
{(H_{\Sigma,\Sigma})}
\row 
{2\nu+1}
\end{equation}
\begin{equation}
\label{eq:afterEGCtwo} 
{(H_{\Sigma,\Sigma})}
\row {\nu+j} 
\circ
{(H_{\Sigma,\Sigma})}
\row {3\nu+j}
=
{(H_{\Sigma,\Sigma})}
\row 
{2\nu+1}
\end{equation} 
\begin{equation}
\label{eq:afterEGCthree} 
{(H_{\Sigma,\Sigma})}
\row 
{j} 
\circ
{(H_{\Sigma,\Sigma})}
\row {\nu+j}
=
{(H_{\Sigma,\Sigma})}
\row 
{\nu+1}
\end{equation}

How to choose $\sigma''$:
Note that $H$ satisfies \cond{EGC}
by Corollary~\ref{cor:ext_GC} hence
$H_{\Sigma'',\Pi''}$ satisfies \cond{EGC} (permuting does not
change \cond{ECG}).
For $j\in[\nu]$, do the following.
Let $i_1$ be the smallest element in $[\nu]$ such that
the inverse of $i_1$ under $\sigma''$ is still undefined.
(For $j=1$, $\sigma''$ is still completely undefined
so we will have $i_1=1$.)
Let $i_2$ be the solution to 
\begin{equation}
\label{tinyrefone}
(H_{\Sigma'',\Sigma''})\row {i_1} \circ (H_{\Sigma'',\Sigma''}) \row {\nu+1} = (H_{\Sigma'',\Sigma''}) \row {i_2}.
\end{equation}
This equation implies that
$$(H_{\Sigma'',\Sigma''})_{i_1,\nu+1} (H_{\Sigma'',\Sigma''})_{\nu+1,\nu+1} = (H_{\Sigma'',\Sigma''})_{i_2,\nu+1}$$
and
$$(H_{\Sigma'',\Sigma''})_{i_1,2\nu+1} (H_{\Sigma'',\Sigma''})_{\nu+1,2\nu+1} = (H_{\Sigma'',\Sigma''})_{i_2,2\nu+1}.$$
Applying 
Equations~(\ref{eq:parttwoA}) and~(\ref{eq:parttwoB}),
the left-hand-side of the first of these equations is $+1$ and
the left-hand-side of the second of these equations is $-1$,
so  
$i_2\in\{\nu+1,\ldots,2\nu\}$.  Also, since no 
other $i_1$ satisfies Equation~(\ref{tinyrefone}) for this value of~$i_2$,
the
inverse of $i_2$ under $\sigma''$ is still undefined
(so there is no problem with defining it now).
Let $i_3$ be the solution to
$$(H_{\Sigma'',\Sigma''})\row {i_1} \circ (H_{\Sigma'',\Sigma''}) \row {2 \nu+1} = (H_{\Sigma'',\Sigma''}) \row {i_3}.$$
This equation implies that
$$(H_{\Sigma'',\Sigma''})_{i_1,\nu+1} (H_{\Sigma'',\Sigma''})_{2\nu+1,\nu+1} = (H_{\Sigma'',\Sigma''})_{i_3,\nu+1}$$
and
$$(H_{\Sigma'',\Sigma''})_{i_1,2\nu+1} (H_{\Sigma'',\Sigma''})_{2\nu+1,2\nu+1} = (H_{\Sigma'',\Sigma''})_{i_3,2\nu+1}.$$
Applying 
Equations~(\ref{eq:parttwoA}) and~(\ref{eq:parttwoB}),
the left-hand-side of the first of these equations is $-1$ and
the left-hand-side of the second of these equations is $+1$,
so  
$i_3 \in \{2\nu+1,\ldots,3\nu\}$  and 
the inverse of $i_3$ under $\sigma''$ is still 
undefined.
Let $i_4$  be the solution to 
$$(H_{\Sigma'',\Sigma''}) \row {i_2} \circ (H_{\Sigma'',\Sigma''}) \row {2 \nu+1} = (H_{\Sigma'',\Sigma''}) \row {i_4}.$$
This equation implies that
$$(H_{\Sigma'',\Sigma''})_{i_2,\nu+1} (H_{\Sigma'',\Sigma''})_{2\nu+1,\nu+1} = (H_{\Sigma'',\Sigma''})_{i_4,\nu+1}$$
and
$$(H_{\Sigma'',\Sigma''})_{i_2,2\nu+1} (H_{\Sigma'',\Sigma''})_{2\nu+1,2\nu+1} = (H_{\Sigma'',\Sigma''})_{i_4,2\nu+1}.$$
Applying 
Equations~(\ref{eq:parttwoA}) and~(\ref{eq:parttwoB}),
the left-hand-side of the first of these equations is $-1$ and
the left-hand-side of the second of these equations is $-1$,
so  
$i_4 \in \{3\nu+1,\ldots,4\nu\}$and 
the inverse of $i_4$ under $\sigma'$ is still undefined.
Let 
$\sigma''(j)=i_1$,
$\sigma''(\nu+j)=i_2$,
$\sigma''(2\nu+j)=i_3$ and
$\sigma''(3\nu+j)=i_4$. 
Note that the choices of $i_1$, $i_2$, $i_3$
and $i_4$ imply the following, which imply Equations~(\ref{eq:afterEGCone}),
(\ref{eq:afterEGCtwo}) and (\ref{eq:afterEGCthree}).
\begin{equation}
\label{eq:afterEGConemod} 
{(H_{\Sigma'',\Sigma''})}
\row {\sigma''(j)} 
\circ
{(H_{\Sigma'',\Sigma''})}
\row {\sigma''(2\nu+j)}
=
{(H_{\Sigma'',\Sigma''})}
\row 
{\sigma''(2\nu+1)}
\end{equation}
\begin{equation}
\label{eq:afterEGCtwomod} 
{(H_{\Sigma'',\Sigma''})}
\row {\sigma''(\nu+j)} 
\circ
{(H_{\Sigma'',\Sigma''})}
\row {\sigma''(3\nu+j)}
=
{(H_{\Sigma'',\Sigma''})}
\row 
{\sigma''(2\nu+1)}
\end{equation} 
\begin{equation}
\label{eq:afterEGCthreemod} 
{(H_{\Sigma'',\Sigma''})}
\row {\sigma''(j)} 
\circ
{(H_{\Sigma,\Sigma})}
\row {\sigma''(\nu+j)}
=
{(H_{\Sigma'',\Sigma''})}
\row 
{\sigma''(\nu+1)}
\end{equation}

Since  
$\sigma''(\nu+1)=\nu+1$,
$\sigma''(2\nu+1)=2\nu+1$,
$\sigma''([\nu])=[\nu]$,
$\sigma''(\{\nu+1,\ldots,2\nu\}) = \{\nu+1,\ldots,2\nu\}$
and
$\sigma''(\{2\nu+1,\ldots,3\nu\}) = \{2\nu+1,\ldots,3\nu\}$,
Equations~(\ref{eq:parttwoA}) and~(\ref{eq:parttwoB})
give us
\begin{equation*}
\forall j \in [\nu],
{(H_{\Sigma,\Sigma})}_{2\nu+1,j}=
{(H_{\Sigma,\Sigma})}_{2\nu+1,2\nu+j}=
+1
\mbox{ and }
{(H_{\Sigma,\Sigma})}_{2\nu+1,\nu+j}=
{(H_{\Sigma,\Sigma})}_{2\nu+1,3\nu+j}=
-1,\end{equation*}
\begin{equation*}
\forall j\in[2\nu],
{(H_{\Sigma,\Sigma})}_{\nu+1,j}=+1
\mbox{ and }
{(H_{\Sigma,\Sigma''})}_{\nu+1,2\nu+j}=-1.\end{equation*}
These, together with
Equations~(\ref{eq:afterEGCone}),
(\ref{eq:afterEGCtwo}), and (\ref{eq:afterEGCthree})
and the symmetry of $H_{\Sigma,\Sigma}$,
give us the result, where $H'$ is the first $\nu$ rows
and columns of $H_{\Sigma,\Sigma}$.
   
\end{proof}

\begin{lemma}
\label{lem:bipolar_backbone}
Let $H$ be a 
normalised 
Hadamard matrix of order $n = 2^k$ which satisfies \cond{GC}.
Let $X= (X_1, \ldots, X_k), Y= (Y_1, \ldots, Y_k)$. 
There are index mappings
$\rho^R:\gft^k\to[n]$ and
$\rho^C:\gft^k\to[n]$  with $\rho^R(0,\ldots,0)=
\rho^C(0,\ldots,0)=1$ and
a permutation $\pi \in S_k$  
such that $H$ is represented by the polynomial $X_{\pi}Y$.
If $H$ is symmetric then $\rho^R=\rho^C$.
$\rho^R$, $\rho^C$ and $\pi$ can be constructed in polynomial time.
\end{lemma}

\begin{proof}

The proof is by induction on~$k$.
The base case is $k=1$ for which $H=H_2$.
In this case, we take the index mapping
$\rho^R$ given by
$\rho^R(0)=1$ and $\rho^R(1)=2$. $\rho^R=\rho^C$ 
and $\pi$ is the identity.

For the inductive step, first suppose that 
$H$ is not symmetric. By Lemma~\ref{lem:assymmetric},
there are permutations~$\Sigma,\Pi\in S_n$ 
with $\Sigma(1)=1$
and $\Pi(1)=1$ and a normalised Hadamard matrix $H'$
satisfying (GC) such that $H_{\Sigma,\Pi}=H_2 \oplus H'$.
These are constructed in polynomial time.
By induction, 
we can construct index mappings
$\rho_{k-1}^R:\gft^{k-1}\to[2^{k-1}]$ and
$\rho_{k-1}^C:\gft^{k-1}\to[2^{k-1}]$  with $\rho_{k-1}^R(0,\ldots,0)=
\rho_{k-1}^C(0,\ldots,0)=1$ and
a permutation $\pi' \in S_{k-1}$  
such that $H'$ is represented by the polynomial 
$$X_{\pi'(1)}Y_{1} \oplus \cdots \oplus X_{\pi'(k-1)} Y_{k-1}.$$

Now take $\rho^R(X_1,\dots,X_k)=
\Sigma (2^{k-1}X_k +\rho^R_{k-1}(X_1,\ldots,X_{k-1}))$
and 
$\rho^C(Y_1,\dots,Y_k)=
 \Pi (2^{k-1}Y_k +\rho^C_{k-1}(Y_1,\ldots,Y_{k-1}))$
and let $\pi\in S_k$ that the permutation that maps
$k$ to itself and applies $\pi'$ to $1,\ldots,k-1$.

Next, suppose that $H$ is symmetric and that it has an entry $-1$
on the diagonal. Using Lemma~\ref{lem:symmetric}
we proceed exactly as before except that 
we are guaranteed (by Lemma~\ref{lem:symmetric}) that $\Pi=\Sigma$
and that $H'$ is symmetric.
Thus, by induction, we are guaranteed that $\rho_{k-1}^C=\rho_{k-1}^R$.
So the construction above gives $\rho^C=\rho^R$.

Finally, suppose that $H$ is symmetric and that it has a positive
diagonal.
Note that $n>2$.
By Lemma~\ref{lem:worsepermuting},
there is a permutations~$\Sigma\in S_n$ with $\Sigma(1)=1$
and   a normalised symmetric Hadamard matrix $H'$
satisfying (GC) such that $H_{\Sigma,\Pi}=H_4 \oplus H'$.
These are constructed in polynomial time.
By induction, 
we can construct an index mapping
$\rho':\gft^{k-2}\to[n]$  with $\rho'(0,\ldots,0)=
1$ and
a permutation $\pi' \in S_{k-2}$  
such that $H'$ is represented by the polynomial 
$$X_{\pi'(1)}Y_{1} \oplus \cdots \oplus X_{\pi'(k-2)} Y_{k-2}.$$

Now take $\rho(X_1,\dots,X_k)=
\Sigma (2^{k-1}X_k + 
2^{k-1}X_{k-1}+\rho'(X_1,\ldots,X_{k-2}))$
and   let $\pi\in S_k$ that the permutation that 
transposes $k$ and $k-1$
and applies $\pi'$ to $1,\ldots,k-2$.

\end{proof}

\begin{proof}[Proof of Lemma \ref{lem:poly_rep}, the Polynomial Representation Lemma]
Let $n=2^k$.
Since $H$ is positive for $\Lambda^R$ and $\Lambda^C$,
choose~$a$ and~$b$ such that
$H_{a,b}=+1$
and (1) $a\in\Lambda^R$ or $\Lambda^R=\emptyset$,
(2) $b\in\Lambda^C$ or $\Lambda^C=\emptyset$, and
(3) If $H$ is symmetric and $\Lambda^R=\Lambda^C$ then $a=b$.
Now let $\Sigma$ be the transposition $(1,a)$
and let $\Pi$ be the transposition $(1,b)$.
Not that $(H_{\Sigma,\Pi})_{1,1}=+1$.
Let $\widehat{H}$ be the matrix defined
by 
$$\widehat{H}_{i,j}
= 
{(H_{\Sigma,\Pi})}_{i,j} {(H_{\Sigma,\Pi})}_{i,1} {(H_{\Sigma,\Pi})}_{1,j}.$$
Note that $\widehat{H}$ is normalised.
Also, it is Hadamard and it satisfies \cond{GC}
since $H_{\Sigma,\Pi}$ is Hadamard and satisfies \cond{GC}.

By Lemma~\ref{lem:bipolar_backbone} we can construct $\widehat{\rho}^R$, $\widehat{\rho}^C$
and ${\pi}$ such that
$\widehat{H}$ is represented by the polynomial 
$\widehat{h}(X,Y) :=X_{{\pi}}Y$.
By the definition of ``represents'', we have
$$ \widehat{H}_{\widehat{\rho}^R(\vec x),\widehat{\rho}^C(\vec y)}=-1 \iff
\widehat{h}(\vec x, \vec y)=1. 
$$
 
Define $g^R(\vec x)=1$ if 
${(H_{\Sigma,\Pi})}_{\widehat{\rho}^R(\vec x),1}=-1$ and
$g^R(\vec x)=0$ otherwise.
Define $g^C(\vec y)=1$ if 
${(H_{\Sigma,\Pi})}_{1,
\widehat{\rho}^C(\vec y)}
=-1$ and
$g^C(\vec y)=0$ otherwise.
Now, note that
$$ (H_{\Sigma,\Pi})_{\widehat{\rho}^R(\vec x),\widehat{\rho}^C(\vec y)}=-1 \iff
\widehat{h}(\vec x, \vec y)
\oplus g^R(\vec x) \oplus g^C(\vec y)
=1. 
$$

Now let $\rho^R(\vec x)=\Sigma(\widehat{\rho}^R(\vec x))$
and let $\rho^C(\vec y)=\Pi(\widehat{\rho}^C(\vec y))$.
Note that 
$H$ is represented by 
$\widehat{h}(\vec x, \vec y)
\oplus g^R(\vec x) \oplus g^C(\vec y)$ with respect to
$\rho^R$ and $\rho^C$.

From Lemma~\ref{lem:bipolar_backbone},
$\widehat{\rho}^R(0,\ldots,0)=1$
so 
${\rho}^R(0,\ldots,0)=a$.
So if $\Lambda^R\neq\emptyset$ then 
$\rho^R(0,\ldots,0)\in\Lambda^R$.
Similarly,
${\rho}^R(1,\dots,1)=b$ so if $\Lambda^C\neq\emptyset$
then 
$\rho^C(0,\ldots,0)\in\Lambda^C$.

Finally, if $H$ is symmetric
then 
$H_{\Sigma,\Pi}$ is symmetric so
$\widehat{H}$ is symmetric so 
Lemma~\ref{lem:bipolar_backbone} guarantees that $\widehat{\rho}^R=
\widehat{\rho}^C$. 
Thus, if
$\Lambda^R=\Lambda^C$, then $a=b$ so 
$\Sigma=\Pi$ so
$g^R=g^C$ and $\rho^R=\rho^C$.

\end{proof}
 
\subsection{Linearity}
 
\begin{proof}[Proof of Lemma \ref{lem:bipolar_lam_linear}, the Linearity Lemma]

Let $H$ be an $n \times n$ Hadamard matrix and $\Lambda^R,\Lambda^C \subseteq  [n]$ subsets of indices.
Let $M,\Lambda$ be the bipartisation of $H$, $\Lambda^R$ and $\Lambda^C$ and let $m=2n$.
Suppose that 
\cond{GC} and \cond{R} are satisfied.
Let $n=2^k$ by Lemma~\ref{lem:bipolar_Hadamard_group}.
We will construct a matrix~$C$ and 
and a reduction 
$\eval{C,I_m,I_{m;\Lambda}} \Tle \eval{M,I_m,I_{m;\Lambda}}$. We will show that
$\eval{C,I_m,I_{m;\Lambda}}$ is \#P-hard unless \cond{L} is satisfied.

The reduction is as follows. 
Let $G=(V,E)$ be an input to $\eval{C,I_m,I_{m;\Lambda}}$.
We construct an input $G'$ to 
$\eval{M,I_m,I_{m;\Lambda}}$ as follows.
Each edge $\{u,v\}\in E$ corresponds to a gadget in $G'$ 
on vertex set $\{u,v,w,w',w''\}$ and edge set $\{\{u,w\},\{v,w\},\{w,w'\},\{w',w''\}\}$,
where $w$, $w''$ and $w''$ are new vertices.
 
Now let us construct the matrix~$C$. Let $\Gamma$ denote the graph with vertices~$u$ and~$v$
and a single edge between them. Clearly, $C_{a,b}$ is equal to the contribution to
$Z_M(\Gamma')$ corresponding to those configurations~$\xi$ 
with $\xi(u)=a$ and $\xi(v)=b$.
Thus, if $c$, $d$, and $e$ denote the choice of spins for 
vertices~$w$, $w'$ and $w''$, respectively we get
\begin{equation}
\label{eq:Csym}
C_{a,b}  =  
\sum_{c=1}^m M_{a,c}M_{b,c}(I_{m;\Lambda})_{c,c} \sum_{d=1}^m \sum_{e=1}^m M_{c,d} M_{d,e}(I_{m;\Lambda})_{e,e}.
\end{equation}
Here we use that the vertices $w,w''$ have odd degree and the vertex $w'$ has
even degree.
From the definition of bipartisation, 
we find that $C_{a,b}=C_{b,a}=0$ for all $a\in[n]$ and $b\in\{n+1,\dots,2n\}$.
Furthermore, for $a,b\in[n]$, 
\begin{align*}
C_{a,b}  & =  
\sum_{c=1}^{n} M_{a,n+c}M_{b,n+c}(I_{m;\Lambda})_{n+c,n+c} \sum_{d=1}^{n} \sum_{e=1}^{n} M_{n+c,d} M_{d,n+e}(I_{m;\Lambda})_{n+e,n+e}
\notag
\\
& =  
\sum_{c,e\in \Lambda^C} H_{a,c}H_{b,c}  \sum_{d=1}^{n}  H_{d,c} H_{d,e}.
\end{align*}

Now, by \cond{R},
there are bijective index mappings 
$\rho^R:\gft^k\to[n]$ and
$\rho^C:\gft^k\to[n]$ 
and a permutation $\pi \in S_k$ such that (w.r.t. $\rho^R$ and $\rho^C$) the matrix $H$ is represented by the polynomial  
$h(X,Y) = X_{\pi}Y \oplus g^R(X) \oplus g^C(Y)$.
Let 
$\tau^R$ be the inverse of $\rho^R$ and $\tau^C$ be the inverse of $\rho^C$. 
Let $L^C=\tau^C(\Lambda^C)$ and $L^R=\tau^R(\Lambda^R)$.
Also, 
let
$\alpha^R=\tau^R(a)$, 
$\beta^R=\tau^R(b)$,
$\gamma^C=\tau^C(c)$, 
$\delta^R=\tau^R(d)$ and 
$\epsilon^C=\tau^C(e)$.
Thus,
\begin{align*}
H_{a,c}H_{b,c}&=(-1)^{h(\alpha^R,\gamma^C)}\cdot(-1)^{h(\beta^R,\gamma^C)}\\
&=(-1)^{\alpha^R_\pi\cdot\gamma^C\oplus g^R(\alpha^R)\oplus g^C(\gamma^C)
\oplus\beta^R_\pi\cdot\gamma^C\oplus g^R(\beta^R)\oplus g^C(\gamma^C)}\\
&={(-1)}^{
g^R(\alpha^R) 
\oplus 
g^R(\beta^R)
\oplus 
\alpha^R_{\pi} \cdot \gamma^C
\oplus
\beta^R_{\pi} \cdot \gamma^C}\\
&=
{(-1)}^{
g^R(\alpha^R) 
\oplus 
g^R(\beta^R)
\oplus \gamma^C \cdot
(\alpha^R_\pi \oplus \beta^R_\pi)}.
\end{align*}
Similarly, we get
\[
H_{d,c}H_{d,e}={(-1)}^{
g^C(\gamma^C) 
\oplus 
g^C(\epsilon^C)
\oplus \delta^R_\pi \cdot
(\gamma^C \oplus \epsilon^C)}.
\]
So, for $a,b\in[n]$, 
\[
C_{a,b} = 
{(-1)}^{g^R(\alpha^R) \oplus g^R(\beta^R)}
\sum_{c,e\in \Lambda^C} 
{(-1)}^{ \gamma^C \cdot (
\alpha^R_\pi \oplus \beta^R_\pi)
\oplus
g^C(\gamma^C)\oplus g^C(\epsilon^C)}
\sum_{d=1}^{n}  
{(-1)}^{ \delta^R_\pi \cdot (\gamma^C \oplus \epsilon^C)}.
\]
Now note that 
\[
\sum_{d=1}^{n}  
{(-1)}^{ \delta^R_\pi \cdot (\gamma^C \oplus \epsilon^C)}
=
\sum_{\delta^R_\pi\in\gft^k } 
{(-1)}^{ \delta^R_\pi \cdot (\gamma^C \oplus \epsilon^C)}
=
\left\lbrace \begin{array}{l l}
		n &, \textrm{ if } \gamma^C = \epsilon^C\\
                0 &, \textrm{ otherwise,}
               \end{array}\right.
\]
so for $a,b\in[n]$, 
\begin{equation}
C_{a,b} = n
{(-1)}^{g^R(\alpha^R) \oplus g^R(\beta^R)}
\sum_{c\in \Lambda^C} 
{(-1)}^{ \gamma^C \cdot (
\alpha^R_\pi \oplus \beta^R_\pi)
}
= n
{(-1)}^{g^R(\alpha^R) \oplus g^R(\beta^R)}
\sum_{\gamma^C\in L^C} 
{(-1)}^{ \gamma^C \cdot (
\alpha^R_\pi \oplus \beta^R_\pi)
}.
\label{eq:topC}
\end{equation}

Similarly,  
\begin{align*}
C_{a+n,b+n}  & =  
\sum_{c=1}^{n} M_{a+n,c}M_{b+n,c}(I_{m;\Lambda})_{c,c} \sum_{d=1}^{n} \sum_{e=1}^{n} M_{c,d+n} M_{d+n,e}(I_{m;\Lambda})_{e,e}
\notag
\\
& =  
\sum_{c,e\in \Lambda^R} H_{c,a}H_{c,b}  \sum_{d=1}^{n}  H_{c,d} H_{e,d},
\end{align*} 
so taking
$\alpha^C=\tau^C(a)$,
$\beta^C=\tau^C(b)$, and
$\gamma^R=\tau^R(c)$, we get
\begin{equation}
\label{eq:botC}
C_{a+n,b+n} = n
{(-1)}^{g^C(\alpha^C) \oplus g^C(\beta^C)}
\sum_{\gamma^R\in L^R} 
{(-1)}^{ 
\gamma^R_{\pi} \cdot
(\alpha^C \oplus \beta^C)
 }.  \end{equation}

Let $\lambda^C=|L^C|$ and $\lambda^R=|L^R|$.
We will now assume that $\eval{C,I_m,I_{m;\Lambda}}$ is not \#P-hard.
Using this assumption, we will show that~$L^C$ and~$L^R$ are
linear subspaces of~$\gft^k$, which implies that \cond{L} is satisfied.
We give the argument for $L^C$. The argument for $L^R$ is symmetric.

If $L^C$ is empty then it is a linear subspace of~$\gft^k$, so assume that it is non-empty.
Condition \cond{R} guarantees that, since $\Lambda^C$ is non-empty, $\rho^C(\vec 0)\in \Lambda^C$.
Hence, $\vec{0} \in L^C$.

Let $\mathcal{L}$ be the subspace of $\gft^k$ spanned by~$L^C$.
$\mathcal{L}$ contains all linear combinations of elements of~$L^C$. We will show that
$L^C=\mathcal{L}$, so $L^C$ is a linear subspace of $\gft^k$.
  
By Equation~(\ref{eq:Csym}), the matrix~$C$ is symmetric.
By Equation~(\ref{eq:topC}), we have
$C_{a,a}=n\lambda^C$ for $a\in [n]$.
Thus, by Lemma~\ref{lem:bipolar_block2_hard} (due to Bulatov and Grohe)
$C_{a,b} \in \{-n\lambda^C,0,n\lambda^c\}$ for all $a,b\in[n]$. Otherwise, 
$\eval{C,I_m,I_{m;\Lambda}}$ is \#P-hard.
Let $\chi = \alpha^R_\pi \oplus \beta^R_\pi$.
Since $C_{a,b} \in \{-n\lambda^C,0,n\lambda^c\}$,  Equation~(\ref{eq:topC}), implies that 
for every such $\chi\in\gft^k$,
$$ 
\sum_{\gamma\in L^C} 
{(-1)}^{ \gamma \cdot \chi}
\in\{-\lambda^C,0,\lambda^C\}.
$$
Since $\vec{0} \in L^C$, one of the items  in the summation is
$(-1)^{\vec 0\cdot\xi}=1$,
so the outcome $-\lambda^C$ is not possible. Therefore, 
we get
\begin{equation}
\sum_{\gamma\in L^C} 
{(-1)}^{ \gamma \cdot \chi}
\in\{0,\lambda^C\}, \mbox{for all $\chi\in\gft^k$.}
 \label{eq:zero_lambda2}
\end{equation}

Let
$\Xi_0 = \{\chi \in \gft^k \mid  \forall\gamma \in L^C,
\chi \cdot \gamma = 0\}$.
If $\chi \in \Xi_0$ then $\chi \cdot \gamma = 0$ for all $\gamma \in \mathcal{L}$.  
Otherwise,   by the linearity of $\mathcal{L}$, 
\[
|\{\gamma \in \mathcal{L} : \chi \cdot \gamma = 0\}| = |\{\gamma \in \mathcal{L} : \chi \cdot \gamma = 1\}|.
\]
Thus
\[
\sum_{\gamma \in \mathcal{L}} {(-1)}^{\chi \cdot \gamma} = 
\left\lbrace \begin{array}{r l}
		|\mathcal{L}| &, \textrm{ if } \chi \in \Xi_0\\
                 0 &, \textrm{ otherwise }
               \end{array}\right.
\]
Hence (the characteristic functions of) the sets $L^C$ and
$\mathcal{L}$ have the same Fourier transform, up to scaling. It
follows\leslie{Should Do: I'd like to see more detail here.
Can anyone think of a convincing reference or way to explain this?
}  that $\mathcal{L} =
L^C$ and $L^C$ is a linear subspace of $\gft^k$ as required.

Finally, note that it is easy, in polynomial time, given $H$,
to construct $C$ and to determine
whether, for all $a,b\in[n]$,
$C_{a,b} \in \{-n\lambda^C,0,n\lambda^c\}$ 
and 
$C_{n+a,n+b} \in \{-n\lambda^R,0,n\lambda^R\}$. 
Thus, it is easy, in polynomial time, to determine whether \cond{L} holds.
\end{proof}

The following fact about linear maps will be useful later. 

\begin{lemma}\label{cor:bipolar_lam_linear}
Let $\phi: \gft^\ell \rightarrow \gft^k$ be a linear map.
There is a surjective map 
$f: \gft^k \rightarrow \gft^{\ell}$ 
and a constant $z\in \Nat$ such that
\begin{itemize}
\item
$f(c_1,\ldots,c_k) \cdot (x_1,\ldots,x_{\ell}) =
(c_1,\ldots,c_k) \cdot \phi(x_1,\ldots,x_\ell)$, and
\item 
$\forall (c'_1,\ldots,c'_{\ell}), z=
\left|
\left\{
(c_1,\ldots,c_k) \mid 
f(c_1,\ldots,c_k) =  (c'_1,\ldots,c'_{\ell})
\right\}
\right|$.
\end{itemize}
\end{lemma}
\begin{proof}
Let $B$ be the $\ell \times k$ matrix defining $\phi$, 
i.e. $\phi(x_1,\ldots,x_\ell) = (x_1,\ldots,x_\ell)B$. Define $f$
by $f(c_1,\ldots,c_k)= (c_1,\ldots,c_k) B^T$.
Then letting $\vec x$ denote the row vector $(x_1,\ldots,x_\ell)$,
\begin{align*}
f(c_1,\ldots,c_k) \cdot (x_1,\ldots,x_\ell)&= 
f(c_1,\ldots,c_k) {\vec x}^T\\
&= (c_1,\ldots,c_k) B^T {\vec x}^T\\
&=
(c_1,\ldots,c_k) {(\vec x B)}^T\\
&= (c_1,\ldots,c_k) \cdot \phi(x_1,\ldots,x_\ell).
\end{align*}
 Fix any 
$\vec {c'}\in\gft^{\ell}$ and any
$\vec c\in\gft^k$   such that 
$f(\vec c) = \vec {c'}$. 
Note that 
\[
f^{-1}(\vec{c'}) = 
\mset{\vec c+\vec x \mid \vec x\in \gft^k,\, 
f(\vec c+\vec x) = \vec {c'}}.
\] 
As $f$ is linear, we have $f(\vec c + \vec x) = f(\vec c) + f(\vec x) = \vec{c'} + f(\vec x)$ 
so 
$f^{-1}(\vec{c'}) = 
\mset{\vec c+\vec x \mid \vec x\in \gft^k,\, 
f(\vec x) = \vec {0}}$. 
Thus, we take $z= 
|\{
\vec x \in \gft^k \mid f(\vec x)=\vec 0
\}|$.
\end{proof}

\subsection{The Degree Condition} 

Let $X=(X_1,\ldots,X_k)$.
Every polynomial in $g(X_1,\ldots,X_k)\in \gft[X_1,\dots,X_k]$ can be
written as a sum of distinct monomials of the form $X_{i_1}\cdot X_{i_2}\cdots X_{i_j}$ for
$1\le i_1<\ldots<i_j\le k$.
Given
a polynomial $g(X)$,
let $\hash(g(X))=|\{\alpha\in\gft^k \mid g(\alpha)=1\}|$.
For
$\alpha,\beta,\gamma\in \gft^k$, let
$$\gquad(X) = g(\alpha \oplus X) \oplus g(\beta \oplus X) \oplus  \gamma \cdot X.$$

\begin{lemma} \label{lem:high_deg_poly}
Let $g \in \gft[X_1, \ldots, X_k]$ be of degree at least $3$. 
Suppose that variables $X_r$, $X_s$ and $X_t$ are contained in a monomial of degree at least~$3$.
Let $\beta=\vec 0$
and let $\alpha\in \gft^k$ be the vector which is all zero except at index~$r$.
Then there are
polynomials 
$h$, $h_{r,s}$, $h_{r,t}$ and $h_r$  
such that $h$ is not identically~$0$ and
\begin{equation}
\gquad(X)  =  X_sX_t h(X \setminus \{r,s,t\}) 
\oplus X_s h_{r,s}(X \setminus \{r,s,t\}) \oplus X_t 
h_{r,t}(X\setminus \{r,s,t\}) \oplus h_{r}(X\setminus \{r,s,t\}) \oplus \gamma \cdot X
\end{equation}
for all $\gamma\in\gft^k$.
\end{lemma}
\begin{proof}

Let $Z$ denote the tuple $X \setminus \{r,s,t\}$.
Let $h'(X)$ be the sum of all monomials of~$g$ that contain $X_r$, $X_s$ and $X_t$.
Let $h(Z)$ be the polynomial satisfying $h'(X) = X_r X_s X_t h(Z)$.
Note that $h(Z)$ is not identically zero.
Choose $h_{r,s}$, $h_{r,t}$, $h_{s,t}$, $h_r$, $h_s$, $h_t$ and $h_\emptyset$
so that 
\begin{eqnarray*}
g(X) & = & X_rX_sX_th(Z) \oplus X_rX_sh_{r,s}(Z) \oplus X_rX_th_{r,t}(Z) \oplus X_sX_th_{s,t}(Z) \\
     &   & \oplus X_rh_{r}(Z) \oplus X_sh_{s}(Z) \oplus X_th_{t}(Z) \oplus h_{0}(Z)
\end{eqnarray*}  
 
Then for $\alpha$ and $\beta$ as defined in the statement of the lemma, we have 
\begin{eqnarray*}
g(\alpha\oplus X) \oplus g(\beta\oplus X) & = & g(\alpha\oplus X) \oplus g(X)  \\ 
& = & \left( (X_r \oplus 1)X_sX_t \oplus X_rX_sX_t\right)h(Z) \\
& & \oplus ((X_r\oplus 1)X_s \oplus X_rX_s)h_{r,s}(Z) \\
& & \oplus ((X_r\oplus 1)X_t \oplus X_rX_t)h_{r,t}(Z) \oplus h_{r}(Z) \\
& = & X_sX_th(Z) \oplus X_sh_{r,s}(Z) \oplus X_th_{r,t}(Z) \oplus h_{r}(Z), \\
\end{eqnarray*}
which finishes the proof.
\end{proof}

\begin{lemma}\label{lem:char_quadratic}
Let $g(X) \in \gft[X_1,\ldots,X_k]$. 
The following are equivalent.
\begin{enumerate}
\item $g$ has degree at most $2$.
\item 
For all $\alpha$ and $\beta$ in $\gft^k$,
\begin{itemize}
\item there is \emph{exactly} one $\gamma \in  \gft^k$ such that 
$\#(\gquad(X)) \in \mset{0,2^k}$, and
\item For all $\gamma' \neq \gamma$, $\#(g_{\alpha,\beta,\gamma'}(X)) = 2^{k-1}$.
\end{itemize}
\end{enumerate}
Also, if $g$ has degree greater than $2$ then 
there are
$\alpha$ and $\beta$ 
in $\gft^k$
for which
there is no $\gamma\in  \gft^k$ such that
$\#(\gquad(X)) \in \{0,2^k\}$.
\end{lemma}

\begin{proof}
Suppose that $g$ has degree at most~$2$.
Let
$g'(X):= g(\alpha\oplus X) \oplus g(\beta \oplus X)$. 
Consider any degree-$2$ term $X_r X_s$ in~$g$.
In $g'$, this term becomes 
$(X_r \oplus \alpha_r)(X_s \oplus \alpha_s) \oplus (X_r \oplus \beta_r)(X_s \oplus \beta_s)$. 
Now$(X_r \oplus \alpha_r)(X_s \oplus \alpha_s) = X_rX_s \oplus X_r\alpha_s  \oplus \alpha_rX_s \oplus \alpha_r\alpha_s$, 
so the term $X_rX_s$ cancels in $g'$. 
We conclude that 
$g'(X)$ is linear in~$X_1,\ldots,X_k$ and (2) holds.

Conversely, suppose that $g$ has degree at least~$3$.
Suppose that variables $X_r$, $X_s$ and $X_t$ are contained in a monomial of degree at least~$3$.
Let $\beta=\vec 0$
and let $\alpha\in \gft^k$ be the vector which is all zero except at index~$r$.
By Lemma~\ref{lem:high_deg_poly},
there are
polynomials 
$h$, $h_{r,s}$, $h_{r,t}$ and $h_r$  
such that $h$ is not identically~$0$ and
$$ 
\gquad(X)  =  X_sX_t h(X \setminus \{r,s,t\}) 
\oplus X_s h_{r,s}(X \setminus \{r,s,t\}) \oplus X_t 
h_{r,t}(X\setminus \{r,s,t\}) \oplus h_{r}(X\setminus \{r,s,t\}) \oplus \gamma \cdot X.
$$
Since $h$ is not identically~$0$, the term 
$X_sX_t h(X \setminus \{r,s,t\})$ does not cancel for any choice of $\gamma$. Hence, there is no 
$\gamma$ such that
$\#(\gquad(X)) \in \mset{0,2^k}$, so (2) does not hold.
\end{proof}

\begin{lemma}\label{lem:exclude_halfness}
Let $g \in \gft[X_1, \ldots, X_k]$ be of degree at least $3$. 
Suppose that variables $X_r$, $X_s$ and $X_t$ are contained in a monomial of degree at least~$3$.
Let $\beta=\vec 0$
and let $\alpha\in \gft^k$ be the vector which is all zero except at index~$r$.
Then 
there is a $\gamma\in\gft^k$ such that 
$$ \#(\gquad(X)) \neq 2^{k-1}.$$
\end{lemma}

\begin{proof}

Suppose, for contradiction, that $ \#(\gquad(X)) = 2^{k-1}$
for every $\gamma\in\gft^k$.
Let $Z$ denote the tuple $X \setminus \{r,s,t\}$.
By Lemma~\ref{lem:high_deg_poly},
there are
polynomials 
$h$, $h_{r,s}$, $h_{r,t}$ and $h_r$  
such that $h$ is not identically~$0$ and
\begin{equation}
\label{eq:fullgquadX}
\gquad(X)  =  X_sX_t h(Z)  \oplus X_s h_{r,s}(Z) \oplus X_t 
h_{r,t}(Z) \oplus h_{r}(Z) \oplus \gamma \cdot X.
\end{equation}
 
Let $\gamma^r\in\gft^{k-1}$ denote the
vector obtained from~$\gamma$ by deleting component~$\gamma_r$. 
Let $\gamma'\in\gft^{k-3}$
denote the vector obtained from~$\gamma$ by deleting components~$\gamma_r$, $\gamma_s$ and $\gamma_t$.
Let
$$g'_{\alpha,\beta,\gamma}(X\setminus \{X_r\}) = 
X_sX_t h(Z)  \oplus X_s h_{r,s}(Z) \oplus X_t 
h_{r,t}(Z) \oplus h_{r}(Z) \oplus \gamma^r \cdot (X\setminus \{X_r\}),$$
so $\gquad(X) = g'_{\alpha,\beta,\gamma}(X\setminus \{X_r\}) \oplus \gamma_r X_r$. The polynomial
$g'_{\alpha,\beta,\gamma}(X\setminus \{X_r\})$ can be simplified   
as in the following table, depending on the possible values of~$X_s$ and~$X_t$.

\bigskip
\begin{center}
\begin{tabular}{|c|c|l|}
\hline
$X_s$ & $X_t$ & $g'_{\alpha,\beta,\gamma}(X\setminus \{X_r\})$ \\
\hline \hline
$0$ & $0$ &   $h_{r}(Z) \oplus \gamma' \cdot Z$\\
$1$ & $0$ &   $h_{r,s}(Z) \oplus h_{r}(Z) \oplus \gamma' \cdot Z \oplus \gamma_s$\\
$0$ & $1$ &   $h_{r,t}(Z) \oplus h_{r}(Z) \oplus \gamma' \cdot Z \oplus \gamma_t$\\
$1$ & $1$ &   $h(Z) \oplus h_{r,s}(Z) \oplus h_{r,t}(Z) \oplus h_{r}(Z) \oplus \gamma' \cdot Z \oplus \gamma_s \oplus \gamma_t$\\
\hline 
\end{tabular} 
\end{center}
Define 
\begin{eqnarray*}
\eta_{0}\;\; & = & \#(h_{r}(Z) \oplus    \gamma'\cdot Z) \\
\eta_{1^+} & = & \#(h_{r,s}(Z) \oplus h_{r}(Z) \oplus \gamma'\cdot Z)\\
\eta_{1^-} & = & \#(h_{r,s}(Z) \oplus h_{r}(Z) \oplus \gamma'\cdot Z \oplus 1)\\
\eta_{2^+} & = & \#(h_{r,t}(Z) \oplus h_{r}(Z) \oplus \gamma'\cdot Z)\\
\eta_{2^-} & = & \#(h_{r,t}(Z) \oplus h_{r}(Z) \oplus \gamma'\cdot Z \oplus 1)\\
\eta_{3^+} & = & \#(h(Z) \oplus h_{r,s}(Z) \oplus h_{r,t}(Z) \oplus h_{r}(Z) \oplus \gamma'\cdot Z)\\
\eta_{3^-} & = & \#(h(Z) \oplus h_{r,s}(Z) \oplus h_{r,t}(Z) \oplus h_{r}(Z) \oplus \gamma'\cdot Z \oplus 1)\\
\end{eqnarray*}

We can express $\#(g'_{\alpha,\beta,\gamma}(X\setminus \{X_r\}))$ in terms of 
$\eta_0$, $\eta_{1^+}$, $\eta_{1^-}$, $\eta_{2^+}$, $\eta_{2^-}$, $\eta_{3^+}$
and
$\eta_{3^-} $, depending on the values of $\gamma_s$ and $\gamma_t$.

\begin{center}
\begin{tabular}{|c|c|l|}
\hline
\smallskip
$\gamma_s$ & $\gamma_t$ & $\#(g'_{\alpha,\beta,\gamma}(X\setminus \{X_r\}))$ \\
\hline \hline
$0$ & $0$ &   $\eta_{0} + \eta_{1^+} + \eta_{2^+} + \eta_{3^+} $\\
$1$ & $0$ &   $\eta_{0} + \eta_{1^-} + \eta_{2^+} + \eta_{3^-}$\\
$0$ & $1$ &   $\eta_{0} + \eta_{1^+} + \eta_{2^-} + \eta_{3^-}$\\
$1$ & $1$ &   $\eta_{0} + \eta_{1^-} + \eta_{2^-} + \eta_{3^+}$\\
\hline 
\end{tabular} 
\end{center}

Since $Z$ is a tuple of $k-3$ variables, each of  
$\eta_0$, $\eta_{1^+}$, $\eta_{1^-}$, $\eta_{2^+}$, $\eta_{2^-}$, $\eta_{3^+}$
and
$\eta_{3^-} $
is between~$0$ and $2^{k-3}$.
Furthermore, we have $\eta_{i^+} + \eta_{i^-} = 2^{k-3}$ for all $i \in [3]$.  
We are assuming $\#(\gquad(X))=2^{k-1}$ for any~$\gamma$, so for any $\gamma$ with $\gamma_r=0$, 
Equation~(\ref{eq:fullgquadX}) implies
$\#(g'_{\alpha,\beta,\gamma}(X\setminus \{X_r\}))
=2^{k-2}$.
Altogether, we obtain the following system of linear equations, which is applicable for any $\gamma$ with
$\gamma_r=0$.
\[
\left(\begin{array}{c c c c c c c}
0& 1& 1 & 0 & 0 & 0 & 0\\
0& 0& 0 & 1 & 1 & 0 & 0\\
0& 0& 0 & 0 & 0 & 1 & 1\\
1& 1& 0 & 1 & 0 & 1 & 0\\
1& 0& 1 & 1 & 0 & 0 & 1\\
1& 1& 0 & 0 & 1 & 0 & 1\\
1& 0& 1 & 0 & 1 & 1 & 0
\end{array}\right)
\left( \begin{array}{c}
        \eta_{0}\\ \eta_{1^+}\\ \eta_{1^-}\\ \eta_{2^+}\\ \eta_{2^-} \\ \eta_{3^+} \\ \eta_{3^-}       
\end{array}\right)
=
\left( \begin{array}{c}
       2^{k-3}\\ 2^{k-3}\\ 2^{k-3} \\ 2^{k-2}\\ 2^{k-2}\\ 2^{k-2} \\ 2^{k-2}
\end{array}\right)
\]

Solving this system yields $\eta_0 = \eta_{1^+} = \ldots = \eta_{3^-} = 2^{k-4}$.
Since $\eta_0=2^{k-4}$,
\begin{equation}
\label{eq:contradict}
\forall \gamma'\in\gft^{k-3},
\#(h_{r}(Z) \oplus  \gamma' \cdot Z) = 
\#(h_{r}(Z) \oplus 1 \oplus \gamma' \cdot Z)= 2^{k-4}.
\end{equation}

We will use Equation~(\ref{eq:contradict}) to derive a contradiction.
Let $W=(W_1,\ldots,W_{k-3})$
and let $Y=(Y_1,\ldots,Y_{k-3})$.
Let $f(W,Y) = h_r(W) \oplus h_r(Y) \oplus W \cdot Y$.
For $\gamma'\in\gft^{k-3}$,
let $f_{\gamma'}(Y)= f(\gamma',Y)$.
by Equation~(\ref{eq:contradict}),
\begin{equation}
\label{eq:rowscontradict}
\forall \gamma'\in\gft^{k-3},
\hash(f_{\gamma'}(Y))=2^{k-4}.
\end{equation}

Note that $f$ represents a symmetric Hadamard matrix $H_f$ of order $2^{k-3}$.\leslie{Should Do: We should come back and add more here. (Does anyone have any ideas?)}
So equation~(\ref{eq:rowscontradict}) says that all rows of $H_f$ have sum~$0$. 
This is impossible because the rows, together with the all-ones vector would then be an
$2^{k-3} + 1$ element basis of a $2^{k-3}$ dimensional vector space. So we have a contradiction.
\end{proof}

\begin{cor}\label{lem:weakened_char_quad}
Let $g(X) \in \gft[X_1, \ldots, X_k]$. 
The following are equivalent.
\begin{enumerate}
\item $g$ has degree at most $2$.
\item For all $\alpha\neq \beta$ in $\gft^k$,
\begin{equation}
\mbox{there is at most one $\gamma\in\gft^k$ such that }
 \label{eq:const}
\#(\gquad(X)) \in \{0,2^k\},\mbox{and }
\end{equation}
\begin{equation}
\mbox{for all $\gamma'\neq\gamma$, }
\label{eq:half}
\#(g_{\alpha,\beta,\gamma'}(X)) = 2^{k-1}.
\end{equation}
\end{enumerate}
\end{cor}

\begin{proof}
If $g$ has degree at most~$2$ then (2) holds by
Lemma~\ref{lem:char_quadratic}.
Suppose that $g$ has degree at least~$3$.
Lemma~\ref{lem:char_quadratic}
provides an $\alpha$ and $\beta$ such that
there is no $\gamma$ such that
$\#(\gquad(X)) \in \{0,2^k\}$.
So to prove the theorem we just have to rule out the
case that 
every $\gamma$ satisfies
$\#(g_{\alpha,\beta,\gamma}(X)) = 2^{k-1}$
for this choice of $\alpha$ and $\beta$, and this is ruled out by
Lemma~\ref{lem:exclude_halfness}.
\end{proof}

\begin{proof}[Proof of Lemma \ref{lem:degree_two}, The Degree Lemma] 
  Let $H$ be an $n \times n$ Hadamard matrix and $\Lambda^R,\Lambda^C
  \subseteq [n]$ subsets of indices.  Let $M,\Lambda$ be the bipartisation of
  $H$, $\Lambda^R$ and $\Lambda^C$ and let $m=2n$.  Suppose that
  \cond{GC},\cond{R} and \cond{L} are satisfied.  For integers~$p$ we will
  construct a matrix~$C^{[p]}$ 
and a reduction $\eval{C^{[p]}} \Tle
  \eval{M,I_m,I_{m;\Lambda}}$.  We will show that if \cond{D} does not hold
  then there is a $p$  such that
  $\eval{C^{[p]}}$ is \hash P-hard.
 
  The reduction is as follows. Let $G=(V,E)$ be an input to $\eval{C^{[p]}}$.
  We construct an input $G'$ to $\eval{M,I_m,I_{m;\Lambda}}$ as follows.  Each
  edge $\{u,v\}\in E$ corresponds to a ``lotus'' gadget in~$G'$.  The vertex
  set of the gadget is $\{u,v,u'_{i},v'_{i},
  u''_{i},v''_{i},x_i,y_{i},z,w \mid i\in[p]\}$.  See Figure
  \ref{fig:lotus} for an illustration of the lotus gadget for $p=1$.  The
  gadget has the following edges, for all $i \in [p]$: 
  $\mset{z,x_i}$, $\mset{w,x_i}$, $\mset{z,y_{i}}$, $\mset{w,y_{i}}$,
  $\mset{u, u'_{i}}$, $\{u'_{i},u''_{i}\}$, $\mset{x_{i},u'_{i}}$,
  $\mset{x_{i},v'_{i}} $, $\mset{v, v'_{i}}$, $\{v'_{i},v''_{i}\}$,
  $\mset{y_{i},u'_{i}}$, and $\mset{y_{i},v'_{i}}$.

  Note that the vertices of the gadget have the following degrees:
  \begin{gather*}
    d(u''_{i}) = d(v''_{i}) = 1\\
  d(u'_{i})= d(v'_{i}) = d(x_i)=d(y_{i}))= 4,\\
  d(z)=d(w)=2p.
  \end{gather*}
  Furthermore, for the ``boundary'' vertices $u,v$ we have
  \[
  d_{G'}(u) = p
  \cdot d_G(u),\quad
  d_{G'}(v) = p
  \cdot d_G(v).
  \]
  We will stipulate that $p$ is even. Then the degree
  of vertices, except for the $u''_{i}$ and $v''_{i}$, is
  even.
\begin{figure}
\begin{center} 
\resizebox{5cm}{!}{
\includegraphics{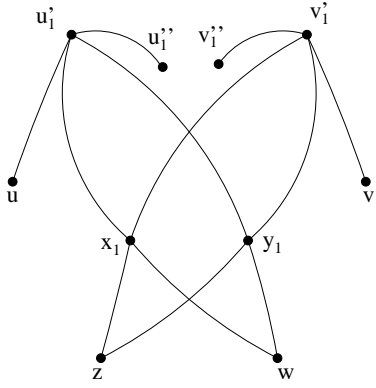}
}
\end{center}
\caption{The Lotus gadget for $p=1$.}
\label{fig:lotus}
\end{figure}

Now let us construct the matrix $C^{[p]}$. Let $\Gamma$ denote the graph with vertices $u$ and $v$ and a single edge between them.
Clearly, $C^{[p]}_{a,b}$ is equal to the contribution to $Z_{M,I_m,I_m;\Lambda}(\Gamma')$
corresponding to those configurations $\xi$ with $\xi(u)=a$ and $\xi(v)=b$.
 
By \cond{R},
there are bijective index mappings 
$\rho^R:\gft^k\to[n]$ and
$\rho^C:\gft^k\to[n]$ 
and a permutation $\pi \in S_k$ such that (w.r.t. $\rho^R$ and $\rho^C$) the matrix $H$ is represented by the polynomial  
$h(X,Y) = X_{\pi}Y \oplus g^R(X) \oplus g^C(Y)$.
Let 
$\tau^R$ be the inverse of $\rho^R$ and $\tau^C$ be the inverse of $\rho^C$. 
Let $L^C=\tau^C(\Lambda^C)$ and $L^R=\tau^R(\Lambda^R)$.
By condition \cond{L} we know that the sizes of $L^R$ and $L^C$ are powers of~$2$.
Let $|L^R|=2^{\ell^R}$ and
let $|L^C|=2^{\ell^C}$. If $\Lambda^R$ is nonempty then
let 
$\phi^R: \gft^{\ell^R} \rightarrow \gft^k$
be a coordinatisation of $\Lambda^R$ with respect to $\rho^R$. Similarly, if
$\Lambda^C$ is nonempty, let
$\phi^C$ be a coordinatisation of $\Lambda^C$ with respect to $\rho^C$.
Let $\phi^C=\phi^R$ if $\Lambda^C=\Lambda^R$ and this is nonempty and $H$ is symmetric.
Note that if $\Lambda^C$ and $\Lambda^R$ are empty then \cond{D} is satisfied.

Let $\Gamma_{i}$ be the subgraph of $\Gamma'$ induced by $\{u, x_i, y_{i},
u'_{i},u''_{i}\}$.  For $\alpha,\gamma,\delta\in\gft^k$, let
$a^R=\rho^R(\alpha)$, $c^R=\rho^R(\gamma)$ and $d^R=\rho^R(\delta)$.  Let
$Z^R(\alpha,\gamma,\delta)$ denote the contribution to
$Z_{M,I_m,I_m;\Lambda}(\Gamma_{i})$ corresponding to those configurations
$\xi$ with $\xi(u)=a^R$ and $\xi(x_i)=c^R$ and $\xi(y_{i})=d^R$, ignoring
contributions due to $I_{m;\Lambda}$ for vertices $u$, $x_i$, and $y_{i}$.
(We ignore these contributions because these vertices will have even degree in
$G'$ so these contributions will cancel when we use $Z(\alpha,\beta,\gamma)$.)
Using $n+a'$ to denote the spin at~$u'_{i}$ (which must be in the range
$\{n+1,\ldots,2n\}$, otherwise the contribution is zero) and $a''$ to denote
the spin at~$u''_{i}$ (which must be in $[n]$), we get
\begin{align*}
Z^R(\alpha,\gamma,\delta) & = 
\sum_{a'=1}^n \sum_{a''=1}^n M_{a^R,n+a'} M_{n+a',a''} M_{c^R,n+a'} M_{d^R,n+a'} (I_{m;\Lambda})_{a'',a''}
\\ & = \sum_{a''\in \Lambda^R} \sum_{a'=1}^n H_{a^R,a'} H_{a'',a'} H_{c^R,a'} H_{d^R,a'}.
\end{align*}

Plugging in the representation of~$H$ where $\rho^R(\phi^R(\mu))$ is the spin $a''\in\Lambda^R$, we get
the following.
\begin{align*}
Z^R(\alpha,\gamma,\delta)  = 
{(-1)}^{g^R(\alpha)\oplus g^R(\gamma) \oplus g^R(\delta)}
\sum_{\mu\in \gft^{\ell^R}} 
{(-1)}^{g^R({\phi^R(\mu)})}
\sum_{{{\alpha'}}\in\gft^k}
{(-1)}^{ {\alpha'} \cdot (
\alpha_\pi \oplus {\phi^R(\mu)}_\pi \oplus {\gamma}_\pi \oplus \delta_\pi
)}.
\end{align*}

Note that
$$
\sum_{{{\alpha'}}\in\gft^k}
{(-1)}^{ {\alpha'} \cdot (
\alpha_\pi \oplus {\phi^R(\mu)}_\pi \oplus {\gamma}_\pi \oplus \delta_\pi
)}
 = \left\lbrace \begin{array}{l l}
 n &, \text{ if } 
{
 \phi^R(\mu)}_\pi = \alpha_\pi \oplus  {\gamma}_\pi \oplus \delta_\pi
 \\
 0 &, \text{ otherwise}
\end{array}\right.$$
Equivalently,
$$
\sum_{{{\alpha'}}\in\gft^k}
{(-1)}^{ {\alpha'} \cdot (
\alpha_\pi \oplus {\phi^R(\mu)}_\pi \oplus {\gamma}_\pi \oplus \delta_\pi
)}
 = \left\lbrace \begin{array}{l l}
 n &, \text{ if } 
{
 \phi^R(\mu)} = \alpha \oplus  {\gamma} \oplus \delta
 \\
 0 &, \text{ otherwise}
\end{array}\right.$$

Thus,  $Z^R(\alpha,\gamma,\delta)=0$ unless
$\alpha \oplus  {\gamma} \oplus \delta  \in L^R$
and in this case,
\begin{equation}
\label{eqZR}
Z^R(\alpha,\gamma,\delta)  = n
{(-1)}^{g^R(\alpha)\oplus g^R(\gamma) \oplus g^R(\delta) \oplus g^R(\alpha \oplus \gamma \oplus \delta)}.\end{equation}

Our strategy for the rest of the proof is the following: The goal is to prove
that either there 
is a $p$
such that $\eval{C^{[p]}}$ is \#P-hard or the
following two conditions are satisfied:
\begin{description}
  \item[Row Condition] Either $\Lambda^R$ is empty or the
polynomial $g^R\circ\phi^R$ has degree at most $2$.
  \item[Column Condition] Either $\Lambda^C$ is empty or the
polynomial $g^C\circ\phi^C$ has degree at most $2$.
\end{description}
Let us turn to the Row Condition first. Suppose that $\Lambda^R$ is nonempty;
otherwise there is nothing to prove. Let $a,b\in \Lambda^R$.
Define $\alpha^R$ and $\beta^R$ in $\gft^{\ell^R}$ so that
$\phi^R(\alpha^R)=\tau^R(a)$ and $\phi^R(\beta^R)=\tau^R(b)$.
Note that contribution to $Z_{M,I_m,I_m;\Lambda}(\Gamma')$
of a configuration 
$\xi$ with $\xi(u)=a$ and $\xi(v)=b$
is zero unless the
spins of vertices $u'_{i}$, $v'_{i}$, $z$ and $w$ are in $\{n+1,\ldots,2n\}$ and
the rest of the spins are in $[n]$.
Then taking 
$\rho^C(\epsilon)+n$ as the spin of $z$ and
$\rho^C(\zeta)+n$ as the spin of~$w$, we get
\begin{align*}
C^{[p]}_{a,b} & = 
\sum_{\epsilon,\zeta \in\gft^k}
\prod_{i=1}^p
\left(
\sum_{\gamma_i,\delta_i\in\gft^k}
Z^R(\phi^R(\alpha^R),\gamma_i,\delta_{i}) Z^R(\phi^R(\beta^R),\gamma_i,\delta_{i})
(-1)^{
({(\gamma_i)}_\pi \oplus {(\delta_{i})}_\pi)\cdot
(\epsilon \oplus \zeta) }
\right)\\
& = 
\sum_{\epsilon,\zeta\in\gft^k}
\left( 
  \sum_{\gamma,\delta\in\gft^k}
Z^R(\phi^R(\alpha^R),\gamma,\delta) Z^R(\phi^R(\beta^R),\gamma,\delta)
(-1)^{
({\gamma}_\pi \oplus {\delta}_\pi)\cdot
(\epsilon \oplus \zeta) }
\right)^p 
\end{align*}
 From Equation~(\ref{eqZR}) we find that if 
we take any 
${\gamma'}$ and ${\delta'}$
such that
${\gamma'} \oplus {\delta'} = \gamma \oplus \delta$
then 
$Z^R(\alpha,\gamma,\delta) Z^R(\beta,\gamma,\delta)=
Z^R(\alpha,\gamma',\delta') Z^R(\beta,\gamma',\delta')
 $ for any $\alpha$ and $\beta$.
Thus, we can simplify the expression using 
$\psi$ to denote $\epsilon \oplus \zeta$ and
$\eta$ to denote
$\gamma \oplus \delta$.
\begin{align*}
C^{[p]}_{a,b} 
& = n
\sum_{\psi\in\gft^k}
\left( 
  \sum_{\gamma,\delta\in\gft^k}
Z^R(\phi^R(\alpha^R),\gamma,\delta) Z^R(\phi^R(\beta^R),\gamma,\delta)
(-1)^{
({\gamma}_\pi \oplus {\delta}_\pi)\cdot
\psi }\right)^p
\\ 
&= n
\sum_{\psi\in\gft^k}
\left( 
  n
  \sum_{\eta\in\gft^k}
  Z^R(\phi^R(\alpha^R),\eta,\vec 0) Z^R(\phi^R(\beta^R),\eta,\vec 0)
  (-1)^{
    {\eta}_\pi \cdot
    \psi }\right)^p
\\ 
&= n^{p+1}
\sum_{\psi\in\gft^k}
{\left(
\sum_{\eta\in\gft^k}
Z^R(\phi^R(\alpha^R),\eta,\vec 0) Z^R(\phi^R(\beta^R),\eta,\vec 0)
(-1)^{
{\eta}_\pi \cdot
\psi }\right)
}^{p}
\end{align*}
Now, by equation~(\ref{eqZR}), the contribution for a given~$\eta$
is~0 unless $\phi^R(\alpha^R) \oplus \eta$ and $\phi^R(\beta^R) \oplus \eta$ are in $L^R$.
But $\phi^R(\alpha^R)$ and $\phi^R(\beta^R)$ are in $L^R$,
so by \cond{L}, the contribution for a given $\eta$ is nonzero
exactly when $\eta\in L^R$.
Thus, we can use equation~(\ref{eqZR}) to simplify, writing 
$\eta$ as $\phi^R(\mu)$.
\begin{align*}
C^{[p]}_{a,b} 
&= n^{p+1}
\sum_{\psi\in\gft^k}
{\left(
\sum_{\eta\in L^R}
n^2 
{(-1)}^{
g^R(\phi^R(\alpha^R)) \oplus g^R(\phi^R(\beta^R)) \oplus g^R(\phi^R(\alpha^R) 
\oplus \eta)
\oplus g^R(\phi^R(\beta^R) \oplus \eta)\oplus
\eta_\pi \cdot \psi }\right)
}^{p}
\\ 
&\hspace{-1truecm}
= n^{3p+1}
{\left({(-1)}^{g^R(\phi^R(\alpha^R)) \oplus g^R(\phi^R(\beta^R))}\right)}^{p}
\sum_{\psi\in\gft^k}
{\left(
\sum_{\mu\in \gft^{\ell^R}}
{(-1)}^{g^R(\phi^R(\alpha^R) \oplus \phi^R(\mu))
\oplus g^R(\phi^R(\beta^R) \oplus \phi^R(\mu))\oplus
\phi^R(\mu)_\pi \cdot \psi }\right)
}^{p}
\end{align*}
Since $p$ is even, we have ${\left({(-1)}^{g^R(\phi^R(\alpha^R)) \oplus
      g^R(\phi^R(\beta^R))}\right)}^{p}=1$. Using the linearity of $\phi^R$
and inverting~$\pi$, we further simplify as follows.
\begin{align*}
C^{[p]}_{a,b} 
& = n^{3p+1}
\sum_{\psi\in\gft^k}
{\left(
\sum_{
\mu\in \gft^{\ell^R}}
{(-1)}^{
 g^R(\phi^R(\alpha^R  \oplus \mu))
\oplus g^R(\phi^R(\beta^R \oplus  \mu))\oplus
\phi^R(\mu) \cdot \psi_{\pi^{-1}} }\right)
}^{p}
\\  
& = n^{3p+1}
\sum_{\chi\in\gft^k}
{\left(
\sum_{
\mu\in\gft^{\ell^R}}
{(-1)}^{
 g^R\phi^R(\alpha^R  \oplus \mu)
\oplus g^R\phi^R(\beta^R  \oplus  \mu)\oplus
\phi^R(\mu) \cdot \chi }\right)
}^{p}
\end{align*} 
Since $\phi^R$ is linear, by Lemma~\ref{cor:bipolar_lam_linear},
there is a surjective map $f:\gft^k\rightarrow \gft^{\ell^R}$ 
and a constant $\kappa^R\in \Nat$ such that
$\phi^R(\mu) \cdot \chi = f(\chi) \cdot \mu$
and for any $\gamma\in \gft^{\ell^R}$ the number of $\chi$
with $f(\chi)=\gamma$ is $\kappa^R$ so we can simplify.
\begin{align*}
C^{[p]}_{a,b} 
& = n^{3p+1}
\kappa^R
\sum_{\gamma\in\gft^{\ell^R}}
{\left(
\sum_{
\mu\in\gft^{\ell^R}}
{(-1)}^{
 g^R\phi^R(\alpha^R  \oplus \mu)
\oplus g^R\phi^R(\beta^R  \oplus  \mu)\oplus
\mu \cdot \gamma }\right)
}^{p}
\end{align*} 
Let 
\[
\widehat{C}^{[p]} = \frac{C^{[p]}}{
n^{3p+1}\cdot
\kappa^R}.
\] 
Clearly $\eval{C^{[p]}} \equiv \eval{\widehat{C}^{[p]}}$.
We will now show that $g^R\circ\phi^R$ has degree at most~$2$ or
there 
is an even $p$ such that
$\eval{\widehat{C}^{[p]}}$ is \hash P-hard.
First note that $\widehat{C}^{[p]}$ is symmetric and
\[
\widehat{C}^{[p]}_{a,a} 
=  
\sum_{\gamma\in\gft^{\ell^R}}
{\left(
\sum_{
\mu\in\gft^{\ell^R}}
{(-1)}^{\mu \cdot \gamma  }\right)
}^{p} =   2^{\ell^R p}.
\]
For $X = (X_1,\ldots,X_{\ell^R})$ and 
$a,b \in \Lambda^R$ and $\gamma\in \gft^{\ell^R}$,
define  
the polynomial 
\renewcommand{\gphi}{g^R\circ\phi^R}
\renewcommand{\gphiquad}{{\tilde g}_{a,b,\gamma}}
\[
\gphiquad(X) = \gphi(\alpha^R \oplus  X) \oplus \gphi(\beta^R \oplus  X)
\oplus \gamma\cdot X.
\]
For all $a,b \in \Lambda^R$ we define:
\begin{eqnarray*}
\m{{C}}_{a,b} &:=& \mset{ \gamma\in\gft^{\ell^R} \mid \#(\gphiquad(X)) \in \mset{0,2^{\ell^R}}} \\
\m{G}_{a,b} &:=& \mset{\gamma\in\gft^{\ell^R} \mid \#(\gphiquad(X)) \notin \mset{0,2^{\ell^R-1},2^{\ell^R}}} \\
\m{H}_{a,b} &:=& \mset{\gamma\in \gft^{\ell^R} \mid \#(\gphiquad(X)) = 2^{\ell^R-1}},
\end{eqnarray*}
where $\#(\gphiquad(X))$ denotes the number of $x\in\gft^{\ell^R}$ such that $\gphiquad(x)=1$.

For every $\gamma \in \m{G}_{a, b}$ define $z_{a,b,\gamma} := \sum_{\mu\in
  \gft^{\ell^R}}(-1)^{\gphiquad(\mu)}$, which, by definition, satisfy $z_{a,b,\gamma}\neq 0$ and $\vert z_{a,b,\gamma} \vert < 2^{\ell^R}$. Let
$z_{a,b}^{\max} = \max_{ \gamma \in \m{G}_{ a,b}} \vert z_{a,b,\gamma} \vert$
and $z_{a,b}^{\min} = \min_{ \gamma \in \m{G}_{ a, b}} \vert  z_{a,b,\gamma}
\vert.$ For
$a,b\in \Lambda^R$, we can simplify the expression for $\widehat{C}^{[p]}_{ a,
  b}$.
\begin{eqnarray*}
\widehat{C}^{[p]}_{a, b} =  \sum_{\gamma\in\gft^{\ell^R}  } \left(\sum_{\mu\in \gft^{\ell^R}}(-1)^{\gphiquad(\mu)}\right)^{p} 
                & = & \left(\sum_{ \gamma \in \m{{C}}_{ a, b}} 2^{\ell^Rp}  + \sum_{ \gamma \in \m{G}_{ a, b}}  (z_{a,b,\gamma})^{p} +  \sum_{ \gamma \in \m{H}_{ a, b}} 0 \right) \\
                & = & \left(\vert\m{{C}}_{ a, b}\vert 2^{\ell^Rp}  + \sum_{ \gamma\in \m{G}_{ a, b}}  (z_{a,b,\gamma})^{p}\right)
\end{eqnarray*}
Since  $p$ is even, $(z_{a,b,\gamma})^{p}$ is positive for all $ \gamma \in \m{G}_{ a, b}$ and thus 
$\widehat{C}^{[p]}_{ a,  b}$ is non-negative for all $ a , b \in \Lambda^R$. 
If $\Lambda^R$ is empty then the relevant condition in \cond{D} is satisfied,
so suppose that it is nonempty.
We will now show that $g^R\circ\phi^R$ has degree at most~$2$ or there exists
an even $p$ such that 
$\widehat{C}^{[p]}$ has a block of rank at least two.

\paragraph*{Case A. There are $a,b\in \Lambda^R$ such that
  $\m{G}_{ a, b} \neq \emptyset$}
Choose such $a,b$. The principal $2 \times 2$ submatrix of $\widehat{C}^{[p]}$, defined by~$a$
and~$b$ has determinant
\begin{equation}
\label{eq:deg-det}
\left\vert\begin{array}{c c}
             \widehat{C}^{[p]}_{ a, a} & \widehat{C}^{[p]}_{ a ,b}\\
            \widehat{C}^{[p]}_{ b, a} & \widehat{C}^{[p]}_{ b ,b}
          \end{array}\right\vert =
\left\vert\begin{array}{c c}
            2^{\ell^Rp}  & \widehat{C}^{[p]}_{ a ,b} \\
            \widehat{C}^{[p]}_{ a ,b} & 2^{\ell^Rp} 
          \end{array}\right\vert = 2^{2\ell^Rp} - (\widehat{C}^{[p]}_{ a, b})^2.
\end{equation}
If the determinant is zero, then
$\dfrac{\widehat{C}^{[p]}_{ a, b}}{2^{\ell^Rp}}=1$.
We consider two cases. If $\m{{C}}_{ a ,b} = \emptyset$, then
\begin{align*}
\frac{\widehat{C}^{[p]}_{ a, b}}{2^{\ell^Rp}}&
= \frac{\left(\sum_{
      \gamma \in \m{G}_{ a, b}}  (z_{a,b,\gamma})^{p}\right)}{2^{\ell^Rp}}\\
&\le \frac{\vert \m{G}_{ a ,b}
  \vert (z_{a,b}^{\max})^{p}}{2^{\ell^Rp}}\\
& \le {2^{\ell^R}}
\left(\frac{z_{a,b}^{\max}}{{2^{\ell^R}}}\right)^{p}\\
&\le
{2^{\ell^R}}\left(\frac{2^{\ell^R}-1}{2^{\ell^R}}\right)^p&\text{(because
  $z_{a,b}^{\max} < 2^{\ell^R}$)}\\
&\le 2^{\ell^R}\cdot e^{- p/2^{\ell^R}}.
\end{align*}
This is less than one for all $p>\ell^R2^{\ell^R}$.
  Hence the determinant
\eqref{eq:deg-det} is nonzero.
Furthermore, as $\m{G}_{ a, b} \neq \emptyset$ we have $\widehat{C}^{[p]}_{
  a, b}\neq 0$ and hence $\widehat{C}^{[p]}$ contains a block of rank at least
two. This implies the $\#\PP$-hardness of $\eval{\widehat{C}^{[p]}}$ by Lemma
\ref{lem:bipolar_block2_hard}.
(Recall that $\widehat{C}^{[p]}_{ a,  b}$ is non-negative since $a,b\in
\Lambda^R$.)

For the other case, suppose
$\vert \m{{C}}_{ a, b}\vert \ge 1$. Then
\begin{align*}
\frac{\widehat{C}^{[p]}_{ a, b}}{2^{\ell^Rp}} 
& = 2^{-\ell^Rp} \left(|\m{{C}}_{ a, b}| 2^{\ell^Rp}  + \sum_{
    \gamma \in \m{G}_{ a ,b}}  (z_{a,b,\gamma})^{p}\right)\\ 
&\ge  2^{-\ell^Rp} \left(\vert\m{{C}}_{ a ,b}\vert 2^{\ell^Rp}
  + \vert \m{G}_{ a,b} \vert  (z_{a,b}^{\min})^{p}\right)\\  
&>  |\m{{C}}_{ a, b}|\ge 1.
\end{align*}
Here, the second-but-last inequality holds, because $z_{a,b}^{\min} > 0$ and (by the
precondition of case A) $\m{G}_{ a, b} \neq \emptyset$.
Hence again we have $\frac{\widehat{C}^{[p]}_{ a, b}}{2^{\ell^Rp}}\neq1$, and
the determinant \eqref{eq:deg-det} is nonzero. As in the
first case this implies the $\#\PP$-hardness of $\eval{\widehat{C}^{[p]}}$.

\paragraph*{Case B. For all $a,b\in \Lambda^R$ it holds that 
  $\m{G}_{ a, b} =\emptyset$}
Then for all $ a, b \in {\Lambda}^R$ we have
\[
\widehat{C}^{[p]}_{ a ,b}  
=  \vert\m{{C}}_{ a ,b}\vert 2^{\ell^Rp}  + \sum_{ \gamma \in \m{G}_{ a ,b}}  z_{a,b,\gamma}^{p}
=  \vert \m{{C}}_{ a, b} \vert 2^{\ell^Rp}.
\]
So the principal $2 \times 2$ submatrix of $\widehat{C}^{[p]}$ defined by $ a,
b $ has determinant
\[
\left\vert\begin{array}{c c}
            \widehat{C}^{[p]}_{ a ,a} & \widehat{C}^{[p]}_{ a, b}\\
            \widehat{C}^{[p]}_{ b ,a} & \widehat{C}^{[p]}_{ b ,b}
          \end{array}\right\vert =
\left\vert\begin{array}{c c}
            2^{\ell^Rp} & 2^{\ell^Rp} \vert \m{{C}}_{ a, b} \vert \\
            2^{\ell^Rp} \vert \m{{C}}_{ a, b} \vert & 2^{\ell^Rp}
          \end{array}\right\vert = 2^{2\ell^Rp}(1 - \vert \m{{C}}_{ a ,b} \vert^2).
\]
This determinant is zero if and only if $\vert \m{{C}}_{ a, b} \vert = 1$, and
the submatrix is part of a block iff $\m{{C}}_{ a ,b} \neq \emptyset$. Hence,
we have $\#\PP$-hardness by Lemma \ref{lem:bipolar_block2_hard}, if there are
$ a, b\in\Lambda^R$ such that $\vert \m{{C}}_{ a, b} \vert \notin \mset{0,1}$.
Assume that for all $ a, b \in \Lambda^R$ we have $| \m{C}_{ a, b} |
\in \mset{0,1}$. Define sets
\begin{align*}
\m{I} &:= \mset{ ( a, b) 
\mid a\in\Lambda^R, b\in \Lambda^R,  |\m{C}_{ a, b}| = 1,   a\neq  b},\\ 
\m{Z} &:= \mset{ ( a, b) 
\mid a\in \Lambda^R, b\in \Lambda^R, |\m{C}_{a,b} | = 0, \;  a\neq  b}.
\end{align*}
Obviously, these form a partition of pairs of distinct elements in
$\Lambda^R$.  In other words, for all $ a\neq b \in {\Lambda^R}$ there is at
most one $ \gamma \in\gft^{\ell^R}$ such that $\#(\gphiquad(X)) \in \mset{0,
  2^{\ell^R}}$.  Furthermore, $\m{G}_{ a ,b} = \emptyset$ implies that for all
other $ {\gamma'} \neq \gamma$ we have $\#(\tilde g_{ a, b, {\gamma'}}(X)) =
2^{\ell^R-1}$.  But Corollary \ref{lem:weakened_char_quad} implies that in
this case $\gphi$ has degree at most two. This finishes Case~B and hence the
proof of the Row Condition.

For the Column Condition, in a symmetric way to how we defined $Z^R(\alpha,\gamma,\delta)$, we let
$Z^C(\alpha,\gamma,\delta)$ 
denote the contribution to $Z_{M,I_m,I_m;\Lambda}(\Gamma_{i})$ corresponding
to those configurations $\xi$ with 
$\xi(u)=n+a^C$, $\xi(x_i)=n+c^C$ and 
$\xi(y_{i})=n+d^C$,  
ignoring  contributions due to $I_{m;{\Lambda}}$ for vertices $u$, $x_i$, and $y_{i}$.
Using this, we can compute $C^{[p]}_{n+a,n+b}$ for $a,b\in \Lambda^C$ and show that,
if $\Lambda^C$ is nonempty, then
either 
$g^C \phi^C$ has degree at most~$2$
or
$\eval{C^{[p]}}$ is \hash P-hard.

Finally, we note that it is straightforward, in polynomial time, to determine whether 
$\eval{C^{[p]}}$ is \hash P-hard or \cond{D} holds.
 \end{proof}

\begin{cor}\label{cor:nonbip}
Let $H$ be a symmetric $n \times n$ Hadamard matrix and $\Lambda^R=\Lambda^C \subseteq  [n]$ identical subsets of indices.
If $H$ is positive for $\Lambda^R$ and $\Lambda^C$
then $\eval{H,I_n,I_{n;\Lambda^R}}$ is polynomial time computable if, and only if, $H$ $\Lambda^R$
and $\Lambda^C$ satisfy the group condition \cond{GC} and conditions \cond R, \cond L, and \cond D.
Otherwise $\eval{H,I_n,I_{n;\Lambda^R}}$ is $\#\PP$-hard.
If $H$ is not positive for $\Lambda^R$ and $\Lambda^C$
then $\eval{H,I_n,I_{n;\Lambda^R}}$ is polynomial time computable if, and only if, $-H$ $\Lambda^R$
and $\Lambda^C$ satisfy the group condition \cond{GC} and conditions \cond R, \cond L, and \cond D.
Otherwise $\eval{H,I_n,I_{n;\Lambda^R}}$ is $\#\PP$-hard.
\end{cor}

\begin{proof}%
By the equivalence of $\eval{H,I_n,I_{n;\Lambda^R}}$ and
$\eval{-H,I_n,I_{n;\Lambda^R}}$ we can assume that $H$ is
positive for $\Lambda^R$ and $\Lambda^C$.
First, suppose that one of the conditions is not satisfied. By Theorem~\ref{thm:bipolar_dichotomy},
$\eval{M,I_m,I_{m;\Lambda}}$ is $\#\PP$-hard.
Since $M$ is bipartite, $\eval{M,I_m,I_{m;\Lambda}}$ remains $\#\PP$-hard
when restricted to connected bipartite instances~$G$.
But for these instances, $Z_{M,I_m,I_{m;\Lambda}}(G)=2 Z_{H,I_n,I_{n;\Lambda^R}}(G)$,
so $\eval{H,I_n,I_{n;\Lambda^R}}$ is $\#\PP$-hard.

It remains to give the proof for the tractability part. For symmetric~$H$ and 
$\Lambda^R=\Lambda^C$ satisfying \cond{GC}, \cond{R}, \cond{L} and \cond{D}, we 
shall show how to compute $Z_{H,I_n,I_{n;\Lambda^R}}(G)$ for an input graph $G$ in polynomial time.
Let $V_o \subseteq V$ denote the set of odd-degree vertices of $G$ and $V_e = V\setminus V_o$. We have
\begin{eqnarray*}
Z_{H,I_n,I_{n;\Lambda^R}}(G) &=& \sum_{\xi:V \rightarrow [n]} \prod_{\{u,v\} \in E} H_{\xi(u),\xi(v)} \prod_{v \in V_o} (I_{n;\Lambda^R})_{\xi(v),\xi(v)} 
= \sum_{\substack{\xi:V \rightarrow [n] \\ \xi(V^o) \subseteq \Lambda^R}} \prod_{\{u,v\} \in E} H_{\xi(u),\xi(v)}
\end{eqnarray*}
Fix a configuration $\xi : V \rightarrow [n]$ and let $\rho=\rho^R=\rho^C$ be the index mapping and $h$ the $\gft$-polynomial representing $H$ as given in condition \cond{R}. Let furthermore $\phi: = \phi^R = \phi^C$ be the coordinatisation of $\Lambda^R$ as given in condition \cond{D}. Let $\tau$ be the inverse of $\rho$ and $L = \tau(\Lambda^R)$. Then $\xi$ induces a configuration  $\varsigma: V \rightarrow \gft^k$ defined by $\varsigma = \tau \circ \xi$ which implies, for all $u,v \in V$ that $h(\varsigma(u),\varsigma(v)) = 1$ iff $H_{\xi(u),\xi(v)} = -1$.
 We can simplify
\begin{equation}\label{eq:Z_by_h_expression}
Z_{H,I_n,I_{n;\Lambda^R}}(G) = \sum_{\substack{\xi:V \rightarrow [n] \\ \xi(V_o) \subseteq \Lambda^R}} \prod_{\{u,v\} \in E} (-1)^{ h(\tau \circ \xi(u),\tau \circ \xi(v))} 
            = \sum_{\substack{\varsigma:V \rightarrow \gft^k \\ \varsigma(V_o) \subseteq L}} (-1)^{\bigoplus_{\{u,v\} \in E} h(\varsigma(u),\varsigma(v))}
\end{equation}

Define, for each $v \in V$ a tuple $X^v = (X^v_1,\ldots, X^v_k)$ and an $\gft$-polynomial
$$
h_G = \bigoplus_{\{u,v\} \in E} h(X^u,X^v).
$$

Let $\var(h_G)$ denote the set of variables in $h_G$ and, 
for mappings $\chi: \var(h_G) \rightarrow \gft$, we use the expression $\chi(X^v) := (\chi(X^v_1),\ldots,\chi(X^v_k))$ as a shorthand.
Define $h_G(\chi) := \bigoplus_{\{u,v\} \in E} h(\chi(X^u),\chi(X^v))$ and note that this is a sum in $\gft$.

For $a \in \gft$ let 
\begin{equation}\label{eq:count_poly_sols}
s_a := \vert \{ \chi: \var(h_G) \rightarrow \gft \mid \chi(X^v) \in L \text{ for all } v \in V_o \text{ and } h_G(\chi) = a\} \vert. 
\end{equation}

Hence, by equation \eqref{eq:Z_by_h_expression},  
$Z_{H,I_n,I_{n;\Lambda^R}}(G) = s_0 - s_1$.
It remains therfore to show how to compute the values $s_a$.
Clearly,
\begin{eqnarray*}
h_G &=& \bigoplus_{\{u,v\} \in E} (X^u)_{\pi} X^v \oplus g(X^u) \oplus g(X^v) 
     = \bigoplus_{\{u,v\} \in E} (X^u)_{\pi} X^v  \oplus \bigoplus_{v \in V_o}  g(X^v)
\end{eqnarray*}
as the term $g(X^v)$ occurs exactly $\deg(v)$ many times in the above expression and thus these terms cancel for all even degree vertices.

By equation \eqref{eq:count_poly_sols} we are interested only in those assignments $\chi$ which satisfy
$\chi(X^v) \in L$ for all $v \in V_o$. With $\vert \Lambda^R \vert = 2^l$ for some appropriate $l$, we introduce variable vectors $Y^v = (Y^v_1,\ldots, Y^v_l)$ for all $v \in V_o$. If $u \in V_o$ or $v\in V_o$ then we can express the term $(X^u)_{\pi} X^v$ in $h_G$ in terms of these new variables. In particular,
let
\begin{eqnarray*}
h''_G &=& 
\phantom{\oplus} \bigoplus_{\substack{\{u,v\} \in E\\u,v \in V_o}} (\phi(Y^u))_\pi \cdot \phi(Y^v)
\oplus
\bigoplus_{\substack{\{u,v\} \in E\\u,w \in V_e}}  (X^u)_\pi \cdot X^v  
 \oplus
\bigoplus_{\substack{\{u,v\} \in E\\u \in V_o,v\in V_e}} (\phi(Y^u) )_\pi \cdot X^v.
\end{eqnarray*}
Let
\begin{eqnarray*}
h'_G &=& h''_G \oplus \bigoplus_{v \in V^o} \oplus g(\phi(Y^v)).
\end{eqnarray*}
Then we see that
\begin{equation}
s_a := \vert \{ \chi: \var(h'_G) \rightarrow \gft \mid h'_G(\chi) = a\} \vert. 
\end{equation}

By condition \cond{D} $g \circ \phi$ is a polynomial of degree at most $2$ and therefore $h'_G$ is a polynomial of degree at most $2$. Furthermore, we have expressed $s_a$ as the number of solutions to a polynomial equation over $\gft$. Therefore, as in the proof of Theorem \ref{thm:bipolar_dichotomy}, the proof now follows by Fact \ref{fact:gft_count}.
\end{proof}

\begin{proof}[Proof of Theorem~\ref{thm:hadamard}]
  Let $H$ be a symmetric $n\times n$ Hadamard matrix and
  $\Lambda^R=\Lambda^C=[n]$. Then $H$ is positive for $\Lambda^R$ and
  $\Lambda^C$. Let $M,\Lambda$ be the bipartisation of
  $H,\Lambda^R,\Lambda^C$.  

  Suppose first that $H$ has no quadratic representation. Then there
  are no index mapping $\rho=\rho^R=\rho^C$ and coordinatisation
  $\phi=\phi^R=\phi^C$ such that conditions (R) and (D) are satisfied.
  Hence by Theorem~\ref{thm:bipolar_dichotomy},
  $\eval{M,I_m,I_{m;\Lambda}}$ is $\#\PP$-hard.  Since $M$ is
  bipartite, $\eval{M,I_m,I_{m;\Lambda}}$ remains $\#\PP$-hard when
  restricted to connected bipartite instances~$G$.  But for these
  instances, $Z_{M,I_m,I_{m}}(G)=2 Z_{H,I_n,I_{n}}(G)$, so
  $\eval{H,I_n,I_{n}}$ is $\#\PP$-hard.
  Suppose next that $H$ has a quadratic representation with index
  mapping $\rho:\gft^k\to[n]$ and polynomial $h(X,Y)$. Instead of
  going through Theorem~\ref{thm:bipolar_dichotomy}, it is easier to
  prove the tractability of $\eval H$ directly along the lines of the
  proof of the tractability part of the theorem. 
We leave the details
  to the reader.
This is similar to the tractability part of the proof of Corollary~\ref{cor:nonbip}.
\end{proof}

\section{The Proofs for Section~\ref{sec:general_case}}
\label{sec:appendix_general_case}

\subsection{Technical Preliminaries}

\begin{lemma}
\label{cor:bipolar_std_conversion}
Let $C \in \algR ^{m \times m}$ be a symmetric matrix and let $\Delta^+$ and $\Delta^-$ be diagonal $m \times m$ matrices. 
Let  $D$ be the component-wise sum $D = \Delta^+ + \Delta^-$ and let $O = \Delta^+ - \Delta^-$.
Let $A$ be the tensor product
$$A = 
\begin{pmatrix}1 & -1\\
-1 & 1\end{pmatrix} \otimes C.$$
Let $\Delta$ be the $2m \times 2m$ matrix
such that, for all $i\in[m]$ and $j\in[m]$,
$\Delta_{i,j} = \Delta^+_{i,j}$,
$\Delta_{i,m+j}=\Delta_{m+i,j}=0$, and
$\Delta_{m+i,m+j}=\Delta^-_{i,j}$.
Then
$$
Z_{C,D,O}(G) = Z_{A,\Delta}(G) \text{ for all graphs } G.
$$ 
\end{lemma}
 
\begin{proof}
 
It is useful to think of $A$ and $\Delta$ in terms of four $m\times m$ tiles as follows.
$$
A = \left( \begin{array}{r r}C & -C \\ -C & C \end{array}\right)
\text{ and }
\Delta =
\left( \begin{array}{c c}\Delta^+ & 0 \\ 0 & \Delta^- \end{array}\right).
$$
We will simplify the expression for $Z_{A, \Delta}(G)$ now.
Let $\xi: V \rightarrow [2m]$ 
be a map such that, for some
$w \in V$, $\xi(w) \in [m]$. Let $\psi$ be the mapping such that for all $v \in V$
$$
 \psi(v) := \xi(v) + \left\lbrace\begin{array}{l l}
                         m&, \text{ if } w=v\\
                         0&, \text{ otherwise. }
                        \end{array}\right.
$$

Then
 
\begin{align*}
\prod_{\{u,v\} \in E} A_{\psi(u), \psi(v)} 
&=  
\prod_{\{w,w\} \in E} A_{\psi(w), \psi(w)}
\prod_{\substack{\{w,v\} \in E\\v\neq w}} A_{\psi(w), \psi(v)} \prod_{\substack{\{u,v\} \in E \\ u,v \neq w}} A_{\psi(u), \psi(v)} \\
&=  
\prod_{\{w,w\} \in E} A_{\xi(w), \xi(w)}
\prod_{\substack{\{w,v\} \in E\\v\neq w}}
-A_{\xi(w), \xi(v)} \prod_{\substack{\{u,v\} \in E \\ u,v \neq w}} A_{\xi(u), \xi(v)}
\end{align*}

which implies that
$$
  \prod_{\{u,v\} \in E} A_{\xi(u), \xi(v)} = (-1)^{\deg(w)} \prod_{\{u,v\} \in E} A_{\psi(u), \psi(v)}
$$
where $\deg(w)$ denotes the degree of $w$ in G (self-loops add two to this degree).
Since  $\prod_{v \in V} \Delta_{\xi(v),\xi(v)} = \Delta_{\xi(w),\xi(w)} \prod_{w\neq v \in V} \Delta_{\xi(v),\xi(v)}$,
we have
\begin{eqnarray*}
Z_{A,\Delta}(G) &=& \sum_{\xi:V \rightarrow [2m]} \prod_{\{u,v\} \in E} A_{\xi(u), \xi(v)} \prod_{v \in V} \Delta_{\xi(v), \xi(v)} \\
                 &=& \sum_{\substack{\xi:V \rightarrow [2m] \\ \xi(w) \in [m]}} \prod_{\{u,v\} \in E} A_{\xi(u), \xi(v)} \left(\Delta_{\xi(w),\xi(w)} + (-1)^{\deg(w)}\Delta_{m+\xi(w), m+\xi(w)}\right) \prod_{w \neq v \in V} \Delta_{\xi(v), \xi(v)} \\
\end{eqnarray*}
As this argument can be applied independently to all $w \in V$
we obtain
\begin{eqnarray*}
Z_{A,\Delta}(G) &=& \sum_{\xi:V \rightarrow [m]} \prod_{\{u,v\} \in E} A_{\xi(u), \xi(v)} \prod_{w \in V}\left(\Delta_{\xi(w),\xi(w)} + (-1)^{\deg(w)}\Delta_{m+\xi(w), m+\xi(w)}\right)\\
&=& \sum_{\xi:V \rightarrow [m]} \prod_{\{u,v\} \in E} C_{\xi(u), \xi(v)} \prod_{\substack{w \in V\\ \deg(w) \text{ even}}}
D_{\xi(w),\xi(w)} \prod_{\substack{w \in V\\ \deg(w) \text{ odd}}} O_{\xi(w), \xi(w)}\\
      &=& Z_{C,D,O}(G).
\end{eqnarray*}

\end{proof}

\begin{cor}\label{cor:rank1_bipolar_FP}
Let $C$ be a symmetric $m\times m$ matrix which contains exclusively blocks of rank $1$. Let $D$ and $O$ be diagonal $m\times m$ matrices.
Then the problem $\eval{C,D,O}$ is polynomial time computable.
\end{cor}

\begin{proof}
  By Lemma~\ref{cor:bipolar_std_conversion} the problem $\eval{C,D,O}$ is
  polynomial time equivalent to a problem $\eval{A,\Delta}$ with $A$ a matrix
  consisting of blocks of row rank at most $1$. Thus the statement of the
  corollary follows from Lemma~\ref{lem:rank1_FP}.
\end{proof}

\subsubsection{Extended Twin Reduction}

\newcommand{\pmtwinred}[1]{\mathcal{T}^{\pm}(#1)}

Unfortunately the Twin Reduction Lemma \ref{lem:twin_red} does fully satisfy our needs. As we are dealing with possible negative rows we will be in a situation, where it is useful to reduce matrices even further, namely by collapsing two rows $A \row i$ and $A \row j$ into one if  $A \row i = \pm A\row j$. 

To achieve this we say that two rows $A\row i$ and $A \row j$ are
\emph{plus-minus-twins} (\emph{pm-twins} for short) iff $A \row i = \pm A \row
j$. This induces an equivalence relation on the rows (and by symmetry on the
columns) of $A$. Let $I_1, \ldots I_k$ be a partition of the row indices of
$A$ according to this relation.  For technical reasons it will be convenient
to partition the sets $I_i$ into the positive and the negative part. That is
for every $i \in [k]$ we define a partition $(P_i,N_i)$ of $I_i$ such that
$P_i\neq\emptyset$ and for all $\nu,\nu' \in P_i$ and $\mu, {\mu'} \in N_i$ we
have $A \row \nu = A \row {\nu'}$, $A \row {\mu} = A \row {\mu'}$ and $A \row
\nu = - A \row \mu$.

The \emph{pm-twin}-resolvent of $A$ is the matrix defined, for all $i,j \in [k]$, by
$$
\pmtwinred A_{i,j} := A_{\mu, \nu} \text{ for some } \mu \in P_i, \nu \in P_j.
$$

This definition is technical and seems to be counter-intuitive, as we are not taking the $N_i$ into account. However its motivation will become clear with the following Lemma and it is still well-defined, even 
though possibly $N_i = \emptyset$ for some $i \in [k]$. 

As before, we define a mapping $\tau: [m] \rightarrow [k]$ defined by $\mu \in I_{\tau(\mu)}$ that is $\tau$ maps $\mu \in [m]$ to the class $I_j$ it is contained in.
Therefore, we have $\pmtwinred A_{\tau(i),\tau(j)} = \pm A_{i,j}$ for all $i,j \in [m]$. We call $\tau$ the \emph{pm-twin-resolution mapping} of $A$.
Define $N = N_1 \cup \ldots \cup N_k$ and $P = P_1 \cup \ldots\cup P_k$. Then in particular 
$$\pmtwinred A_{\tau(i),\tau(j)} = \phantom{-}A_{i,j} \text{ for all } (i,j) \in (P\times P) \cup (N \times N)$$
$$\pmtwinred A_{\tau(i),\tau(j)} = - A_{i,j} \text{ for all } (i,j) \in (P\times N) \cup (N \times P)$$

\begin{lemma}[Extended Twin Reduction Lemma]\label{lem:ext_twin_red}
Let $A$ be a symmetric $m \times m$ matrix and $\Delta$ a diagonal $m \times m$ matrix of vertex weights. Let $(P_1,N_1),\ldots,(P_k,N_k)$ be a partition of the row indices of $A$ according to the pm-twin-relation.

Then 
$$
 Z_{A,\Delta}(G) = Z_{\pmtwinred A,D,O}(G) \text{ for all graphs } G
$$
where $D$ and $O$  are diagonal $k \times k$ matrices defined by 
$$
D_{i,i} = \sum_{\nu \in P_i} \Delta_{\nu,\nu} + \sum_{\mu \in N_i} \Delta_{\mu,\mu} \quad \text{ and } \quad
O_{i,i} = \sum_{\nu \in P_i} \Delta_{\nu,\nu} - \sum_{\mu \in N_i} \Delta_{\mu,\mu} \text{ for all } i \in [k]. 
$$
\end{lemma}
\begin{proof}
  Define $J_i = P_i$ and $J_{k+i} = N_i$ for all $i \in [k]$. W.l.o.g. we may
  assume that if there is a minimal $l \in [k]$ such that $J_{k+l} =
  \emptyset$ then for all $j \ge l$ we have $J_{k+j} = \emptyset$ (this can be
  achieved by appropriate relabelling of the $P_i$ and $N_i$). Let $l := k+1$
  if all $J_{k+i}$ are non-empty.  Then $J_{1}, \ldots, J_{k+l-1}$ are the
  equivalence classes of $A$ according to the twin-relation. Therefore, the
  Twin Reduction Lemma \ref{lem:twin_red} implies that for the diagonal
  $(k+\ell-1)\times(k+\ell-1)$ diagonal matrix $ \Delta''$ defined by
  $\Delta''_{j,j}=\sum_{\nu\in J_j}\Delta_{\nu,\nu}$
we have
$$
Z_{A,\Delta}(G) = Z_{\twinred A,\Delta''}(G) \text{ for all graphs } G.
$$
Let $n' : = k+l-1$ and note that by the definition of the sets $J_i$, $\twinred A$ is the upper left $n' \times n'$ submatrix of the $2k \times 2k$ matrix
$$
M = \left( \begin{array}{r r} \pmtwinred A & -\pmtwinred A \\ -\pmtwinred A & \pmtwinred A \end{array}\right)
  = \left( \begin{array}{r r} 1 & -1 \\ -1 & 1 \end{array}\right) \otimes \pmtwinred A.
$$
that is $\twinred A = M_{[n'][n']}$. Define a $2k \times 2k$ diagonal matrix $\Delta'$ such that
$\Delta'_{i,i} = \Delta''_{i,i}$ for all $i \in [n']$ and $\Delta'_{i,i} = 0$ for all $n' < i \le 2k$. Then 
$$
Z_{M,\Delta'}(G) = Z_{\twinred A,\Delta''}(G) \text{ for all graphs } G.
$$
Moreover, by the definition of $\Delta''$, the matrix $\Delta'$ satisfies, for all  $i \in [k]$,
\begin{equation}\label{eq:def_twin_red_diag_result}
\Delta'_{i,i} = \sum_{\nu \in P_i} \Delta_{\nu,\nu} \quad \text{ and } \quad \Delta'_{k+i, k+i} = \sum_{\nu \in N_i} \Delta_{\nu,\nu}.
\end{equation}
Now, by 
Lemma~\ref{cor:bipolar_std_conversion},
$Z_{M,\Delta'}(G)=Z_{\pmtwinred A,D',O'}$
where $D'$ and $O'$ are $k\times k$ matrices such that
$D'_{i,i} = \Delta'_{i,i} + \Delta'_{k+i,k+i}$
and $O'_{i,i} = \Delta'_{i,i} - \Delta'_{k+i,k+i}$.
But by
Equation \eqref{eq:def_twin_red_diag_result}, we see 
that $D'=D$ and $O'=O$.
\end{proof}

\begin{lemma}[Row-Column Negation Lemma] \label{lem:bipolar_rowcol_neg}
Let $C$ be a symmetric $m \times m$ matrix and $D,O$ diagonal $m\times m$ matrices of vertex weights.

Let $i \in [m]$ and define $C'$ as the matrix obtained from $C$ by multiplying row and column $i$ with $-1$. Let $O'$ be the matrix obtained from $O$ by negating the diagonal entry $O_{i,i}$. Then 
$$
 Z_{C,D,O}(G) = Z_{C',D,O'}(G) \text{ for all graphs } G.
$$
\end{lemma}
\begin{proof}
Let $G=(V,E)$ be a graph and $V_o,V_e$ the sets of odd (even) degree vertices in $V$. Recall that
$$
Z_{C,D,O}(G) = \sum_{\xi: V \rightarrow [m]} \prod_{\{u,v\} \in E} C_{\xi(u),\xi(v)} \prod_{v \in V_e} D_{\xi(v),\xi(v)}\prod_{v \in V_o} O_{\xi(v),\xi(v)}
$$

Fix some mapping $\xi: V \rightarrow [m]$. We will prove the Lemma by showing that
$$
 \prod_{\{u,v\} \in E} C_{\xi(u),\xi(v)} \prod_{v \in V_e} D_{\xi(v),\xi(v)}\prod_{v \in V_o} O_{\xi(v),\xi(v)} =
\prod_{\{u,v\} \in E} C'_{\xi(u),\xi(v)} \prod_{v \in V_e} D_{\xi(v),\xi(v)}\prod_{v \in V_o} O'_{\xi(v),\xi(v)}.
$$

Define $W := \xi^{-1}(i)$ and let $W_e := V_e \cap W$ and $W_o := V_o \cap W$ denote the even and odd degree vertices in $W$.
By the definition of $O'$ we have
$$
\prod_{v \in V_o} O'_{\xi(v),\xi(v)} = (-1)^{\vert W_o \vert} \prod_{v \in V_o} O_{\xi(v),\xi(v)}.
$$
Furthermore, for all edges $\{u,v\} \in E$ we have that $C_{\xi(u), \xi(v)} = C'_{\xi(u),\xi(v)}$ if and only if either both $u,v \in W$ or $u,v \notin W$. If exactly one of the vertices is in $W$ then $C_{\xi(u), \xi(v)} = -C'_{\xi(u), \xi(v)}$.
Therefore, if we denote by $e(W,V\setminus W)$ the number of edges $e=\{u,v\}$ in $G$ such that exactly one vertex is in $W$, we have
$$
\prod_{\{u,v\} \in E} C'_{\xi(u),\xi(v)} = (-1)^{e(W,V\setminus W)}\prod_{\{u,v\} \in E} C_{\xi(u),\xi(v)}.
$$

To finish the proof it thus suffices to prove that 
$$
 e(W,V\setminus W) \equiv \vert W_o \vert\; (\text{mod }2).
$$
The proof will
be given by induction on the number $\vert W \vert$ of vertices in $W$. The
case that $W = \emptyset$ is trivial. Assume therefore that there is a vertex
$w \in W$ and let $U := W \setminus \{w\}$.  By the induction hypothesis, we
have $e(U,V\setminus U) \equiv \vert U_o \vert (\text{mod }2)$.  If $w$ has
even degree then $\vert W_o \vert = \vert U_o \vert$ and $w$ either has an odd
number of neighbours both in $U$ and $V \setminus U$ or it has an even number
of neighbours in both sets.  If otherwise $w$ has $\vert W_o \vert = 1 + \vert
U_o \vert$ and the parity of the number of neighbours of $w$ in $U$ is
opposite to that of the number of neighbours in $V\setminus U$. This finishes
the proof.
\end{proof}

\subsubsection{Pinning vertices}

In the proof of Lemma \ref{lem:component_pinning} it will be convenient to ``pin''
certain vertices of the input graph $G$ to prescribed spins. We will develop the tools which are necessary for this now. 
These results extend analogous techniques used in \cite{dyegre00} and \cite{bulgro05}.

Let $A$ be an $m \times m$ matrix and $D$ a diagonal $m \times m$ matrix of
positive vertex weights. 
In the following, a \emph{labelled graph} is a triple $G=(V,E,z)$, where
$(V,E)$ is a graph and $z\in V$.
For a labelled graph $G=(V,E,z)$ and a $k\in[m]$, we let
$$
Z_{A,D}(k,G) = (D_{k,k})^{-1}\sum_{\substack{\xi: V \rightarrow [m] \\
                                  \xi(z) = k}} \prod_{\{u,v\}\in E}A_{\xi(u),\xi(v)}\cdot\prod_{v\in V}D_{\xi(v),\xi(v)}
$$
The \emph{product $G H $} of two labelled graphs $G$ and $H$ is formed by
taking the disjoint union of the graphs and then 
identifying the labelled vertices.
Let $H^s$ denote the product of $H$ with itself taken $s$ times.
Note that $Z_{A,D}(k,G H) = Z_{A,D}(k,G)Z_{A,D}(k,H)$ for all labelled graphs $G$ and~$H$.

Recall that a twin-free matrix $A$ is a matrix such that $A_i \neq A_j$ for all row indices $i \neq j$. Furthermore an \emph{automorphism} of $(A,D)$ is a bijection $\alpha: [m] \rightarrow [m]$ such that $A_{i,j} = A_{\alpha(i),\alpha(j)}$ and $D_{i,i} = D_{\alpha(i),\alpha(i)}$ for all $i \in [m]$. The following lemma follows by a result of Lov\'asz (Lemma 2.4 in \cite{lov06}).
\begin{lemma}\label{lem:lov06}
Let $A \in \Real^{m \times m}$ be twin free, $D \in \Real^{m \times m}$ a diagonal matrix of positive vertex weights and $i,j \in [m]$.
If for all labelled graphs $G$ we have 
$$
Z_{A,D}(i,G) = Z_{A,D}(j,G)
$$
then there is an automorphism $\alpha$ of $(A,D)$ such that $j = \alpha(i)$.
\end{lemma}

We furthermore need some standard result about interpolation, which we use in the form as stated in \cite{dyegre00} Lemma 3.2:

\begin{lemma}\label{lem:dg00_interpolate}
Let $w_1,\ldots, w_r$ be known distinct non-zero constants. Suppose that we know the values $f_1, \ldots f_r$ such that
$$
f_i = \sum_{j=1}^r c_j w_j^i \text{ for all } i \in [r].
$$
Then the coefficients $c_1,\ldots,c_r$ are uniquely determined and can be computed in polynomial time.
\end{lemma}

\begin{lemma}
[Pinning Lemma]
\label{lem:fixation}
Let $A\in \algR^{m\times m}$ be a symmetric matrix and $\Delta \in \algR^{m \times m}$ a diagonal matrix of positive real entries. Then for every labelled graph $G$ and every $k \in [m]$, we can compute 
$Z_{A,\Delta}(k,G)$ in polynomial time using an $\eval{A,\Delta}$ oracle.
\end{lemma}
\begin{proof}
Let the matrices $B$ and $D$ be the result of twin-reduction (Lemma \ref{lem:twin_red}) when applied to $A$ and $\Delta$. In particular, $B$ is twin-free and
$Z_{A,\Delta}(G) = Z_{B,D}(G)$ for all graphs $G$. Therefore, using the
oracle, we can  compute $Z_{B,D}(G)$ in polynomial time (for input $G$).

Consider a graph
$G=(V,E)$ with a labelled vertex~$z$ and a particular spin $k \in [m]$.
we will show how to compute $Z_{B,D}(k,G)$ using an oracle for $Z_{B,D}$.

Call 
spins $i,j \in [m]$ \emph{equivalent} if there is an automorphism $\alpha$ of $(B,D)$ such 
that $j = \alpha(i)$. Partition $[m]$ into equivalence classes $I_1, \ldots, I_c$ according to this definition.
For every spin~$j$ in equivalence class~$I_i$, let 
$c_j$ denote the size of the equivalence class --- $c_j = |I_i|$.
For every equivalence class $i\in[c]$ let $k_i$ denote a particular spin $k_i\in I_i$.

For any two equivalent spins $a$ and $a'$  
we have $Z_{B,D}(a,F) = Z_{B,D}(a',F)$ for every graph $F$. Therefore,
\begin{equation}\label{eq:std_sum_fixation}
Z_{B,D}(G) = \sum_{i = 1}^c c_{k_i} Z_{B,D}(k_i,G)
\end{equation}
We will now prove the following claim. The result follows by taking
$S=\bigcup_{i\in[c]} \{k_i\}$.

\begin{claim}\medskip\noindent
Given a set~$S$ of inequivalent spins and a spin $k\in S$ we
can compute $Z_{B,D}(k,G)$ in polynomial time
using an oracle for
computing $\sum_{k\in S} c_k Z_{B,D}(k,G)$.
\end{claim}

\proof
The proof is by induction on $|S|$.
The base case $|S|=1$ is straightforward, so assume $|S|>1$.  
We will show how to compute $Z_{B,D}(k,G)$ (for any spin $k\in S$) using
an oracle for $\sum_{k\in S} c_k Z_{B,D}(k,G)$.
Fix distinct spins $i$ and $j$ in $S$.
By Lemma \ref{lem:lov06},  there is a labelled graph $G_{i,j}$ such that 
\begin{equation}\label{eq:distinguish_pf}
Z_{B,D}(i,G_{i,j}) \neq Z_{B,D}(j,G_{i,j}).
\end{equation}
Note that the construction of $G_{i,j}$ takes $O(1)$ time since $G_{i,j}$ does not
depend on any input graph~$G$.
Partition $S$ into classes $J_1,\ldots ,J_t$ such that $\nu,{\nu'} \in J_\mu$ iff 
$Z_{B,D}(\nu,G_{i,j}) = Z_{B,D}({\nu'},G_{i,j})$. 
We will show below how to compute $\sum_{k\in J_\mu} c_k Z_{B,D}(k,G)$ (for any $\mu\in [t]$) using
an oracle for $\sum_{k\in S} c_k Z_{B,D}(k,G)$.
Once we've done that, we can finish as follows.
For a fixed $k\in S$, suppose $k\in J_\mu$. Note that $|J_\mu|<S$
since one of spins~$i$ and~$j$ is not in~$J_\mu$.
By induction, we can compute $Z_{B,D}(k,G)$ using the
newly-constructed oracle to compute $\sum_{k\in J_\mu} c_k Z_{B,D}(k,G)$.

To finish, we now show how to compute $\sum_{k\in J_\mu} c_k Z_{B,D}(k,G)$
using
an oracle for $\sum_{k\in S} c_k Z_{B,D}(k,G)$.
For every $\mu\in[t]$, let $s_\mu$ be a spin in $J_\mu$.
Let $w_\mu=Z_{B,D}(s_\mu,G_{i,j})$.
Let
\begin{align*}
f_r &= \sum_{k\in S} c_k Z_{B,D}(k,G G_{i,j}^r)\\
&=
\sum_{\mu\in [t]}\sum_{k\in J_\mu}  c_k Z_{B,D}(k,G G_{i,j}^r)\\
&= \sum_{\mu\in[t]} \sum_{k \in J_\mu} c_k Z_{B,D}(k,G) {(Z_{B,D}(k,G_{i,j}))}^r.\\
&= \sum_{\mu\in[t]} { w_\mu}^r
\sum_{k \in J_\mu} c_k Z_{B,D}(k,G).
\end{align*}
Note that we can compute $f_r$ in polynomial time using the oracle.
Now by  Lemma~\ref{lem:dg00_interpolate}
we can recover $\sum_{k\in J_\mu} c_k Z_{B,D}(k,G)$ for every $\mu$ apart from   the
one with $w_\mu=0$ (if there is a $\mu$ with $w_\mu=0$). But we can recover this one, if it exists, by subtraction
since
\[\sum_{k\in J_\mu} c_k Z_{B,D}(k,G) = 
\sum_{k\in S} c_k Z_{B,D}(k,G) - 
\sum_{\nu\neq \mu} \sum_{k\in J_\nu} c_k Z_{B,D}(k,G).
\]%
\end{proof}

The following Corollary will be helpful in the proof of Lemma  \ref{lem:bipolar_bipartite_tensor_decomp}

\begin{cor}\label{cor:fixation_bipolar}
Let $C\in \algR^{m\times m}$ be a symmetric matrix and $D,O \in \algR^{m \times m}$ diagonal matrices such that the diagonal of $D$ is positive and that of $O$ non-negative such that $D-O$ is non-negative. Then, for every labelled graph $G$ and every $k \in [m]$, we can compute $Z_{C,D,O}(k,G)$ in polynomial time using an $\eval{C,D,O}$ oracle.
\end{cor}
\begin{proof}
Let $\Delta^+$ and $\Delta^-$ be diagonal $m\times m$ matrices with
$\Delta^+_{i,i} = (D_{i,i}+O_{i,i})/2$ and $\Delta^-_{i,i}=(D_{i,i}-O_{i,i})/2$.
Let 
Let  
$$A = 
\begin{pmatrix}1 & -1\\
-1 & 1\end{pmatrix} \otimes C.$$
Let $\Delta$ be the $2m \times 2m$ matrix
such that, for all $i\in[m]$ and $j\in[m]$,
$\Delta_{i,j} = \Delta^+_{i,j}$,
$\Delta_{i,m+j}=\Delta_{m+i,j}=0$, and
$\Delta_{m+i,m+j}=\Delta^-_{i,j}$.
Then by Lemma~\ref{cor:bipolar_std_conversion}
$$
Z_{C,D,O}(G) = Z_{A,\Delta}(G) \text{ for all graphs } G.
$$ 
   
Let $I = \{ i\in [2m] \mid \Delta_{i,i} \neq 0\}$. Since $D+O$ and
$D-O$ are non-negative, we have that the matrix $\Delta_{II}$ has a positive diagonal. By inspection we have $$Z_{A,\Delta}(G) = Z_{A_{II},\Delta_{II}}(G) \text{ for all graphs } G.$$
By 
the Pinning 
Lemma \ref{lem:fixation} we can compute the value $Z_{A_{II},\Delta_{II}}(k,G)$ by an algorithm with oracle access to $\eval{A_{II}, \Delta_{II}}$. Now, $Z_{A_{II},\Delta_{II}}(k,G) = Z_{C,D,O}(k,G)$ for every $k \in [m]$. This finishes the proof.
\end{proof}

\subsubsection{Tensor Product Decomposition}

The following technical Lemma which will be used in the proof of Lemma \ref{lem:decomp}.

\begin{lemma}\label{lem:tensor_decomp}
Given symmetric $r \times r$ matrices $A$ and $D$ and $m\times m$ matrices $A',D'$. Then 
$$Z_{A\otimes A',D \otimes D'}(G) = Z_{A,D}(G) \cdot Z_{A',D'}(G) \text{ for every graph } G. $$
\end{lemma}
\begin{proof}
We consider the indices of $A\otimes A'$ and $D \otimes D'$ as pairs $(i,j) \in[r]\times[m]$ such that, e.g.
$$
(A\otimes A')_{(i,i')(j,j')} = A_{i,j} \cdot A'_{i',j'}
$$

Let $\pi : [r]\times[m] \rightarrow [r]$ and $\rho: [r]\times[m] \rightarrow [m]$ be the canonical projections  i.e. for every $(i,j) \in [r] \times[m]$ we have $ \pi(i,j) = i$ and $ \rho(i,j) = j$.

Thus
\begin{eqnarray*}
Z_{A\otimes A',D \otimes D'}(G) & = & \sum_{\xi : V \rightarrow [r]\times[m]} \prod_{uv \in E} (A\otimes A')_{\xi(u),\xi(v)} \prod_{v \in V} (D\otimes D')_{\xi(v),\xi(v)}\\
& = & \sum_{\xi : V \rightarrow [r]\times[m]}\prod_{uv \in E} A_{\pi(\xi(u)),\pi(\xi(v))} A'_{\rho\xi(u),\rho\xi(v)} 
         \prod_{v \in V} D_{\pi(\xi(v)),\pi(\xi(v))}D'_{\rho\xi(v),\rho\xi(v)}\\
& = & \sum_{\substack{\xi : V \rightarrow [r] \\
                      \xi' : V \rightarrow [m]}} \prod_{uv \in E} A_{\xi(u),\xi(v)}A'_{\xi'(u),\xi'(v)} 
         \prod_{v \in V} D_{\xi(v),\xi(v)}D'_{\xi'(v),\xi'(v)}\\
& = & Z_{A,D}(G)\cdot Z_{A',D'}(G)
\end{eqnarray*}
\end{proof}

It is not hard to see that this kind of decomposition can be performed for parity-distinguishing partition functions 
as well, as the following lemma shows.

\begin{lemma}\label{lem:bipolar_tensor_decomp}
  Suppose that $A'$ is a symmetric $m' \times m'$ matrix and $D'$ and $O'$ are
  diagonal $m'\times m'$ matrices.  Suppose that $A''$ is a symmetric $m''
  \times m''$ matrix and $D''$ and $O''$ are diagonal $m''\times m''$
  matrices.  Then, for every graph $G$,
  \[
  Z_{A'\otimes A'',D' \otimes D'',O' \otimes O''}(G) = 
  Z_{A',D',O'}(G) \cdot Z_{A'',D'',O''}(G).
  \]
\end{lemma}

\begin{proof}
Let $A=A'\otimes A''$, $D=D'\otimes D''$ and $O=O'\otimes O''$.
We consider the indices of $A$, 
$D$ and $O$  as pairs $(i,j) \in[m']\times[m'']$ such that, for example,
$$
(A)_{(i',i'')(j',j'')} = A'_{i',j'} \cdot A''_{i'',j''}.
$$

Let $\pi' : [m']\times[m''] \rightarrow [m']$ and $\pi'': [m']\times[m''] \rightarrow [m'']$ be the canonical projections  i.e. for every $(i,j) \in [m'] \times[m'']$ we have $ \pi'(i,j) = i$ and $ \pi''(i,j) = j$.

With $V_o \subseteq V$ the set of even degree vertices and $V_e = V \setminus V_o$ we have
\begin{eqnarray*}
Z_{A,D,O}(G) & = & \sum_{\xi : V \rightarrow [m']\times[m'']} \prod_{\{u,v\} \in E} A_{\xi(u),\xi(v)} \prod_{v \in V_e} D_{\xi(v),\xi(v)}\prod_{v \in V_o} O_{\xi(v),\xi(v)}\\
\end{eqnarray*}
With 
\begin{eqnarray*}
\prod_{v \in V_e} D_{\xi(v),\xi(v)} &=& \prod_{v \in V_e} D^{'}_{\pi'(\xi(v)),\pi'(\xi(v))}\prod_{v \in V_e} D^{''}_{\pi''(\xi(v)),\pi''(\xi(v))} \\
\prod_{v \in V_o} O_{\xi(v),\xi(v)} &=& \prod_{v \in V_O} O^{'}_{\pi'(\xi(v)),\pi'(\xi(v))}\prod_{v \in V_o} O^{''}_{\pi''(\xi(v)),\pi''(\xi(v))}
\end{eqnarray*}
and
\begin{eqnarray*}
\prod_{\{u,v\} \in E} A_{\xi(u),\xi(v)} &=& \prod_{\{u,v\} \in E} A'_{\pi'(\xi(u)),\pi'(\xi(v))}\prod_{\{u,v\} \in E} A''_{\pi''(\xi(u)),\pi''(\xi(v))}
\end{eqnarray*}
we therefore have
\begin{eqnarray*}
Z_{A,D,O}(G) & = & \left(\sum_{\psi' : V \rightarrow [m']} \prod_{\{u,v\} \in E} A'_{\psi'(u),\psi'(v)} \prod_{v \in V_e} D^{'}_{\psi'(v),\psi'(v)}\prod_{v \in V_o} O^{'}_{\psi'(v),\psi'(v)}\right)\\
                 & & \cdot\left(\sum_{\psi'' : V \rightarrow [m'']} \prod_{\{u,v\} \in E} A''_{\psi''(u),\psi''(v)} \prod_{v \in V_e} D^{''}_{\psi''(v),\psi''(v)}\prod_{v \in V_o} O^{''}_{\psi''(v),\psi''(v)}\right)\\
& = & Z_{A',D^{'},O^{'}}(G)\cdot Z_{A'',D^{''},O^{''}}(G)
\end{eqnarray*}

\end{proof}

\begin{cor}\label{lem:1may}
  Let $B'$ be a symmetric $m' \times m'$ block and let $D^{R'}$ and $O^{R'}$
  be diagonal $m' \times m'$ matrices. Let $B''$ be a symmetric $m'' \times
  m''$ block and let $D^{R''}$ and $O^{R''}$ be diagonal $m'' \times m''$
  matrices.  Let $D^R = D^{R'}\otimes D^{R''}$ and $O^R = O^{R'}\otimes
  O^{R''}$ and $B = B'\otimes B''$.  If $\eval{B'',D^{R''},O^{R''}}$ is
  polynomial time computable
then 
$$
\eval{B,D^R,O^R}
\Tequiv 
\eval{B',D^{R'},O^{R'}}.
$$
\end{cor}

\begin{proof}
For every graph $G$,
Lemma~\ref{lem:bipolar_tensor_decomp} gives
$$Z_{B'\otimes B'',D^{R'} \otimes D^{R''},O^{R'} \otimes O^{R''}}(G) = 
Z_{B',D^{R'},O^{R'}}(G) \cdot Z_{B^{''},D^{R''},O^{R''}}(G).$$
If $\eval{B'',D^{R''},O^{R''}}$ is polynomial time computable then this gives
\[
\eval{B,D^R,O^R}
\Tequiv 
\eval{B',D^{R'},O^{R'}}.
\]
\end{proof}

\begin{lemma}\label{lem:bipolar_bipartite_tensor_decomp}
Let $B'$ be an $m' \times n'$ block, $D^{R'}$ and $O^{R'}$ be diagonal $m' \times m'$ matrices and $D^{C'}$ and $O^{C'}$ be diagonal $n' \times n'$ matrices. Let $B''$ be an $m'' \times n''$ block, $D^{R''}$ and $O^{R''}$ be diagonal $m'' \times m''$ matrices and $D^{C''}$ and $O^{C''}$ be diagonal $n'' \times n''$ matrices.
Let 
$$
D' = \left( \begin{array}{c c}
            D^{R'} & 0 \\
            0 & D^{C'}
           \end{array}\right)
\text{ and }
D'' = \left( \begin{array}{c c}
            D^{R''} & 0 \\
            0 & D^{C''}
           \end{array}\right)
\text{ and }
D = \left( \begin{array}{c c}
            D^{R'} \otimes D^{R''} & 0 \\
            0 & D^{C'} \otimes D^{C''}
           \end{array}\right)
$$
and let $O$ and $O',O''$ be constructed from $O^R,O^C$ and $O^{R'},O^{C'}$ in the analogous way. Let $A, A', A''$ be the connected bipartite matrices with underlying blocks $B := B'\otimes B''$, $B'$ and $B''$ respectively. 

If $\eval{A'',D'',O''}$ is polynomial time computable and 
$D+O$ and $D-O$ have only non-negative entries then
$$
\eval{A,D,O} \Tequiv \eval{A',D',O'}.
$$
\end{lemma}
\begin{proof}
  Note that $Z_{A,D,O}(G) = 0$ unless $G$ is bipartite. Therefore we will
  assume in the following that all graphs $G$ are bipartite and that $(U,W)$
  is a partition of the vertex set $V$ into two independent
  sets.
Assume first that $G$ is connected - the
  case of non-connected graphs will be handled later. Note that $A$ is a
  square matrix of order $m + n$ for $m = m'm''$ and $n = n'n''$. For diagonal
  $r \times r$ matrices $D,O$ a set $X \subseteq V$ and a configuration $\xi :
  X \rightarrow [r]$ define
$$
\nodew_{D,O}(X,\phi) : = \prod_{\substack{x \in X \\ \text{deg}(x) \text { even}}} D_{\xi(x),\xi(x)}
                   \prod_{\substack{x \in X \\ \text{deg}(x) \text { odd}}} O_{\xi(x),\xi(x)}.
$$

By the above definitions we have, 
\begin{eqnarray*}
Z_{A,D,O}(G) &=& \sum_{\substack{\xi: U \rightarrow [m+n] \\ \psi: W \rightarrow [m+n]}}
               \prod_{\{u,w\} \in E} A_{\xi(u),\psi(w)} 
               \nodew_{D,O}(U,\xi)\nodew_{D,O}(W,\psi) \\
\end{eqnarray*}

And therefore, since $G$ is connected
\begin{eqnarray*}%
Z_{A,D,O}(G) &=& \phantom{ + }\sum_{\substack{\xi: U \rightarrow [m] \\ \psi: W \rightarrow [n]}}
               \prod_{\{u,w\} \in E} B_{\xi(u),\psi(w)} 
               \nodew_{D^R,O^R}(U,\xi)\nodew_{D^C,O^C}(W,\psi) \\
            & &  + \sum_{\substack{\xi: U \rightarrow [n] \\ \psi: W \rightarrow [m]}}
               \prod_{\{u,w\} \in E} B_{\psi(w),\xi(u)} 
               \nodew_{D^C,O^C}(U,\xi)\nodew_{D^R,O^R}(W,\psi) 
\end{eqnarray*}

Define
\begin{equation}\label{eq:bip_right}
Z^{\rightarrow}_{A,D,O}(G) :=\sum_{\substack{\xi: U \rightarrow [m] \\ \psi: W \rightarrow [n]}}
               \prod_{\{u,w\} \in E} B_{\xi(u),\psi(w)} 
               \nodew_{D^R,O^R}(U,\xi)\nodew_{D^C,O^C}(W,\psi) \\
\end{equation}
and
\begin{equation}\label{eq:left}
Z^{\leftarrow}_{A,D,O}(G) := \sum_{\substack{\xi: U \rightarrow [n] \\ \psi: W \rightarrow [m]}}
               \prod_{\{u,w\} \in E} B_{\psi(w),\xi(u)} 
               \nodew_{D^C,O^C}(U,\xi)\nodew_{D^R,O^R}(W,\psi)
\end{equation}
That is
\begin{equation}\label{eq:z_both_dir}
Z_{A,D,O}(G) = Z^{\rightarrow}_{A,D,O}(G) + Z^{\leftarrow}_{A,D,O}(G) 
\end{equation}
For matrices $A',D',O'$ and $A'',D'',O''$ we define the analogous expressions ($Z^{\leftarrow}_{A',D',O'}(G)$, etc.).

We consider the indices of $B'\otimes B''$ as pairs. That is row indices are $(i',i'') \in [m']\times[m'']$ and column indices become $(j',j'') \in [n']\times [n'']$.
$$
(B'\otimes B'')_{(i',i'')(j',j'')} = B'_{i',j'} \cdot B''_{i'',j''}
$$

Let $\rho' : [m']\times[m''] \rightarrow [m']$, $\rho'' : [m']\times[m'']
\rightarrow [m'']$ and $\gamma': [n']\times[n''] \rightarrow [n']$,$\gamma'':
[n']\times[n''] \rightarrow [n'']$ be the canonical projections.
That is for $(i',i'') \in [m'] \times[m'']$ we have $ \rho'(i',i'') = i'$, $\rho''(i',i'') = i''$  and for $(j',j'') \in [n']\times [n'']$ we have $\gamma'(j',j'') = j'$ and $\gamma''(j',j'') = j''$.
Therefore, for all $\xi: U \rightarrow [m]$ and $\psi: W \rightarrow [n]$
we have
\begin{eqnarray*}
\prod_{\{u,w\} \in E} B_{\xi(u),\psi(w)} 
&=& \prod_{\{u,w\} \in E} B'_{\rho'\circ\xi(u),\gamma'\circ\psi(w)} 
    \cdot \prod_{\{u,w\} \in E} B''_{\rho''\circ\xi(u),\gamma''\circ\psi(w)}
\end{eqnarray*}
and
\begin{eqnarray*}
\nodew_{D^R,O^R}(U,\xi) &=& \nodew_{D^{R'},O^{R'}}(U,\rho'\circ\xi)\nodew_{D^{R''},O^{R''}}(U,\rho''\circ\xi)\\
\nodew_{D^C,O^C}(W,\psi) &=& \nodew_{D^{C'},O^{C'}}(W,\gamma'\circ\psi)\nodew_{D^{C''},O^{C''}}(W,\gamma''\circ\psi)\\
\end{eqnarray*}
Hence, we can rewrite equation \eqref{eq:bip_right}:
\begin{eqnarray*}
Z^{\rightarrow}_{A,D,O}(G) &=& \left(\sum_{\substack{\xi': U \rightarrow [m'] \\ \psi': W \rightarrow [n']}}
               \prod_{\{u,w\} \in E} B'_{\xi'(u),\psi'(w)} 
               \nodew_{D^{R'},O^{R'}}(U,\xi')\nodew_{D^{C'},O^{C'}}(W,\psi') \right) \\
            & &\left(\sum_{\substack{\xi'': U \rightarrow [m''] \\ \psi'': W \rightarrow [n'']}}
               \prod_{\{u,w\} \in E} B''_{\xi''(u),\psi''(w)} 
               \nodew_{D^{R''},O^{R''}}(U,\xi'')\nodew_{D^{C''},O^{C''}}(W,\psi'') \right)\\
            &=& Z^{\rightarrow}_{A',D',O'}(G)\cdot Z^{\rightarrow}_{A'',D'',O''}(G)
\end{eqnarray*}
With an analogous argument this extends to $Z^{\leftarrow}_{A,D,O}(G)$. We therefore have
\begin{eqnarray}
Z^{\leftarrow}_{A,D,O}(G)  &=& Z^{\leftarrow}_{A',D',O'}(G)\cdot Z^{\leftarrow}_{A'',D'',O''}(G) \label{eq:z_left}\\
Z^{\rightarrow}_{A,D,O}(G) &=& Z^{\rightarrow}_{A',D',O'}(G)\cdot Z^{\rightarrow}_{A'',D'',O''}(G) \label{eq:z_right}
\end{eqnarray}

\begin{claim}\label{cl:bip_left_right_sep}  The values
$Z^{\rightarrow}_{A,D,O}(G)$  and $Z^{\leftarrow}_{A,D,O}(G)$ can be computed in polynomial time for every graph $G$ by an algorithm with oracle access to $\eval{A,D,O}$.
\end{claim}
\begin{clproof}
Let $G = (U,W,E)$ be a given connected bipartite graph and label a vertex $u \in U$. Then
$$
Z^{\rightarrow}_{A,D,O}(G) = \sum_{k=1}^m Z_{A,D,O}(k,G).
$$
and the values $Z_{A,D,O}(k,G)$ can be computed using the $\eval{A,D,O}$ oracle by Corollary \ref{cor:fixation_bipolar}.

The analogous argument labelling a vertex $w\in W$ yields the result for $Z^{\leftarrow}_{A,D,O}(G)$.
\end{clproof}

We will show first, that $\eval{A,D,O} \Tle \eval{A',D',O'}$. 
Let $G$ be a given connected graph.
By equations \eqref{eq:z_both_dir} \eqref{eq:z_left} and \eqref{eq:z_right} we have
\begin{eqnarray*}
Z_{A,D,O}(G) &=& Z^{\rightarrow}_{A,D,O}(G) + Z^{\leftarrow}_{A,D,O}(G) \\ 
             &=& Z^{\rightarrow}_{A',D',O'}(G)Z^{\rightarrow}_{A'',D'',O''}(G) +
                 Z^{\leftarrow}_{A',D',O'}(G)Z^{\leftarrow}_{A'',D'',O''}(G) 
\end{eqnarray*}
By Claim~\ref{cl:bip_left_right_sep} we can compute the values
$Z^{\rightarrow}_{A',D',O'}(G)$ and $Z^{\leftarrow}_{A',D',O'}(G)$ using the
$\eval{A',D',O'}$ oracle. The values $Z^{\rightarrow}_{A'',D'',O''}(G)$ and
$Z^{\leftarrow}_{A'',D'',O'}(G)$ can be computed by
Claim~\ref{cl:bip_left_right_sep} using the fact that $\eval{A'',D'',O''}$ is
polynomial time computable by the condition of the Lemma.

To see that $\eval{A',D',O'} \Tle \eval{A,D,O}$ note that by
Claim~\ref{cl:bip_left_right_sep} be can compute 
\[
Z^{\rightarrow}_{A,D,O}(G)=Z^{\rightarrow}_{A',D',O'}(G)Z^{\rightarrow}_{A'',D'',O''}(G)
\]
and 
\[
Z^{\leftarrow}_{A,D,O}(G)=Z^{\leftarrow}_{A',D',O'}(G)Z^{\leftarrow}_{A'',D'',O''}(G)
\] 
using an
$\eval{A,D,O}$ oracle. And by Claim~\ref{cl:bip_left_right_sep} using the fact
that $\eval{A'',D'',O''}$ is polynomial time computable, we can compute
$Z^{\rightarrow}_{A'',D'',O''}(G)$ and $Z^{\leftarrow}_{A'',D'',O''}(G)$, hence $Z^{\rightarrow}_{A',D',O'}(G)$ and $Z^{\leftarrow}_{A',D',O'}(G)$, and
finally
$$
Z_{A',D',O'}(G)=Z^{\rightarrow}_{A',D',O'}(G) + Z^{\leftarrow}_{A',D',O'}(G).
$$

The proof for non-connected $G$ follows from the above using the fact that 
$$
Z_{A,D,O}(G) = \prod_{i=1}^c Z_{A,D,O}(G_i)
$$
with $G_1, \ldots, G_c$ being the connected components of $G$.
\end{proof}

\subsection{The Proof of Lemma \ref{lem:component_pinning}}

\begin{proof}[Proof of Lemma \ref{lem:component_pinning}]
Let $G$ be a given graph note that if $G = (V,E)$ is not connected with $G_1,\ldots,G_k$ being the components of $G$ then we have
$$
Z_{A}(G) = \prod_{i=1}^k \sum_{j=1}^c Z_{A_j}(G_i)
$$
This proves (2). To prove (1) note that for hardness %
we may restrict ourselves to connected $G$.

Therefore, for some $i \in [c]$ fix a component $A_i$ of $A$ and let $I \subseteq [m]$ be the set of row/columns indices such that $A_i = A_{II}$. Let $G = (V,E)$ be a connected graph and call some vertex $z \in V$ the labelled vertex of $G$. Then by the connectedness of $G$ we have
$$
Z_{A_i}(G) = \sum_{k \in I} Z_{A}(k,G)
$$
The proof now follows by 
the Pinning 
Lemma \ref{lem:fixation}.
\end{proof}

\subsection{The Proof of Lemma \ref{lem:decomp}}
\newcommand{\bpf}{parity-distinguishing }
In order to prove Lemma~\ref{lem:decomp},
it will be convenient to transition 
from partition functions to \bpf partition functions. How this translation can be performed will be described in Lemma \ref{lem:conditions_C1_C2}. Once we have determined some conditions on the shape of the
resulting partition functions the proof of Lemma \ref{lem:decomp} will become straightforward.

\paragraph*{Shape Conditions.} 
Given an evaluation problem $\eval{C,D,O}$ with $D,O$ diagonal matrices of vertex weights and $C$ a connected bipartite matrix with underlying block $B$.
We define conditions on the shape of $C$ and $D,O$. These conditions will be used incrementally, that is, we will rely on \cond{C$(i+1)$} only if \cond{C1}-\cond{C$i$} are assumed to hold.
\begin{condition}{(C1)}
 There are $r,m,n \in \Nat$, a 
non-singular $r \times r$-matrix $H$ with entries in $\{-1,1\}$ and vectors $v \in {\algR}^m_{>0}$, $w \in {\algR}^n_{>0}$ of pairwise distinct entries such that
       $$
          B = vw^T \otimes H = \left(\begin{array}{c c c}
               v_1w_1 H & \ldots & v_1w_n H\\
               \vdots  & \ddots & \vdots \\
               v_mw_1 H & \ldots & v_mw_n H 
            \end{array}\right).
       $$
\end{condition}

\medskip\noindent 
If $B$ satisfies \cond{C1},
for convenience, we consider the
indices of the entries in $B$ as pairs such that $B_{(\mu,i),(\nu,j)} = v_\mu
w_\nu H_{i,j}$, for $\mu \in [m], \nu \in [n]$ and $i,j \in [r]$. We call the
submatrices $v_\mu v_\nu H $ the \emph{tiles} of $B$.

The diagonal entries of the matrices $D$ and $O$ are vertex weights which by the shape of $C$
$$
C = \left(\begin{array}{c c} 0 & B \\ B^T & 0 \end{array}\right)
$$
will be considered with respect to $B$. As $B$ is a $rm \times rn$ matrix, we group the entries of $O$ and $D$ into $rm\times rm$ submatrices $D^R,O^R$ corresponding to the rows of $B$ and $rn \times rn$ submatrices $D^C,O^C$ corresponding to the columns of $B$ so as to obtain
$$
D = \left(\begin{array}{c c}D^R & 0 \\ 0 & D^C\end{array} \right)
\text{ and }
O = \left(\begin{array}{c c}O^R & 0 \\ 0 & O^C\end{array} \right).
$$
Furthermore, according to the tiles of $B$ the matrix $D^R$ can be grouped into to $m$ \emph{tiles} $D^{R,\mu}$ (for all $\mu\in[m]$) each of which is an $r \times r$ diagonal matrix. Analogously we group the matrix $D^C$ into $n$ submatrices $D^{C,\nu}$ for all $\nu \in [n]$ and we obtain
$$
D^R = \left(\begin{array}{c c c}D^{R,1} & \ldots & 0 \\ \vdots & \ddots & \vdots \\
                                0 & \ldots & D^{R,m}\end{array} \right)
\text{ and }
D^C = \left(\begin{array}{c c c}D^{C,1} & \ldots & 0 \\ \vdots & \ddots & \vdots \\
                                0 & \ldots & D^{C,n }\end{array} \right).
$$
The matrices $O^R$ and $O^C$ are grouped analogously.
If $B$ is symmetric then $D^R=D^C$ and $O^R=O^C$.
We define four more conditions

\begin{condition}{(C2)}
 $D$ is a diagonal matrix of positive vertex weights, $O^{R,1},O^{C,1}$ and $D+O$ and $D-O$ are non-negative.
\end{condition}
\begin{condition}{(C3)}
 The matrix $H$ is a Hadamard matrix. %
\end{condition}
\begin{condition}{(C4)}

 For all $\mu \in [m],\nu \in [n]$ there are an $\alpha^R_\mu, \alpha^C_\nu$ such that $D^{R,\mu} = \alpha^R_\mu I_r$ and $D^{C,\nu} = \alpha^C_\nu I_r$. 

\end{condition}
\begin{condition}{(C5)}
 There are sets $\Lambda^R,\Lambda^C \subseteq [r]$ such that
 
 for all $\mu \in [m],\nu \in [n]$ there is a $\beta^R_\mu, \beta^C_\nu$ such that 
$O^{R,\mu} = \beta^R_\mu I_{r;\Lambda^R}$ and $O^{C,\nu} = \beta^C_\nu
I_{r;\Lambda^C}$.

\end{condition}

Before we transform a given problem $\eval{A}$ into the form $\eval{C,D,O}$ in Lemma \ref{lem:conditions_C1_C2} we will exclude some cases from our consideration. That is, we show in the following Lemma that $\eval{A}$ is $\#\PP$-hard unless the block $B$ underlying $A$ satisfies $\rank \abs B = 1$.

\begin{lemma}\label{lem:to_abs_rank_1}
Let $A$ be a symmetric connected bipartite matrix with underlying block $B$. Then 
at least one of the following outcomes occurs.
\begin{description}
\item[Outcome 1]
$\eval{A}$ is \#P-hard. If $B$ is symmetric, then
$\eval{B}$ is \#P-hard.
\item[Outcome 2] 
For some $m,n\in \Nat$ there are vectors $v \in \algR^m$ and $w \in \algR^n$ satisfying $0 < v_1 < \ldots < v_m$ and $0 < w_1 < \ldots < w_n$
and permutations~$\Sigma$ and~$\Pi$ such that
$$
B_{\Sigma,\Pi} =\left(\begin{array}{c c c}
               v_1w_1 S^{1 1} & \ldots & v_1w_n S^{1 n}\\
               \vdots  & \ddots & \vdots \\
               v_mw_1 S^{m 1} & \ldots & v_mw_n S^{m n}
            \end{array}\right)
$$
where, for $i\in[m]$ and $j\in[n]$, 
$S^{ij}$ is a
$\{-1,1\}$-matrix of some order $m_i \times n_j$.
If $B$ is symmetric then $\Sigma=\Pi$.
\end{description}
\end{lemma}
\begin{proof}
By Lemma~\ref{lem:bipolar_block2_hard},
$\eval{A}$  is
 $\#\PP$-hard unless $\rank \abs B = 1$. 
Similarly, if $B$ is symmetric then
$\eval{B}$ is $\#\PP$-hard unless $\rank \abs B = 1$. 

We conclude that $\abs B$ = $xy^T$ for some non-negative real vectors $x,y$.
If $B$ is symmetric then we can take $y=x$.
To see, suppose $\hat x$ and $\hat y$ are vectors such that $\hat x {\hat y}^T$ is symmetric
and let $x_i = y_i = \sqrt{\hat{x}_i \hat{y}_i}$. Note
that $x_i y_j = \sqrt{\hat{x}_i \hat{y}_i \hat{x}_j \hat{y}_j} = \hat{x}_i
\hat{y}_j$.%

Note that the vectors $x$ and $y$ contain no zero entries. This follows from the fact that $\abs B$ is a block because $B$ is. Hence if some entry of $x$ satisfies $x_i = 0$ then $A\row i = x_ix^T = 0$ and therefore $B$ has a decomposition.

Let $v \in \algR^m$ be the vector of ascendingly ordered distinct entries of $x$%
. That is, $v_i < v_j$ for all $i<j$ and, for each $x_i$, there is a $j \in [m]$ s.t. $x_i = v_j$. Similarly, let $w$ be the vector of
ascendingly ordered distinct entries of $y$.
\end{proof}

\begin{lemma}\label{lem:symnonsing_submat_of_sym_mat}
Let $A$ be a symmetric $n \times n$ matrix of rank $r$ and $I \subseteq [n]$ a set of indices with $\vert I \vert = r$. If $A_{I*}$ has rank $r$ then the matrix $A_{II}$ is non-singular.
\end{lemma}
\begin{proof}
As $\rank A_I = \rank A$ the rows of $A$ with indices in $\bar{I}$ depend linearly on those from $I$. By symmetry this holds for the columns as well and is still true in $A_I$. Hence $\rank A = \rank A_{II}$.
\end{proof}

\begin{lemma}\label{lem:conditions_C1_C2}
Let $A$ be a symmetric connected bipartite matrix with underlying block $B_A$ of rank $r$.
Then at least one of the following outcomes occurs.
\begin{description}
\item[Outcome 1]
$\eval{A}$ is \#P-hard. If $B_A$ is symmetric, then
$\eval{B_A}$ is \#P-hard.
\item[Outcome 2]
There 
is a connected bipartite matrix~$C$, whose underlying block $B$ is size $m r \times n r$ for some $m$ and $n$, and diagonal matrices $D$ and $O$ 
which satisfy conditions \cond{C1} and \cond{C2}, such that
$$
\eval{C,D,O} \Tequiv \eval{A}.
$$
The matrices $D$ and $O$ consist of $mr \times mr$ submatrices $D^R,O^R$ and $nr \times nr$ submatrices $D^C,O^C$ such that
$$
D = \left(\begin{array}{c c}D^R & 0 \\ 0 & D^C\end{array} \right)
\text{ and }
O = \left(\begin{array}{c c}O^R & 0 \\ 0 & O^C\end{array} \right).
$$
$C,D$ and $O$ can be computed in time polynomial in the size of $A$.
If $B_A$  is symmetric then so is $B$. Also $D^R = D^C$, $O^R = O^C$
and 
$$
\eval{B,D^R,O^R} \Tequiv \eval{B_A}.
$$ 
\end{description}
\end{lemma}
\begin{proof}
Suppose that the matrix~$A$ does not give Outcome~1 in Lemma~\ref{lem:to_abs_rank_1}.
Let $\Sigma$ and $\Pi$ be the permutations from Lemma~\ref{lem:to_abs_rank_1}
and let $\Phi$ be the permutation on the rows of~$A$ that applies~$\Sigma$
to the rows of~$B_A$ and applies~$\Pi$ to the columns. Let $\widetilde{A}=
A_{\Phi,\Phi}$. Note that $\eval{A}\equiv\eval{\widetilde{A}}$.
Also, the block underlying~$\widetilde{A}$ is ${(B_A)}_{\Sigma,\Pi}$,
which we denote~$\widetilde{B}$. Note that~$\widetilde{B}$ is symmetric if 
$B_A$ is symmetric, since $\Sigma=\Pi$ in that case
and $\eval{B_A}\equiv \eval{\widetilde{B}}$.
By Lemma \ref{lem:to_abs_rank_1} there are~$m,n\in \mathbb{N}$ such that
$$
\widetilde{B} =\left(\begin{array}{c c c}
               v_1w_1 S^{1 1} & \ldots & v_1w_n S^{1 n}\\
               \vdots  & \ddots & \vdots \\
               v_mw_1 S^{m 1} & \ldots & v_mw_n S^{m n} 
            \end{array}\right)
$$
for vectors $v \in \algR^m,\; w \in \algR^n$ of positive pairwise distinct reals and $\{-1,1\}$-matrices $S^{\kappa \lambda}$ of order $m_\kappa \times n_\lambda$.
Let
$$
 S =\left(\begin{array}{c c c}
               S^{1 1} & \ldots & S^{1 n}\\
               \vdots  & \ddots & \vdots \\
               S^{m 1} & \ldots & S^{m n} 
            \end{array}\right)
$$

For convenience, we consider the indices of the entries in $\widetilde{B}$ as pairs such that $\widetilde{B}_{(\kappa,i),(\lambda,j)} = v_{\kappa}w_{\lambda}S^{\kappa \lambda}_{i,j}$, for $(\kappa,\lambda) \in [m]\times[n]$ and $(i,j) \in [m_\kappa]\times [n_\lambda]$. Entries and submatrices of $S$ will be treated in the same way.

First we shall see that we may assume that every pair of rows (or columns) of $S$ is either orthogonal, or they are (possibly negated) copies of each other.
\begin{claim}\label{cl:s_shape} 
Outcome~1 occurs unless for all $\kappa,\lambda \in [m]$ and $i \in [m_\kappa], j \in [m_\lambda]$ 
\begin{equation}\label{eq:row_cond_S}
\begin{array}{r r}
\text{ either} & \scalp{S^{\kappa \nu}\row i,S^{\lambda\nu}\row j} = 0 \text{ for every } \nu \in [n] 
\\
 \text{ or there is a } s \in\{-1, +1\} \text{ such that } & S^{\kappa\nu}\row i = s S^{\lambda\nu}\row j \text{ for every } \nu \in [n]. \\
\end{array}
\end{equation}
The analogues holds for the columns of $S$: for all $\kappa,\lambda \in [n]$ and $i\in [n_\kappa], j \in [n_\lambda]$ 
\begin{equation}\label{eq:col_cond_S}
\begin{array}{r r}
\text{ either} & \scalp{S^{\mu \kappa}\col i,S^{\mu\lambda}\col j} = 0 \text{ for every } \mu \in [m] \\
 \text{ or there is a } s \in\{-1,+1 \} \text{ such that } & S^{\mu\kappa}\col i = s S^{\mu\lambda}\col j \text{ for every } \mu \in [m]. \\
\end{array}
\end{equation}
\end{claim}
\begin{clproof}
Let $p \in \Nat$ be odd. By $p$-thickening and subsequent $2$-stretching we obtain a reduction $$\eval{A'} \Tle \eval{\widetilde{A}}$$ for a matrix $A' = (\widetilde{A}^{(p)})^2$ which contains submatrices $\widetilde{B}^{(p)}(\widetilde{B}^{(p)})^T$ and $(\widetilde{B}^{(p)})^T\widetilde{B}^{(p)}$.
The same reduction gives
$\eval{{(\widetilde{B}^{(p)})}^2}\Tle \eval{\widetilde{B}}$
if $\widetilde{B}$ is symmetric.
We will give the proof of equation \eqref{eq:row_cond_S} by focusing on $\widetilde{B}^{(p)}(\widetilde{B}^{(p)})^T$. The analogous argument on $(\widetilde{B}^{(p)})^T\widetilde{B}^{(p)}$ yields equation \eqref{eq:col_cond_S}.

Let $\widetilde{C} = \widetilde{B}^{(p)}(\widetilde{B}^{(p)})^T$. For $\kappa,\lambda \in [m]$ and $i \in [m_\kappa], j \in [m_\lambda]$ we have:
\begin{equation}\label{eq:sign_struc_Cshape}
\widetilde{C}_{(\kappa,i),(\lambda,j)} = \sum_{(\nu,k)} \widetilde{B}^{(p)}_{(\kappa,i),(\nu,k)}\widetilde{B}^{(p)}_{(\lambda,j),(\nu,k)} = v^{p}_{\kappa}v^{p}_{\lambda}\sum_{\nu=1}^n w^{2p}_{\nu} \scalp{S^{\kappa\nu}\row i,S^{\lambda\nu}\row j}.
\end{equation}
Note that by $2$-thickening we have a reduction $\eval{A''} \Tle \eval{\widetilde{A}}$ for a matrix $A'' = (A')^{(2)}$. 
This also gives a reduction $\eval{\widetilde{C}^{(2)}} \Tle \eval{\widetilde{B}}$ if $\widetilde{B}$ is symmetric.
The matrix $A''$ has only non-negative entries and contains the submatrix $\widetilde{C}^{(2)}$. The result of Bulatov and Grohe \cite{bulgro05} implies that 
$\eval{\widetilde{C}^{(2)}}$ and
$\eval{A''}$ are $\#\PP$-hard, in which case, Outcome~1 occurs,
if $\widetilde{C}^{(2)}$ contains a block of row rank at least $2$. We shall determine the conditions under which this is not the case.

A $2\times 2$ principal submatrix of $\widetilde{C}^{(2)}$, defined by $(\kappa,i),(\lambda,j)$ has determinant
\begin{eqnarray*}
\textup{det}_{(\kappa,i),(\lambda,j)} := \left\vert \begin{array}{r r} 
              \widetilde{C}^{(2)}_{(\kappa,i),(\kappa,i)} & \widetilde{C}^{(2)}_{(\kappa,i),(\lambda,j)} \\ 
              \widetilde{C}^{(2)}_{(\lambda,j),(\kappa,i)} & \widetilde{C}^{(2)}_{(\lambda,j),(\lambda,j)} \end{array} \right\vert 
&=& (\widetilde{C}_{(\kappa,i),(\kappa,i)}\widetilde{C}_{(\lambda,j),(\lambda,j)})^2 - (\widetilde{C}_{(\kappa,i),(\lambda,j)})^4
\end{eqnarray*}
We have
\begin{eqnarray*}
\widetilde{C}^{(2)}_{(\kappa,i),(\kappa,i)} &=& v^{4p}_{\kappa}\left(\sum_{\nu=1}^n w^{2p}_{\nu} \scalp{S^{\kappa\nu}\row i,S^{\kappa\nu}\row i}\right)^2 = v^{4p}_{\kappa}\left(\sum_{\nu=1}^n w^{2p}_{\nu} n_\nu\right)^2
\end{eqnarray*}
and therefore
\begin{eqnarray*}
\textup{det}_{(\kappa,i),(\lambda,j)} &=& v^{4p}_{\kappa}v^{4p}_{\lambda} \left(\left(\sum_{\nu=1}^n w^{2p}_{\nu} n_\nu\right)^4 - \left(\sum_{\nu=1}^n w^{2p}_{\nu} \scalp{S^{\kappa\nu}\row i,S^{\lambda\nu}\row j}\right)^4\right)
\end{eqnarray*}

This determinant is zero iff there is an $s \in \{-1,1\}$ such that $\scalp{S^{\kappa\nu}\row i,S^{\lambda\nu}\row j} = s n_\nu$ for all $\nu \in [n]$ which implies $S^{\kappa\nu}\row i = s S^{\lambda\nu}\row j$ for all $\nu \in [n]$.
By equation \eqref{eq:sign_struc_Cshape} and Lemma \ref{lem:bij_coeff_power} we further have $\widetilde{C}^{(2)}_{(\kappa,i),(\lambda,j)} = 0$ for arbitrarily large $p$ iff
$\scalp{S^{\kappa\nu}\row i,S^{\lambda\nu}\row j} = 0$ for all $\nu \in [n]$.
\end{clproof}

Assume from now on that 
Equations~(\ref{eq:row_cond_S}) and~(\ref{eq:col_cond_S}) hold.
The next claim states that
the rank of each tile of $S$ equals the rank of $S$ itself (which is equal to $r$, the rank of 
$\widetilde{B}$ wich is the rank of~$B_A$).

\begin{claim} \label{cl:all_tiles_same_rank}
$\rank{S} = \rank {S^{\kappa \lambda}}$ for all $(\kappa,\lambda) \in [m]\times[n]$.
\end{claim}
\begin{clproof}
Equation \eqref{eq:row_cond_S} implies that $\rank{ S^{\kappa\mu}} = \rank {S^{\kappa\nu}}$ for all $\kappa \in [m]$ and $\mu, \nu \in [n]$. Combining this with equation \eqref{eq:col_cond_S} we obtain $\rank{ S^{\kappa\mu}} = \rank {S^{\lambda\nu}}$ for all $\kappa,\lambda \in [m]$ and $\mu, \nu \in [n]$.

Therefore it suffices to show that 
$r = \rank {S} = \rank {S^{11}}$ holds.
Let $S^{\absent 1}$ denote the matrix
$$S^{\absent 1} = \left(\begin{array}{c} S^{11} \\ \vdots \\ S^{m1} \end{array} \right).$$
Let $I$ be a set of row indices with 
$\vert I \vert = \rank S = r$ such that the set $\mset{S\row i \mid i \in I}$
is linearly independent.
By equation \eqref{eq:row_cond_S} we have $\scalp{S\row i,S\row j} = 0$ and $\scalp{S^{\absent 1}\row i,S^{\absent 1}\row j} = 0$ for all $i \neq j \in I$. 
Hence, $S^{\absent 1}$ has 
rank $r$. 
As $S^{11}$ is a $m_1 \times n_1$ matrix there is a set $J\subseteq [n_1]$ s.t. the columns of $S^{\absent 1}$ with indices in $J$ form a rank $r$ set. 
Equations \eqref{eq:col_cond_S} implies $\scalp{S^{11}\col i,S^{11}\col j} = 0$ for all $i \neq j \in J$. This proves the claim.
\end{clproof}

Claim~\ref{cl:all_tiles_same_rank} has strong implications on $S$ ( and $\widetilde{B}$). It implies that for all $(\kappa,\lambda) \in [m] \times [n]$ there are sets $K_{(\kappa,\lambda)},L_{(\kappa,\lambda)}$ of cardinality $r$ such that $S^{\kappa\lambda}_{K_{(\kappa,\lambda)}L_{(\kappa,\lambda)}}$ is non-singular. By equation \eqref{eq:row_cond_S} we take, without loss of generality,
$K_{(\kappa,\lambda)} = K_{(\kappa,\lambda')}$ for all $\kappa \in [m]$ and $\lambda,\lambda' \in [n]$. Analogously, equation \eqref{eq:col_cond_S} implies $L_{(\kappa,\lambda)} = L_{(\kappa',\lambda)}$ for all $\kappa,\kappa' \in [m]$ and $\lambda \in [n]$. Therefore,   there are sets of indices $K_1,\ldots,K_m$ and $L_1,\ldots,L_n$ each of cardinality $r$ such that  the matrix 
\begin{equation}\label{eq:select_non-sing_subtiles}
S^{\kappa\lambda}_{K_{\kappa}L_{\lambda}} \text{ is non-singular for all }(\kappa,\lambda) \in [m] \times [n].
\end{equation}

If $B_A$ is symmetric then $\widetilde{B}$ is symmetric and
we may assume, by Lemma \ref{lem:symnonsing_submat_of_sym_mat}, that $K_\kappa = L_\kappa$ for all $\kappa \in [m]$. But there is more we can infer from Claim~\ref{cl:s_shape}, namely the above non-singular subtiles of each tile are (up to row-column negations and permutations) equal:
\begin{claim}\label{cl:all_tiles_equal}
For all $\kappa \in [m]$ and $\lambda \in [n]$ the sets $K_\kappa$ and $L_\lambda$ have orderings
$$ K_\kappa = \{k_{\kappa,1}, \ldots,k_{\kappa,r}\} \text{ and } L_\lambda = \{\ell_{\lambda,1}, \ldots,\ell_{\lambda,r}\}$$
and there are families $\{\tau^R_{\kappa}:[r] \rightarrow \{-1,1\}\}_{\kappa \in [m]}$ and $\{\tau^C_{\lambda}:[r] \rightarrow \{-1,1\}\}_{\lambda \in [n]}$ of mappings such that:
$$
  S^{11}_{k_{1,a}\ell_{1,b}} = \tau^R_{\kappa}(a)\tau^C_{\lambda}(b) S^{\kappa \lambda}_{k_{\kappa,a},\ell_{\lambda,b}} \text{ for all } (\kappa,\lambda) \in [m]\times[n], \; a,b \in [r].
$$
If $\widetilde{B}$ is symmetric then $S$ is symmetric and 
$K_\kappa=L_\kappa$ and
$\tau^R_{\kappa}=\tau^C_{\kappa}$ for
all $\kappa\in[m]$.
\end{claim}
\begin{clproof}
As $S^{11}_{K_1L_1} = S_{K_1L_1}$ and $\rank{S^{11}} = \rank{S}$, equation \eqref{eq:row_cond_S} implies that every row in $S$ is either a copy or a negated copy of a row in $S_{K_1\absent}$. Fix an arbitrary ordering 
$K_1 = \{k_{1,1}, \ldots,k_{1,r}\}$. As $S^{\kappa 1}\row {K_{\kappa}}$ has rank $r$ for all
$\kappa \in [m]$ there is an ordering $\{k_{\kappa,1},\ldots, k_{\kappa,r} \}$ 
and, for every $a\in[r]$, an $s_a\in\{-1,+1\}$ such that
$S^{11}\row {k_{1,a}} = s_a S^{\kappa 1}\row {k_{\kappa,a}}$. 
Let $\tau^R_{\kappa}(a)=s_a$. Then
$ S^{11}\row {k_{1,a}} = \tau^R_{\kappa}(a) S^{\kappa 1} \row {k_{\kappa,a}} \text{ for all } a \in [r].$
Equation \eqref{eq:row_cond_S} implies that this extends to
$$
S^{1\lambda}\row {k_{1,a}} = \tau^R_{\kappa}(a) S^{\kappa \lambda} \row {k_{\kappa,a}} \text{ for all } a \in [r], \kappa \in [m], \lambda \in [n].
$$

An analogous argument on the columns of $S$ using equation \eqref{eq:col_cond_S}, yields  orderings of the sets 
$L_\lambda$ and mappings $\tau_\lambda$ such that
$$
S^{\kappa 1}\col {\ell_{1,b}} = \tau^C_{\lambda}(b) S^{\kappa \lambda} \col {\ell_{\lambda,b}} \text{ for all } b \in [r], \kappa \in [m], \lambda \in [n].
$$
Combining both finishes the proof of Claim~\ref{cl:all_tiles_equal}.
\end{clproof}

For $\kappa\in[m]$, let $\pi^R_\kappa$ be a permutation of $[m_\kappa]$
which satisfies $\pi^R_\kappa(a)=k_{\kappa,a}$ for $a\in[r]$.
For $\lambda\in[n]$, let $\pi^C_\lambda$ be a permutation of $[n_\lambda]$
which satisfies $\pi^C_\lambda(a)=\ell_{\lambda,a}$ for all $a\in[r]$.

Let $\hat S^{\kappa\lambda}$ be the result of the permutations $\pi^R_{\kappa}$ and $\pi^C_{\lambda}$ when applied to $S^{\kappa\lambda}$ that is $\hat S^{\kappa\lambda} := (S^{\kappa\lambda})_{\pi^R_{\kappa}, \pi^C_{\lambda}}$.
Let $\hat{B}$ be the matrix defined by 
$\hat{B}_{(\kappa,i),(\lambda,j)} = v_\kappa w_\lambda
\hat{S}^{\kappa \lambda}_{i,j}$ and let
$\hat{S}$ be the matrix defined by 
$\hat{S}_{(\kappa,i),(\lambda,j)} = \hat{S}^{\kappa \lambda}_{i,j}$. Let $\hat{A}$ be the
bipartite matrix with underlying block $\hat B$. 
Note that 
$\eval{\hat{A}}\equiv \eval{\widetilde{A}}\equiv \eval{A}$.
The definition of these permutations implies that $\hat B$ is symmetric if $\widetilde{B}$ is symmetric
(which is true if $B_A$ is symmetric).
In this case,
$\eval{\hat{B}}\equiv \eval{\widetilde{B}} \equiv \eval{B_A}$.
Equation \eqref{eq:select_non-sing_subtiles} simplifies to
\begin{equation}\label{eq:non_sing_subtiles}
 \hat S^{\kappa \lambda}_{[r][r]} \text{ is non-singular for all} (\kappa,\lambda) \in [m] \times [n]
\end{equation}
and Claim~\ref{cl:all_tiles_equal} implies furthermore that
\begin{equation}
\label{eq:C15}
 \hat S^{11}_{a,b} = \tau^R_{\kappa}(a)\tau^C_{\lambda}(b) \hat S^{\kappa \lambda}_{a,b} \text{ for all } (\kappa,\lambda) \in [m]\times[n], \; a,b \in [r].
\end{equation}

We consider the twin-relation on $\hat A$ now. As $\hat A$ is bipartite, the equivalence classes of this relation induce collections of equivalence classes separately for the rows and columns of $\hat B$. Furthermore, as 
$\hat{B}_{(\kappa,i),(\lambda,j)} = v_\kappa w_\lambda \hat{S}^{\kappa \lambda}_{i,j}$
and the values $v_i$ are pairwise distinct and positive, two rows corresponding to different $v_i$ values are not twins. This is similarly true for the columns of $\hat B$. Hence, the equivalence classes 
of rows 
can be grouped into collections $\m I_1, \ldots, \m I_m$ and 
the equivalence clases of columns can be grouped into collections
$\m J_1, \ldots, \m J_n$ such that, for every $\kappa \in [m]$, the collection $\m I_\kappa$ contains the equivalence classes of rows in the submatrix 
$$
T^{\kappa\absent} := \left( \begin{array}{c c c}
         v_{\kappa}w_1 \hat S^{\kappa 1} & \ldots & v_{\kappa}w_n \hat S^{\kappa n}
       \end{array}\right)
$$
of $\hat B$. By equation \eqref{eq:row_cond_S} and equation \eqref{eq:non_sing_subtiles} every row in $T^{\kappa\absent}$ is either a copy or a negated copy of a row in $(T^{\kappa\absent})_{[r]\absent}$. Moreover, every two $i\neq j \in [r]$ belong to different equivalence classes by equation \eqref{eq:non_sing_subtiles}.

We may therefore assume, without loss of generality, that the collection $\m I_\kappa$ consists of classes $P^{\kappa\absent}_1, \ldots,P^{\kappa\absent}_r$ and $N^{\kappa\absent}_{1},\ldots, N^{\kappa\absent}_{r}$ such that $i \in P^{\kappa\absent}_i$ for all $i \in [r]$.
Furthermore, the sets $N^{\kappa\absent}_i$ account for the possible negated copies of rows in $(T^{\kappa\absent})_{[r]\absent}$ and therefore some of these sets may be empty. But for all $i \in [r]$ if $N^{\kappa\absent}_i$ is non-empty then all $a \in N^{\kappa\absent}_i$ are indices of negated copies of rows from $P^{\kappa\absent}_i$.

Similarly, the collection $J_\lambda$ of equivalence classes of columns
corresponds to the submatrix
 $$
T^{\absent\lambda} := \left( \begin{array}{c}
         v_{1}w_{\lambda} \hat S^{1 \lambda} \\ \vdots \\ v_{m}w_\lambda \hat S^{m \lambda}
       \end{array}\right)
$$
of $\hat B$. By equation \eqref{eq:col_cond_S} every column in $T^{\absent \lambda}$ is either a copy or a negated copy of a column in $(T^{\absent \lambda})_{\absent [r]}$. Moreover, by equation \eqref{eq:non_sing_subtiles} every two $i\neq j \in [r]$ belong to different equivalence classes of the twin relation.

We may assume that the collection $\m J_\lambda$ consists of classes $P^{\absent\lambda}_1, \ldots,P^{\absent\lambda}_r$ and $N^{\absent\lambda}_1,\ldots, N^{\absent\lambda}_r$ such that $i \in P^{\absent\lambda}_i$ for all $i \in [r]$. The sets $N^{\absent\lambda}_i$ account for the possible negated copies of columns in $(T^{\absent\lambda})_{\absent [r]}$ and therefore some of these sets may be empty. But for all $i \in [r]$ if $N^{\absent\lambda}_i$ is non-empty then all $a \in N^{\absent\lambda}_i$ are indices of negated copies of columns from $P^{\absent\lambda}_i$.

Note  that if $\hat B$ is symmetric the above definitions directly imply that   $m = n$ and, for all $\mu \in [m]$,  $\m I_\mu = \m J_\mu$. Also,
we can take $P^{\mu \absent}_i = P^{\absent \mu}_i$ and $N^{\mu \absent}_i = N^{\absent \mu}_i$ for all $i \in [r]$.

Application of the Extended Twin Reduction Lemma \ref{lem:ext_twin_red} according to these equivalency classes therefore yields an evaluation problem 
$\eval{\hat{C},D,\hat{O}} \Tequiv \eval{\hat A} (\Tequiv \eval{A})$ 
such that the block $\hat B'$ underlying $\hat{C}$ satisfies
$$
\hat B' =\left(\begin{array}{c c c}
               v_1w_1 \hat S^{1 1}_{[r][r]} & \ldots & v_1w_n \hat S^{1 n}_{[r][r]}\\
               \vdots  & \ddots & \vdots \\
               v_mw_1 \hat S^{m 1}_{[r][r]} & \ldots & v_mw_n \hat S^{m n}_{[r][r]} 
            \end{array}\right).
$$
That is, $\hat B'$ is an $mr \times nr$ matrix and $D$ and $\hat{O}$ are diagonal matrices of vertex weights of order $mr + nr$. Grouping these vertex weights according to the rows and columns of $\hat B'$ to which they correspond, we obtain
$$
D = \left(\begin{array}{c c}D^R & 0 \\ 0 & D^C\end{array} \right)
\text{ and }
\hat{O} = \left(\begin{array}{c c}\hat{O}^R & 0 \\ 0 & \hat{O}^C\end{array} \right).
$$
for $mr \times mr$ diagonal matrices $D^R,\hat{O}^R$ and $nr \times nr$ diagonal matrices $D^C,\hat{O}^C$. Their structure corresponding to the tiles of $\hat B'$ in turn is
$$
D^R = \left(\begin{array}{c c c}D^{R,1} & \ldots & 0 \\ \vdots & \ddots & \vdots \\
                                0 & \ldots & D^{R,m}\end{array} \right)
\text{ and }
D^C = \left(\begin{array}{c c c}D^{C,1} & \ldots & 0 \\ \vdots & \ddots & \vdots \\
                                0 & \ldots & D^{C,n }\end{array} \right).
$$
which holds analogously for $\hat{O}$ such that the $D^{R,\mu},\hat{O}^{R,\mu},D^{C,\nu},\hat{O}^{C,\nu}$ for all $\mu\in[m], \nu \in [n]$ are $r\times r$ diagonal matrices. The definition of these matrices according to the Extended Twin Reduction Lemma \ref{lem:ext_twin_red} is then, for all $\mu \in [m],\; \nu \in [n],\;i,j\in [r]$, given by
\begin{equation}\label{eq:D_O_definition}
\begin{array}{c c c}
 D^{R,\mu}_{i,i} = \vert P^{\mu \absent}_i \vert + \vert N^{\mu \absent}_i \vert 
& \text{and}& D^{C,\nu}_{j,j} = \vert P^{\absent \nu}_j \vert + \vert N^{\absent \nu}_j \vert \\
 \hat{O}^{R,\mu}_{i,i} = \vert P^{\mu \absent}_i \vert - \vert N^{\mu \absent}_i \vert
& \text{and}& \hat{O}^{C,\nu}_{j,j} = \vert P^{\absent \nu}_j \vert - \vert N^{\absent \nu}_j \vert \\
\end{array} 
\end{equation}

If $B_A$ is symmetric then $\hat{B}$ is symmetric and $D^R=D^C$. Also,
$\hat{B}'$ is also symmetric and
$\eval{\hat{B}',D^R,\hat{O}^R} \equiv \eval{\hat{B}}$.

Clearly, the matrix $D$ is a diagonal matrix of vertex weights whose diagonal is positive as the sets $P^{\kappa\absent}_i$ and $P^{\absent\lambda}_i$ are non-empty by definition for all $\kappa \in [m], \lambda \in [n]$ and $i \in [r]$.

By Equation~(\ref{eq:C15}),
for all $(\kappa,\lambda) \in [m]\times[n]$, the matrix $\hat S^{\kappa \lambda}_{[r][r]}$ is -- up to negations of rows and columns -- just a copy of the matrix $\hat S^{11}_{[r][r]}$.
However, the diagonal entries of $\hat{O}$ given by equation \eqref{eq:D_O_definition} may be negative in some cases.
To satisfy condition \cond{C2} we therefore define mappings $\rho :[r] \rightarrow \{-1,1\}$ and $\gamma: [r] \rightarrow \{-1,1\}$ by
$$
  \rho(i) = \left\lbrace \begin{array}{r l}
                                    -1&, \text{if } \hat{O}^{R,1}_{i,i} < 0 \\
                                     1&, \text{ otherwise}
                                   \end{array}\right.
\qquad \text{ and } \qquad
  \gamma(j) = \left\lbrace \begin{array}{r l}
                                    -1&, \text{if } \hat{O}^{C,1}_{j,j} < 0 \\
                                     1&, \text{ otherwise}
                                   \end{array}\right.
$$
We will use these mappings below to ``transfer'' the signs of diagonal entries of $\hat{O}^{R,1}$ and $\hat{O}^{C,1}$ to $\hat{B}'$.
Note that $\rho=\gamma$ if $B_A$ is symmetric since $\hat{O}^R=\hat{O}^C$ in this case.
Define matrices $\check S^{\kappa \lambda}_{[r][r]}$ by applying row and column negations 
according to these mappings, that is
 \begin{equation}\label{eq:row_col_neg_S}
\check S^{\kappa \lambda}_{a,b} =
\rho(a)\gamma(b)
\tau^R_{\kappa}(a)\tau^C_{\lambda}(b)  
\hat S^{\kappa \lambda}_{a,b} 
\text{ for all } (\kappa,\lambda) \in [m]\times[n], \; a,b \in [r].
\end{equation}
  
By equation \eqref{eq:C15}, we have the following for all
$(\kappa,\lambda) \in [m]\times[n]$ and  $a,b \in [r]$:
$$
 \rho(a)\gamma(b)\hat S^{11}_{a,b} = \rho(a)\gamma(b)\tau^R_{\kappa}(a)\tau^C_{\lambda}(b) 
\hat S^{\kappa \lambda}_{a,b} 
= \check S^{\kappa \lambda}_{a,b}.
$$
Thus 
$$\tau^R_{1}(a)\tau^C_{1}(b)
\check S^{11}_{a,b}
=
 \rho(a)\gamma(b)
\tau^R_{1}(a)\tau^C_{1}(b) \tau^R_{1}(a)\tau^C_{1}(b) 
\hat S^{11}_{a,b}  
= \check S^{\kappa \lambda}_{a,b}.
$$
But, by their definition in Claim~\ref{cl:all_tiles_equal}, the mappings $\tau^R_1$ and $\tau^C_1$ 
satisfy $\tau^R_1(i) = \tau^C(i) = 1$ for all $i \in [r]$. 
So the above equation gives
$
\check S^{11}_{a,b}
= \check S^{\kappa \lambda}_{a,b}
$ for all $(\kappa,\lambda) \in [m]\times [n]$ and $a,b \in [r]$ so $\check S^{1 1}_{[r][r]} = \check S^{\kappa \lambda}_{[r][r]}$ for all $(\kappa,\lambda) \in [m]\times [n]$. Define $H := \check S^{1 1}_{[r][r]}$.
Let 
$B$ be the matrix
defined by 
$B_{(\kappa,i),(\lambda,j)}
= v_\kappa w_\lambda H_{i,j}$  so
\begin{align*}
B_{(\kappa,i),(\lambda,j)}
&= v_\kappa w_\lambda 
{\check S^{1 1}}_{i,j}
\\
&= v_\kappa w_\lambda 
{\check S^{\kappa \lambda}}_{i,j}
\\
&= v_\kappa w_\lambda 
\rho(i)\gamma(j)
\tau^R_{\kappa}(i)\tau^C_{\lambda}(j)  
\hat S^{\kappa \lambda}_{i,j} 
\\
&= 
\rho(i)\gamma(j)
\tau^R_{\kappa}(i)\tau^C_{\lambda}(j)  
\hat{B}_{(\kappa,i),(\lambda,j)}.
\end{align*}

Let
$C$
be the symmetric bipartite matrix with underlying block $B$.
For $\kappa\in[m]$, $\lambda\in[n]$ and $i,j\in[r]$,
let 
$$
 O^{R,\kappa}_{i,i} =  \rho(i)
\tau^R_{\kappa}(i)
\hat{O}^{R,\kappa}_{i,i} \quad \text{ and } \quad
 O^{C,\lambda}_{j,j} = \gamma(j)
\tau^C_{\lambda}(j) 
\hat{O}^{C,\lambda}_{j,j}.
$$
Let $O^R$ be the diagonal matrix with tiles $O^{R,\kappa}$ for $\kappa\in[m]$ and
$O^C$ be the diagonal matrix with tiles $O^{C,\lambda}$ for $\lambda\in[n]$.
Let $O$ be the matrix
$$
O = 
\left(\begin{array}{c c} O^R & 0 \\ 0 & 
 O^C\end{array} \right)
$$
 
Since (as noted above) $\tau^R_1(i) = \tau^C(i) = 1$ for all $i \in [r]$, the matrices $ O^{R,1}$ and $ O^{C,1}$ are non-negative.

The Row-Column Negation Lemma \ref{lem:bipolar_rowcol_neg} implies 
$$
\eval{\hat{C},D,\hat{O}} \Tequiv \eval{C, D, O}.
$$
The block $B$ satisfies \cond{C1} 
and the matrices $D$ and $O$ satisfy \cond{C2}. 
The definitiond of $D$ and $\hat{O}$ in equation \eqref{eq:D_O_definition} and the definition of $O$ implies that
$D+O$ and
$D-O$ are non-negative as required.
If $B_A$ is symmetric then $\widetilde{B}$ and $S$ are symmetric so 
$\tau^R_{\kappa}=\tau^C_{\kappa}$ and $\pi^R_{\kappa} =\pi^C_{\kappa}$ so $\hat{S}$ and $\hat{B}$ are symmetric.
Since $\rho=\gamma$ , $B$ is also symmetric.
So the
Row-Column Negation Lemma \ref{lem:bipolar_rowcol_neg} implies 
$$
\eval{\hat{B}',D^R,\hat{O}^R} \Tequiv \eval{B, D^R, O^R}.
$$

Furthermore it is easy to see that all operations performed to form $C, D,  O$ from the matrix $A$ are polynomial time computable. This finishes the proof.
\end{proof}

The remainder of this section relies on a gadget which consists of arrangements of paths of length $2$. These paths affect the matrices $C,D,O$ 
in a similar way to 
$2$-stretching. It is therefore convenient to have a look at the effect this operation has. 
Clearly $2$-stretching yields $\eval{CDC,D,O} \Tle \eval{C,D,O}$.
If $B$ is symmetric it also yields
$\eval{B D^R B, D^R, O^R} \Tle \eval{B,D^R,O^R}$.

Assume that $C$ and $D,O$ satisfy conditions \cond{C1} and \cond{C2}. Recall that $B = vw^T \otimes H$ holds for the block $B$ underlying $C$. Furthermore the matrix $CDC$ contains the submatrices $BD^CB^T$ and $B^TD^RB$  and 
$$
BD^CB^T = \left(\begin{array}{c c c}
               v_1v_1 H(\sum_{\nu=1}^n w^2_{\nu} D^{C,\nu}) H^T & \ldots & v_1v_m H(\sum_{\nu=1}^n w^2_\nu D^{C,\nu})H^T\\
               \vdots  & \ddots & \vdots \\
               v_mv_1 H(\sum_{\nu=1}^n w^2_{\nu} D^{C,\nu}) H^T & \ldots & v_mv_m H(\sum_{\nu=1}^n w^2_\nu D^{C,\nu})H^T\\
            \end{array}\right)
$$

with analogous analysis of $B^TD^RB$ we have
\begin{equation}\label{eq:CDC_shape}
          BD^CB^T = vv^T \otimes \left(H(\sum_{\nu=1}^n w^2_\nu D^{C,\nu})H^T\right) \; \text{ and } \; B^TD^RB = ww^T \otimes \left(H^T (\sum_{\mu=1}^m v^2_\mu D^{R,\mu})H\right)
\end{equation}

\begin{figure}
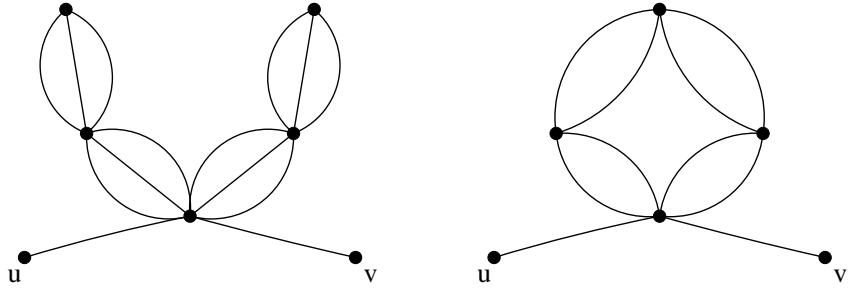

\centering
\begin{minipage}{.48\textwidth}
\centering
\resizebox{5cm}{!}{
\includegraphics{gadget_template_t1.pstex}%
}
\end{minipage}
\begin{minipage}{.40\textwidth }
\resizebox{5cm}{!}{
\includegraphics{gadget_template_t2.pstex}%
}
\end{minipage}
\caption{The gadget templates $T(1,3,2)$ and $T(2,2,1)$}
\label{fig:gadget_template}
\end{figure}

\begin{minipage}{.48\textwidth }
\centering

\end{minipage}

We define a \emph{reduction template} $T(t,p,q)$ which will be used in the proofs of Lemmas \ref{lem:two_properties} and \ref{lem:final_red_to_bipolar}.
Let $P(t,p)$ be a graph constructed as follows.  Start with an edge with a distinguished
endpoint~$a$.  Then perform in succession a $t$-thickening, then a two stretch, 
and finally a $p$-thickening.
(Informally,  there is a vertex $b$  connected to $a$ by $t$~many length $2$ paths such that all edges in 
those paths have multiplicity~$p$.)

The reduction $T(t,p,q)$ works as follows. In a given graph $G = (V,E)$, we $2$-stretch each edge $e \in E$ and call the middle vertex $v_e$. We attach $q$ disjoint copies of $P(t,p)$ by identifying their terminal vertices with $v_e$. Figure \ref{fig:gadget_template} illustrates the construction.

Recall that $M\circ N$ denotes the Hadamard product of matrices $M$ and $N$.

\begin{lemma}\label{lem:gadget_analysis}
Suppose $C$ and $D,O$ satisfy \cond{C1} and \cond{C2}.  
At least one of the following outcomes occurs.
\begin{description}
\item[Outcome 1]
$\eval{C,D,O}$ is $\#\PP$-hard. If $B$ is symmetric, then 
$\eval{B,D^R,O^R}$ is $\#\PP$-hard.
\item[Outcome 2]
For $t,p,q \in \Nat $ and $p' = 2p+1$ and $q' = 2q$ there are $r \times r$ matrices $\Theta = \Theta(t,p')$ and $\Xi = \Xi(t,p')$ defined by
\begin{eqnarray*}
\Theta &=&  (\gamma^R_{p'})^t \cdot \sum_{\mu=1}^m  v_\mu^{tp'} \cdot \left\lbrace \begin{array}{l l}
                                    D^{R,\mu} &, \text{ if $t$ is even} \\
                                    O^{R,\mu} &, \text{ if $t$ is odd}
                                   \end{array}\right. \\
\Xi &=& (\gamma^C_{p'})^t \cdot \sum_{\nu=1}^n  w_\nu^{tp'} \cdot \left\lbrace \begin{array}{l l}
                                    D^{C,\nu} &, \text{ if $t$ is even} \\
                                    O^{C,\nu} &, \text{ if $t$ is odd}
                                   \end{array}\right.
\end{eqnarray*}
for positive
 constants $\gamma^R_{p'}$ and $\gamma^C_{p'}$ depending on $p'$.

The reduction $T(t,p',q')$ yields $\eval{C\Delta C,D,O}\Tle \eval{C,D,O}$ 
for a diagonal matrix 
$$
\Delta = \Delta(t,p',q') = \left(\begin{array}{c c}\Delta^R & 0 \\ 0 & \Delta^C \end{array}\right)
\text{ and a matrix }
C\Delta C = \left(\begin{array}{c c}B\Delta^CB^T & 0 \\ 0 & B^T\Delta^RB \end{array}\right).
$$ 
$\Delta^R$ is a diagonal $rm \times rm$ matrix of $r \times r$ tiles 
$
\Delta^{R,\mu} = v_\mu^{tp'q'} D^{R,\mu} \circ \Theta^{(q')} \text { for all } \mu \in [m].
$
$\Delta^C$ is a diagonal $rn \times rn$ matrix of $r \times r$ tiles 
$
\Delta^{C,\nu} = w_\nu^{tp'q'} D^{C,\nu} \circ \Xi^{(q')} \text{ for all }\nu \in [n].
$
If $B$ is symmetric then the same reduction yields
$$\eval{B \Delta^R B, D^R, O^R} \leq \eval{B,D^R,O^R}.$$
\end{description}
\end{lemma}
\begin{proof}
Let $p',q'$ be as above.
\begin{claim}
Either 
Outcome 1 occurs  or there are constants $\gamma^R_{p'}$ and $\gamma^C_{p'}$ depending on $p'$ such that
\begin{equation}\label{eq:BDB_are_diagonals}
B^{(p')}D^C(B^{(p')})^T = (vv^T)^{(p')} \otimes \gamma^R_{p'} I_r 
\; \text{ and } \; (B^{(p')})^TD^RB^{(p')} = (ww^T)^{(p')} \otimes \gamma^C_{p'}I_r
\end{equation}
\end{claim}
\begin{clproof}
  We have $\eval{C^{(p')}DC^{(p')},D,O} \Tle \eval{C,D,O}$ by 
$p'$-thickening followed by $2$-stretching.
If $B$ is symmetric this also yields
$\eval{B^{(p')}D^R B^{(p')},D^R,O^R} \Tle \eval{B,D^R,O^R}$.
The matrix $C^{(p')}DC^{(p')}$ contains
  submatrices $X := B^{(p')}D^C(B^{(p')})^T$ and $Y :=
  (B^{(p')})^TD^RB^{(p')}$. We show the first part of equation
  \eqref{eq:BDB_are_diagonals} by an argument based on the matrix $X$. The
  second part then follows analogously using $Y$.
(Recall from \cond{C1} that $D^R=D^C$ when $B$ is symmetric, in which case $X=Y$.)

Define $\Pi = \sum_{\nu=1}^n w^2_\nu D^{C,\nu}$. By equation \eqref{eq:CDC_shape} we have $X = (vv^T)^{p'} \otimes \left(H \Pi H^T\right)$. 
Therefore, if $\abs{H \Pi H^T}$ contains a block of row rank at least two then $X$ does.

As $H$ is a $\{-1,1\}$-matrix we have $(H\Pi H^T)_{i,i} = \tr{\Pi}$ for all $i\in [r]$ and the trace of $\Pi$ is positive.
Furthermore $\vert (H\Pi H^T)_{i,j}\vert < \tr{\Pi}$ for all $j \neq i$ by the non-singularity of $H$.
Hence, we obtain a block of rank at least $2$ in $\abs{H \Pi H^T}$, if there is a non-zero entry $(H\Pi H^T)_{i,j}$ for some $i\neq j \in [r]$. The proof follows with $\gamma^R_{p'} = \tr{\Pi}$.
\end{clproof}

For convenience, let $T= T(t)$ denote the matrix $D$, if $t$ is even, and $O$ otherwise.

Recall the reduction template, let $(\mu,i),(\kappa,k) \in [m+n]\times [r]$ denote the spins of $v_e$ and $b$.

The diagonal $(\mu,i)$ entries of $\Delta$ correspond to the partition function of the reduction template with vertex $v_e$ fixed to $(\mu,i)$. Therefore,  for $\mu \in [m]$ 
\begin{eqnarray*}
\Delta^{R,\mu}_{i,i} &=& D^{R,\mu}_{i,i} \left(\sum_{\kappa=1}^m\sum_{k = 1}^{r} T^{R,\kappa}_{k,k} 
                 (C^{(p')}DC^{(p')})_{(\mu,i),(\kappa,k)}^t  \right)^{q'} \\
                    &=& D^{R,\mu}_{i,i} \left(\sum_{\kappa=1}^m\sum_{k = 1}^{r} T^{R,\kappa}_{k,k} 
                 (B^{(p')}D^C(B^{(p')})^T)_{(\mu,i),(\kappa,k)}^t  \right)^{q'} \\
                    &=& v_{\mu}^{tp'q'}D^{R,\mu}_{i,i}  \left((\gamma^R_{p'})^t \cdot \sum_{\kappa=1}^m v_{\kappa}^{tp'} T^{R,\kappa}_{i,i} \right)^{q'} \\
\end{eqnarray*}
where the last equation follows from Claim~1. Similarly, for $\nu \in [n]$
\begin{eqnarray*}
\Delta^{C,\nu}_{ii} &=& D^{C,\nu}_{i,i} \left(\sum_{\kappa=1}^n\sum_{k = 1}^{r} T^{C,\kappa}_{k,k} 
                 (C^{(p')}DC^{(p')})_{(\nu,i),(\kappa,k)}^t  \right)^{q'}\\
                    &=& D^{C,\nu}_{i,i} \left(\sum_{\kappa=1}^n\sum_{k = 1}^{r} T^{C,\kappa}_{k,k} 
                 ((B^{(p')})^TD^RB^{(p')})_{(\nu,i),(\kappa,k)}^t  \right)^{q'}\\
                    &=& w_{\nu}^{tp'} D^{C,\nu}_{i,i} \left((\gamma^C_{p'})^t \sum_{\kappa=1}^n w_{\kappa}^{tp'} T^{C,\kappa}_{i,i} 
                 \right)^{q'}.
\end{eqnarray*}
With $\Theta$ and $\Xi$ defined as in the statement of the Lemma the proof follows.
\end{proof}

\begin{lemma}\label{lem:two_properties}
Let $C$ and $D,O$ satisfy \cond{C1} and \cond{C2}.
At least one of the following outcomes occurs.
\begin{description}
\item[Outcome 1]
$\eval{C,D,O}$ is $\#\PP$-hard. If $B$ is symmetric, then 
$\eval{B,D^R,O^R}$ is $\#\PP$-hard.
\item[Outcome 2]
Conditions \cond{C3} and \cond{C4} are satisfied.
\end{description}
\end{lemma}

\begin{proof}
The $\#\PP$-hardness part will be shown using a gadget construction $T(2,p',q')$ with $p' = 2p+1$ and $q' = 2q$ for $p,q \in \Nat$.
By Lemma \ref{lem:gadget_analysis} this yields a reduction $\eval{C\Delta C,D,O} \Tle \eval{C,D,O}$ such that $C\Delta C$ contains submatrices $B\Delta^CB^T$ and $B^T\Delta^R B$.  
If $B$ is symmetric then $\eval{B\Delta^R B,D^R,O^R} \Tle \eval{B,D^R,O^R}$
Focusing on $B^T\Delta^RB$ we will prove \cond{C3} and the part of \cond{C4} which claims that $D^{R,\mu} = \alpha^{R}_\mu I_r$. The proof for $D^{C,\nu} = \alpha^{C}_\nu I_r$ then follows by analogous arguments based on $B\Delta^CB^T$.

Recall that by
the proof of equation \eqref{eq:CDC_shape} we have
$B^T\Delta^R B = (ww^T) \otimes (H^T\Delta'H)$ for an $r \times r$ diagonal matrix $\Delta'$ defined by
\begin{eqnarray*}
\Delta' = \sum_{\mu=1}^m v^2_\mu \Delta^{R,\mu} &=& \left(\sum_{\mu=1}^m v^{2p'q' + 2}_\mu D^{R,\mu}\right)\circ {\Theta^{[p]}}^{(q')}\\
\end{eqnarray*}
\begin{equation}\label{eqn:theta}
\text{ with } {\Theta^{[p]}} = \Theta(2,p',q') = (\gamma^R_{p'})^2\cdot \sum_{\mu=1}^m  v_\mu^{2p'} \cdot D^{R,\mu}.
\end{equation}

If $\abs{H^T\Delta' H}$ contains a block of rank at least~$2$ then
$\abs{B \Delta^R B}$ does.
So, if $\abs{H^T\Delta' H}$ contains a block of rank at least~$2$, then
Outcome 1 occurs by Lemma \ref{lem:bipolar_block2_hard}.

By the definition of $H^T\Delta' H$, we have $(H^T\Delta' H)_{i,i} = \tr{\Delta'}$ for all $i \in [r]$ and this trace is positive by the definition of $\Delta'$.
Therefore, every principal $2 \times 2$ submatrix of $\abs{H^T\Delta' H}$ has the form
$$
\left(\begin{array}{c c}
       \tr{\Delta'}& \vert (H^T\Delta' H)_{i,j} \vert \\
       \vert (H^T\Delta' H)_{j,i} \vert &  \tr{\Delta'}
      \end{array}\right)
$$
As $H$ is non-singular $\vert (H^T\Delta' H)_{i,j} \vert < \tr{\Delta'}$ for all $i \neq j \in [r]$ and therefore, every such submatrix has non-zero determinant. Furthermore, such a submatrix is part of a block if $(H^T\Delta' H)_{i,j} \neq 0$. Therefore we have  Outcome 1 if we can show that $(H^T\Delta' H)_{i,j} \neq 0$ for some $i \neq j \in [r]$ and some $p,q \in \Nat$.

Assume therefore that $(H^T\Delta' H)_{i,j} = 0$ for all $i \neq j \in [r]$ and all $p,q \in \Nat$. The remainder of the proof is to show that in this case conditions \cond{C3} and \cond{C4} are satisfied.

Let $\vartheta_{p,q,i} = \sum_{\mu=1}^m v^{2p'q' + 2}_\mu D^{R,\mu}_{i,i}$ for all $i \in [r]$. Note that
$\Delta'_{i,i} = \vartheta_{p,q,i} {\Theta^{[p]}}_{i,i}^{q'}$. 
We define an equivalence relation $\sim$ on $[r]$ by letting $i\sim j$ if and only if
$D_{i,i}^{R,\mu} = D_{j,j}^{R,\mu}$ for all $\mu\in[m]$. Let $\mathbf{I}$ be the set of equivalence classes.%
We will use the notation $D^{R,\mu}_{I}$ to denote
the value $D_{i,i}^{R,\mu}$ for $i\in I$.

Recall that the values $v_\mu$ in the definition of ${\vartheta}_{p,q,i}$ are pairwise distinct and non-negative. 
Lemma \ref{lem:bij_coeff_power} 
implies the following, for  all $i,j \in [r]$:
\begin{equation}\label{eq:sig_equality}
(\text{For all $p$ and $q$,}
{\vartheta}_{p,q,i} = {\vartheta}_{p,q,j} )
\text{ iff }  
i\sim j.
\end{equation}
 
\begin{equation}\label{eq:sigma_theta_eq}
(\text{For all $p$,}
{\Theta^{[p]}}_{i,i} = {\Theta^{[p]}}_{j,j} )
\text{ iff }  
i\sim j.
\end{equation}

We use the notation $\vartheta_{p,q,I}$ to denote the value 
$\vartheta_{p,q,i}$ for $i\in I$.
Similarly, we use the notation $\Theta_{p,I}$ to denote the
value $\Theta^{[p]}_{i,i}$ for $i\in I$.

For $i,j \in [r]$ define sets 
$\mathcal{P}_{ij} = \{k \in [r] \mid H_{k,i}H_{k,j} > 0\} \text{ and } \mathcal{N}_{ij} = \{k \in [r] \mid H_{k,i}H_{k,j} < 0\}.$

Then we have
\begin{eqnarray*}
(H^T\Delta' H)_{i,j} = \sum_{k=1}^r H_{k,i}H_{k,j} \Delta'_{k,k} &=& \sum_{k=1}^r H_{k,i}H_{k,j} {\vartheta}_{p,q,k} \left({\Theta^{[p]}}_{k,k}\right)^{q'} \\
                 &=& \left(\sum_{k \in \mathcal{P}_{ij}} {\vartheta}_{p,q,k} ({\Theta^{[p]}}_{k,k})^{q'}  - 
                                \sum_{l \in \mathcal{N}_{ij}} {\vartheta}_{p,q,l} ({\Theta^{[p]}}_{l,l})^{q'} \right).
\end{eqnarray*}

Then
\begin{eqnarray*}
(H^T\Delta' H)_{i,j} &=& \sum_{I \in \mfam{I}} \left(\sum_{k \in I \cap \mathcal{P}_{ij}} {\vartheta}_{p,q,k} ({\Theta^{[p]}}_{k,k})^{q'}  - \sum_{l \in I \cap \mathcal{N}_{ij}} {\vartheta}_{p,q,l} ({\Theta^{[p]}}_{l,l})^{q'} \right)\\
                 &=& \sum_{I \in \mfam{I}} {\Theta}_{p,I}^{q'}\left(\sum_{k \in I \cap \mathcal{P}_{ij}} {\vartheta}_{p,q,I}   - \sum_{l \in I \cap \mathcal{N}_{ij}} {\vartheta}_{p,q,I}  \right)\\
                 &=&\sum_{I \in \mfam{I}} {\vartheta}_{p,q,I}{\Theta}_{p,I}^{q'} \left(\vert I \cap \mathcal{P}_{i,j}\vert - \vert I \cap \mathcal{N}_{i,j}\vert \right)\\
\end{eqnarray*}

\begin{claim}\label{cl:H_is_blockwise_ortho}
Suppose that 
$(H^T\Delta' H)_{i,j} = 0$ for all $i \neq j \in [r]$ and all $p,q \in \Nat$.
Then there is a $J\in\mathbf{I}$ such that
$\vert J \cap \mathcal{P}_{ij}\vert = \vert J \cap \mathcal{N}_{ij}\vert$ for all
$i \neq j \in [r]$.
\end{claim}

\begin{clproof} 

Choose $p$ sufficiently large that there is a unique $J\in \mathbf{I}$
maximising
$\sum_{\mu=1}^m v^{2p'}_\mu D^{R,\mu}_{J}$
For this $p$, and for any $q\in \Nat$ and $I\in \mathbf{I}$,
we have $0<\vartheta_{p,q,I}< \vartheta_{p,q,J}$ and
$0<\Theta_{p,I} < \Theta_{p,J}$.

Now consider $i\neq j\in [r]$.
For all $I\in\mathbf{I}$, let
$c_I = \vert I \cap \mathcal{P}_{ij}\vert - \vert I \cap \mathcal{N}_{ij}\vert$.

Since   $(H^T\Delta' H)_{i,j} = 0$, for all $q\in \Nat$,
\begin{eqnarray}\nonumber
 0  & = & \sum_{I \in \mfam{I}} c_I{\vartheta}_{p,q,I}{\Theta}_{p,I}^{2q} \\
\nonumber
&=& c_J{\vartheta}_{p,q,J}{\Theta}_{p,J}^{2q} + \sum_{I \in \mfam{I}\setminus\{J\}} c_I{\vartheta}_{p,q,I}{\Theta}_{p,I}^{2q} \\
 &=& c_J + \sum_{I \in \mfam{I}\setminus\{J\}} c_I\dfrac{{\vartheta}_{p,q,I}}{{\vartheta}_{p,q,J}}
\left(\dfrac{{\Theta}_{p,I}}{{\Theta}_{p,J}}\right)^{2q}.
\label{eq:april22}
\end{eqnarray}
As $q$ tends to infinity, the sum tends to~$0$ so $c_J=0$.
\end{clproof}

Assume now that $(H^T\Delta' H)_{i,j} = 0$ for all $i \neq j \in [r]$ and $p,q\in\Nat$.
Fix $J\in\mathbf{I}$ such that
$\vert J \cap \mathcal{P}_{ij}\vert = \vert J \cap \mathcal{N}_{ij}\vert$ for all 
$i \neq j \in [r]$.
Recall that $H\row J$ denotes the the submatrix of $H$ consisting of the rows of $H$ with indices in $J$.
For each pair $i \neq j \in [r]$,
the fact that 
$\vert J \cap \mathcal{P}_{ij}\vert = \vert J \cap \mathcal{N}_{ij}\vert$ 
implies $\scalp{(H\row J)\col i,(H\row J)\col j} = 0$.
Hence, the columns in $H\row J$ are pairwise orthogonal.
Since the rank of $H$ is $r$, this implies that $|J|=r$.
Now since the rows of $H^T$ are pairwise orthogonal, we have
$H^T H = r I_r$
so the inverse of $H^T$ is $r^{-1} H$.
As right inverses of matrices are also left inverses, 
we have $r^{-1} H H^T = I_r$ and therefore $H$ is a Hadamard matrix and we have proved condition \cond{C3}.

Finally, $J=[r]$ implies that   
$D^{R,\mu}_{i,i} = D^{R,\mu}_{j,j}$ 
for all $i,j\in[r]$.
Equivalently, $D^{R,\mu} = \alpha^{R}_\mu I_r$ for some appropriate $\alpha^{R}_\mu$. This proves \cond{C4}.
\end{proof}

We call a diagonal matrix $D$ \emph{pre-uniform} if there is a non-negative $d$ such that all diagonal entries $D_{i,i}$ of $D$ satisfy $D_{i,i} \in \mset{0, d}$. An important technical tool in the last step of our proof of conditions \cond{C1}-\cond{C5} will be the following Lemma.
\begin{lemma}[Pre-Uniform Diagonal Lemma]
\label{lem:pre_unif_diag}
Let $H$ be a non-singular $r \times r$ $\{-1,1\}$-matrix and $D$ be an $r \times r$ diagonal matrix with non-negative entries in $\Real$.
If $D$ is not pre-uniform, then there is a $p \in \Nat$
such that $\abs{HD^{(p)}H^T}$ contains a block of row rank at least $2$.
\end{lemma}
\begin{proof}
Note that, if the diagonal of $D$ is constantly zero then $D$ is pre-uniform. Assume therefore that there is some positive diagonal entry in $D$. Define $B := HD^{(p)}H^T$, $K := \mset{k\in [r] \,\vert\;D_{k,k} > 0}$ and $s := \vert K \vert$.
Hence, for $i,j \in [r]$, 
\begin{equation}\label{eq:h_IK}
\begin{array}{c c c c c c c}
 B_{ij}  &=&  \sum_{k=1}^{r} H_{i,k}H_{j,k}(D_{k,k})^{p} &=&  \sum_{k \in K} H_{i,k}H_{j,k}(D_{k,k})^{p} &=& (H\col K D^{(p)}(H\col K)^T)_{i,j} 
\end{array} 
\end{equation}

That is, for every $I \subseteq [r]$, we have $B_{I,I} = H_{I,K}D^{(p)}_{K,K}(H_{I,K})^T$. 
Fix a set $I \subseteq [r]$ such that $\vert I \vert = s$ and the matrix $H_{I,K}$ has rank $s$.
Since $H_{I,K}$ is non-singular, every $2 \times 2$ principal submatrix of $B_{I,I}$ has non-zero determinant. To see this, note that, by equation \eqref{eq:h_IK} we have $B_{i,i} = \tr {D^{(p)}_{K,K}}$ for all $i \in I$ and this trace is positive. Then every such principal $2 \times 2$ submatrix has determinant
$$
\left\vert\begin{array}{c c}
       \tr {D^{(p)}_{K,K}}& \vert (H_{I,K}D^{(p)}_{KK}(H_{I,K})^T)_{i,j} \vert \\
       \vert (H_{I,K}D^{(p)}_{K,K}(H_{I,K})^T)_{j,i} \vert &  \tr {D^{(p)}_{K,K}}
      \end{array}\right\vert
$$
and by the non-singularity of $H_{I,K}$ we have $\vert (H_{I,K}D^{(p)}_{K,K}(H_{I,K})^T)_{i,j} \vert < \tr {D^{(p)}_{K,K}}$ 
(compare equation \eqref{eq:h_IK}). Hence the above determinant is non-zero.

Assume that, for all $p \in \Nat$, there are no non-trivial blocks in $B_{I,I}$, i.e. $B_{i,j} = 0$ for all $i \neq j \in I$. We will show that this implies that $D$ is pre-uniform.

For $i,j \in I$ define the sets 
$\mathcal{P}_{i,j} := \mset{k\in K \, \vert \, H_{i,k}H_{j,k} = 1 }$ and 
$\mathcal{N}_{i,j} := \mset{k\in K \, \vert \, H_{i,k}H_{j,k} = -1 }$. That is, $\mathcal{P}_{i,j}$ and $\mathcal{N}_{i,j}$ form a partition of $K$.
Therefore, for $i,j \in I$ we have
$$B_{i,j} = \sum_{k=1}^n H_{i,k}H_{j,k}D_{k,k}^{p} = \sum_{k \in \mathcal{P}_{i,j}}D_{k,k}^p - \sum_{k \in \mathcal{N}_{i,j}}D_{k,k}^p.$$

Partition $K$ into equivalence classes $J$ such that $i,j \in K$ are in the same equivalence class iff 
$D_{i,i} = D_{j,j}$. 
Let $\mathcal{J}$ be the set of these equivalence classes and for each $J \in \mathcal{J}$
 define 
$D_{J} : = D_{j,j}$ 
for some $j \in J$.
We have
$$B_{i,j} = \sum_{J \in \mathcal{J}} \sum_{k \in J \cap \mathcal{P}_{i,j}}(D_{k,k})^p - \sum_{k \in J \cap \mathcal{N}_{i,j}}(D_{k,k})^p = \sum_{J \in \mathcal{J}} (\vert J \cap \mathcal{P}_{i,j}\vert - \vert J \cap \mathcal{N}_{i,j}\vert) (D_{J})^p .$$

As the $D_{J}$ are positive and pairwise distinct Lemma \ref{lem:bij_coeff_power} implies that,
(for all $p$ we have $B_{i,j} = 0$) iff
($\vert J \cap \mathcal{P}_{i,j}\vert = \vert J \cap \mathcal{N}_{i,j}\vert$ for all $J$).
By our assumption that this is true for all $i \neq j \in I$ we see that the $s \times \vert J \vert$ matrix $H_{I,J}$ is orthogonal which implies $|J| =s$.
In particular, $J = K$ and $D_{K,K}$%
\end{proof}

\begin{lemma}\label{lem:final_red_to_bipolar}
Let $C$ and $D,O$ satisfy conditions \cond{C1} - \cond{C4}. 
At least one of the following outcomes occurs.
\begin{description}
\item[Outcome 1]
$\eval{C,D,O}$ is $\#\PP$-hard. If $B$ is symmetric, then 
$\eval{B,D^R,O^R}$ is $\#\PP$-hard.
\item[Outcome 2]
Condition \cond{C5} is satisfied.
\end{description}
\end{lemma}
 
\begin{proof}
We will use reduction template $T(1,p',q')$ with $p' = 2p+1$ and $q' = 2q$ for $p,q \in \Nat$. By Lemma \ref{lem:gadget_analysis} this yields a reduction $\eval{C\Delta C, D,O} \Tle \eval{C,D,O}$ such that 
$ C\Delta C $ contains submatrices $B\Delta^C B^T$ and $B^T\Delta^R B$. 
If $B$ is symmetric then it yields the reduction
$\eval{B\Delta^R B, D^R,O^R} \Tle \eval{B,D^R,O^R}$.
We base our argument on $B^T\Delta^R B$ to prove that $O^{R,\mu} = \beta^R_{\mu} I_{r;\Lambda^R}$ for all $\mu \in [m]$ and some $\beta^R_{\mu}$ and $\Lambda^R \subseteq [r]$. The analogous argument on $B\Delta^C B^T$ then yields the result for the submatrices of $O^C$.

Recall that by equation \eqref{eq:CDC_shape} we have
$B^T\Delta^R B = (ww^T) \otimes (H^T\Delta'H)$ for an $r \times r$ diagonal matrix $\Delta'$.
With %
\begin{equation}\label{eq:diagonal_theta_shape}
{\Theta^{[p]}}= \Theta(1,p') = \gamma^R_{p'} \cdot \sum_{\mu=1}^m  v_\mu^{p'} \cdot O^{R,\mu}
\end{equation}
the $r \times r$ diagonal matrix $\Delta'$ is defined by
\begin{eqnarray*}
\Delta' = \sum_{\mu=1}^m v^2_\mu \Delta^{R,\mu}  &=& \left(\sum_{\mu=1}^m v^{p'q' + 2}_\mu D^{R,\mu}\right)\circ {\Theta^{[p]}}^{(q')}
 =  \left(\sum_{\mu=1}^m v^{p'q' + 2}_\mu \alpha_\mu I_r\right)\circ {\Theta^{[p]}}^{(q')}
\end{eqnarray*}
The last equality holds by condition \cond{C4}. Taking
\begin{equation}
 \vartheta := \sum_{\mu=1}^{m} v^{p'q' + 2}_\mu \alpha_\mu \text{ we have } \Delta' = \vartheta {\Theta^{[p]}}^{(q')}. \label{eq:delta_prime_shape} 
\end{equation}

If $\abs{H^T\Delta' H}$ contains a block of rank at least~$2$ then
$\abs{B^T\Delta^R B}$ does. So, if $\abs{H^T\Delta' H}$ contains a block of rank at least~$2$, then
Outcome 1 occurs by Lemma \ref{lem:bipolar_block2_hard}.

By the definition of $H^T\Delta' H$, we have $(H^T\Delta' H)_{i,i} = \tr{\Delta'}$ for all $i \in [r]$ and this trace is non-negative by the definition of $\Delta'$.
Therefore, every principal $2 \times 2$ submatrix of $\abs{H^T\Delta' H}$ has the form
$$
\left(\begin{array}{c c}
       \tr{\Delta'}& \vert (H^T\Delta' H)_{i,j} \vert \\
       \vert (H^T\Delta' H)_{j,i} \vert &  \tr{\Delta'}
      \end{array}\right)
$$
As $H$ is non-singular $\vert (H^T\Delta' H)_{i,j} \vert < \tr{\Delta'}$ for all $i \neq j \in [r]$ and therefore, every such submatrix has non-zero determinant, if $\tr{\Delta'}$ is positive. Furthermore, such a submatrix is part of a block if $(H^T\Delta' H)_{i,j} \neq 0$ and $\tr{\Delta'} \neq 0$. Therefore we have  Outcome 1 if we can show that $(H^T\Delta' H)_{i,j} \neq 0$ and $\tr{\Delta'} \neq 0$ for some $i \neq j \in [r]$ and some $p,q \in \Nat$.

Assume therefore that either $(H^T\Delta' H)_{i,j} = 0$ or $\tr{\Delta'} = 0$ for all $i \neq j \in [r]$ and all $p,q \in \Nat$. The remainder of the proof is to show that in this case condition \cond{C5} is satisfied.

Recall that by equation \eqref{eq:delta_prime_shape} the value $\vartheta$ is positive for all $p,q \in \Nat$. Therefore $\Delta'_{i,i} = 0$ iff ${\Theta^{[p]}}_{i,i} = 0$. 
\begin{claim}\label{cl:diag_zero}
There is a $p_0 \in \Nat$ such that for all $p \ge p_0$ and all $i \in [r]$ we have
$$ {\Theta^{[p]}}_{i,i} = 0 \text{ iff }  
(O^{R,\mu}_{i,i} = 0 \text{ for all } \mu \in [m]). $$
\end{claim}
\begin{clproof}
For each $i \in [r]$ application of Lemma \ref{lem:bij_coeff_new} to equation \eqref{eq:diagonal_theta_shape} yields that there is a $p_i$ such that for all $p \ge p_i$ we have $${\Theta^{[p]}}_{i,i} = 0 \text{ iff }  
(O^{R,\mu}_{i,i} = 0 \text{ for all } \mu \in [m]).$$
The Claim follows with $p_0 := \max\{p_1,\ldots,p_r\}$.
\end{clproof}

\begin{claim}\label{cl:imply_pre_unif} Let $p \in \Nat$. If $(H^T\Delta' H)_{i,j} = 0$ for all $i \neq j \in [r]$ and all $q \in \Nat$ then ${\Theta^{[p]}}^{(2)}$ is pre-uniform.
\end{claim}
\begin{clproof}
Define $\Pi = (\Theta^{[p]})^{(2)}$. Then all entries of $\Pi$ are non-negative and $\Pi^{(q)} = (\Theta^{[p]})^{(q')}$.
With $H^T\Delta'H = \vartheta(H^T{\Theta^{[p]}}^{(q')}H) = \vartheta(H^T\Pi^{(q)}H)$ the claim follows by the Pre-Uniform Diagonal Lemma \ref{lem:pre_unif_diag}.
\end{clproof}
\begin{claim}\label{cl:diag_eq}
There is a $p_{=} \in \Nat$ such that for all $p \ge p_{=}$ and all $i,j \in [r]$ we have 
$${\Theta^{[p]}}^2_{i,i} = {\Theta^{[p]}}^2_{j,j} \text{ iff }
( O^{R,\mu}_{i,i} = O^{R,\mu}_{j,j} \text{ for all } \mu \in [m]).$$
\end{claim}
\begin{clproof}
The backward direction holds for all $p \in \Nat$. %
Fix $i,j \in [r]$. By equation \eqref{eq:diagonal_theta_shape} the equality ${\Theta^{[p]}}^2_{i,i} = {\Theta^{[p]}}^2_{j,j}$ implies
\begin{eqnarray*}
 \left\vert \sum_{\mu=1}^{m}v^{p'}_\mu O^{R,\mu}_{i,i}\right\vert &=&  \left\vert\sum_{\mu=1}^{m}v^{p'}_\mu O^{R,\mu}_{j,j}\right\vert.
\end{eqnarray*}

By Lemma~\ref{lem:bij_coeff_new} we either have a $p_{i,j}$ such that for all $p \ge p_{i,j}$
$$
{\Theta^{[p]}}^2_{i,i} = {\Theta^{[p]}}^2_{j,j} \text{ iff } O^{R,\mu}_{i,i} = O^{R,\mu}_{j,j} \text{ for all } \mu \in [m]
$$
or there is a $p^-_{i,j}$ such that for all $p \ge p^-_{i,j}$
$$
{\Theta^{[p]}}^2_{i,i} = {\Theta^{[p]}}^2_{j,j} \text{ iff } O^{R,\mu}_{i,i} = -O^{R,\mu}_{j,j} \text{ for all } \mu \in [m].
$$
However the second possibility would particularly imply that $O^{R,1}_{i,i} < 0$ for some $i \in [r]$ which was precluded by condition \cond{C2}. Therefore the first possibility holds with $p_{i,j}$. The Claim now follows with
$p_{=} = \max\{ p_{i,j} \mid i,j \in [r]\}$.
\end{clproof}

These claims now enable us to finish the proof. Note first that condition \cond{C5} is satisfied with $\Lambda^R = \emptyset$ if $O^{R,\mu}_{i,i} = 0$ for all $\mu \in [m]$ and $i \in [r]$.

Assume therefore that $O^R$ has non-zero diagonal entries. Fix values $p_0,p_{=} \in \Nat$ according to Claims \ref{cl:diag_zero} and \ref{cl:diag_eq} and define $p = \max\{p_0,p_{=}\}$. This implies $\tr{\Delta'} \neq 0$. To see this note that there is some $i \in [r]$ and some $\mu \in [m]$ such that $O^{R,\mu}_{i,i} \neq 0$ which by our choice of $p$ implies ${\Theta^{[p]}}_{i,i} \neq 0$. 

By our assumption, $\tr{\Delta'}\neq 0$ implies $(H^T\Delta' H)_{i,j} = 0$ for all $i \neq j \in [r]$ and all $q \in \Nat$ which by Claim~\ref{cl:imply_pre_unif} yields pre-uniformity of ${\Theta^{[p]}}^{(2)}$.

Define $\Lambda^R := \mset{i \in [r] \,\vert \, {\Theta^{[p]}}_{i,i} \neq 0}$. 
By the pre-uniformity of ${\Theta^{[p]}}^{(2)}$ Claim~\ref{cl:diag_eq} implies that, for each $\mu \in [m]$ and every $i \in \Lambda^R$ there is a $\beta^R_\mu$ such that $O^{R,\mu}_{i,i} = \beta^R_\mu$.
Furthermore, Claim~\ref{cl:diag_zero} implies that for each $\mu \in [m]$ and every $i \in [r] \setminus \Lambda^R$ we have $O^{R,\mu}_{i,i} = 0$. This finishes the proof.
\end{proof}

\subsubsection{Putting everything together.}

We are now able to prove Lemma \ref{lem:decomp}

\begin{proof}[Proof of Lemma \ref{lem:decomp}]

{\bf Bipartite $A$.} 
Consider first the case in which $A$ is bipartite. By Lemmas %
\ref{lem:conditions_C1_C2}, \ref{lem:two_properties} and Lemma \ref{lem:final_red_to_bipolar}, the evaluation problem $\eval{A}$ is $\#\PP$-hard unless $\eval{A} \Tequiv \eval{C,D,O}$ for matrices $C,D,O$ satisfying conditions \cond{C1}-\cond{C5}.

$C$ is a symmetric bipartite matrix with underlying block $B$.
Conditions \cond{C1}-\cond{C5} imply that $B = vw^T \otimes H$, $D^R = D^{R''} \otimes I_r$, $D^C = D^{C''} \otimes I_r$ and $O^R = O^{R''} \otimes I_{r;\Lambda^R}$,  $O^C = O^{C''} \otimes I_{r;\Lambda^C}$ for diagonal $m \times m$ matrices $D^{R''}$ and $O^{R''}$ defined by $D^{R''}_{\mu,\mu} = \alpha^R_\mu$ and $O^{R''}_{\mu,\mu} = \beta^R_\mu$ for all $\mu \in [m]$. The $n \times n$ diagonal matrices $D^{C''}$ and $O^{C''}$ are defined analogously in terms of $\alpha^C_\nu$ and $\beta^C_\nu$.
Then for 
$$
D'' = \left(\begin{array}{c c} D^{R''} & 0 \\ 0 & D^{C''} \end{array}\right)
\text{ and }
O'' = \left(\begin{array}{c c} O^{R''} & 0 \\ 0 & O^{C''} \end{array}\right)
\text{ and }
C''  = \left(\begin{array}{c c} 0 & vw^T \\ wv^T & 0 \end{array}\right)
$$
the problem $\eval{C'',D'',O''}$ is polynomial time computable by Corollary \ref{cor:rank1_bipolar_FP}.
Note that 
$D+O$ and
$D-O$ are non-negative by condition \cond{C2}. Hence with
$M, \Lambda$ being the bipartisation of $H$,$\Lambda^R$ and $\Lambda^C$ we have
$\eval{C,D,O} \Tequiv \eval{M,I_{2r}, I_{2r;\Lambda}}$ by Lemma \ref{lem:bipolar_bipartite_tensor_decomp}.

{\bf Non-Bipartite $A$}
Now suppose that $A$ is not bipartite.
Let $M$ be the bipartisation of $A$. Recall that this is a matrix of the form
$$
M = \left( \begin{array}{c c} 0 & A \\ A & 0 \end{array}\right).
$$
By Lemmas 
\ref{lem:conditions_C1_C2}, \ref{lem:two_properties} and Lemma \ref{lem:final_red_to_bipolar} the evaluation problem $\eval{A}$ is $\#\PP$-hard unless 
there are matrices $C,D,O$ 
with block $B$ underlying $C$
satisfying conditions \cond{C1}-\cond{C5}
such that
$\eval{A} \Tequiv \eval{B,D^R,O^R}$

Conditions \cond{C1}-\cond{C5} imply that $B = vv^T \otimes H$, 
$D^R = D^{R''} \otimes I_r$, 
and $O^R = O^{R''} \otimes I_{r;\Lambda^R}$ 
  for diagonal $m \times m$ matrices $D^{R''}$ and $O^{R''}$ defined by $D^{R''}_{\mu,\mu} = \alpha^R_\mu$ and $O^{R''}_{\mu,\mu} = \beta^R_\mu$ for all $\mu \in [m]$.

Then  
the problem $\eval{v v^T,D^{R''},O^{R''}}$ is polynomial time computable by Corollary \ref{cor:rank1_bipolar_FP}.
Hence 
we have
$\eval{B,D^R,O^R} \Tequiv \eval{A,I_{r}, I_{r;\Lambda^R}}$ 
by Corollary \ref{lem:1may}.
 
{\bf Finishing the Proof:}
It remains to state the polynomial time computability. Note that conditions \cond{C2} \cond{C5} are straightforwardly checkable in polynomial time and for \cond{C1} this follows from Lemma \ref{lem:conditions_C1_C2}.
\end{proof}

\bibliographystyle{plain}
\bibliography{partftn}

\end{document}